\documentclass[a4paper,11pt]{article}
\pdfoutput=1 

\usepackage{jcappub} 
\usepackage{mathtools}
\usepackage[dvipsnames]{xcolor}

\newcommand{\cm}{\ensuremath{\,\textrm{cm}}}
\newcommand{\s}{\ensuremath{\,\textrm{s}}}

\newcommand{\msun}{M_\odot}

\usepackage{array}
\newcolumntype{P}[1]{>{\centering\arraybackslash}p{#1}}
\newcolumntype{M}[1]{>{\centering\arraybackslash}m{#1}}
\usepackage{graphicx}
\usepackage{longtable}
\usepackage{verbatim}
\usepackage[utf8]{inputenc}

\usepackage{amsmath}
\usepackage{amsfonts}
\usepackage{amsxtra}
\usepackage{hyperref}
\usepackage{amssymb}
\usepackage{dsfont}
\usepackage{graphicx,epsfig}
\usepackage{natbib}
\usepackage{xspace}

\usepackage{cleveref}
\crefname{section}{§}{§§}
\usepackage[T1]{fontenc} 
\graphicspath{ {Figures/} }

\title{\boldmath Unidentified Gamma-ray Sources as Targets for Indirect Dark Matter Detection with the \textit{Fermi}-Large Area Telescope}

\author[a,b]{Javier Coronado-Bl\'azquez}
\author[a,b]{Miguel A. S\'anchez-Conde}
\author[c]{Alberto Dom\'inguez}
\author[a,b]{Alejandra Aguirre-Santaella}
\author[d,e]{Mattia Di Mauro}
\author[d,f]{N\'estor Mirabal}
\author[c]{Daniel Nieto}
\author[g]{Eric Charles}

\affiliation[a]{Instituto de Física Teórica UAM-CSIC,\\Universidad Autónoma de Madrid, C/ Nicolás Cabrera, 13-15, 28049 Madrid, Spain}
\affiliation[b]{Departamento de Física Teórica, M-15,\\Universidad Autónoma de Madrid, E-28049 Madrid, Spain}
\affiliation[c]{Grupo de Altas Energías, Facultad de Ciencias F\'isicas, Universidad Complutense de Madrid, Plaza de las Ciencias 1, 28040 Madrid, Spain}
\affiliation[d]{NASA Goddard Space Flight Center, Greenbelt, MD 20771, USA}
\affiliation[e]{CRESST, Catholic University of America, Department of Physics, Washington DC 20064, USA}
\affiliation[f]{CRESST/CSST/Department of Physics, UMBC, Baltimore, MD 21250, USA}
\affiliation[g]{W. W. Hansen Experimental Physics Laboratory, Kavli Institute for Particle Astrophysics and Cosmology, Department of Physics and SLAC National Accelerator Laboratory, Stanford University, Stanford, CA 94305, USA}

\emailAdd{javier.coronado@uam.es}
\emailAdd{miguel.sanchezconde@uam.es}

\abstract{One of the predictions of the $\Lambda$CDM cosmological framework is the hierarchical formation of structure, giving rise to dark matter (DM) halos and subhalos. When the latter are massive enough they retain gas (i.e., baryons) and become visible. This is the case of the dwarf satellite galaxies in the Milky Way (MW). Below a certain mass, halos may not accumulate significant amounts of baryons and remain completely dark. However, if DM particles are {\it Weakly Interacting Massive Particles} (WIMPs), we expect them to annihilate in subhalos, producing gamma rays which can be detected with the \textit{Fermi} satellite. Using the three most recent point-source \textit{Fermi Large Area Telescope} (LAT) catalogs (3FGL, 2FHL and 3FHL), we search for DM subhalo candidates among the unidentified sources, i.e., sources with no firm association to a known astrophysical object. We apply several selection criteria based on the expected properties of the DM-induced emission from subhalos, which allow us to significantly reduce the list of potential candidates. Then, by characterizing the minimum detection flux of the instrument and comparing our sample to predictions from the Via Lactea II (VL-II) N-body cosmological simulation, we place conservative and robust constraints on the $\langle\sigma v\rangle-m_{DM}$ parameter space. For annihilation via the  $\tau^+\tau^-$ channel, we put an upper limit of $4\times 10^{-26}~(5\times 10^{-25})~\cm^3\s^{-1}$ for a mass of 10 (100) GeV. A critical improvement over previous treatments is the repopulation we made to include low-mass subhalos below the VL-II mass resolution.
With more advanced subhalo candidate filtering the sensitivity reach of our method can potentially improve these constraints by a factor 3 (2) for $\tau^+\tau^-$ ($b \bar{b}$) channel.}

\begin{document}
\maketitle
\flushbottom

\section{Introduction}
\label{sec:intro}

Mounting evidence has been found that about 85\% of all matter in the Universe is non-baryonic, this is the so-called dark matter (DM) \cite{Planck15,GarrettDuda09,Bertone+05,Freese09}.  The composition of this DM remains unknown, and is one of the most important open questions in modern physics.

N-body numerical simulations reveal that DM structures form hierarchically in a bottom-up scenario, with DM particles first collapsing into small gravitationally bound systems known as halos, and then forming more massive halos through a history of mergers. As a consequence, individual DM halos contain a very large number of smaller subhalos \cite{Madau2008}.

At large scales, these simulations have been able to test with great success the predictions of the $\Lambda$CDM cosmological model. Nevertheless, when dealing with individual DM halos and their corresponding subhalos populations, such as our Galaxy, the situation is more uncertain, as the simulation resolution is limited and does not resolve the full range of possible subhalo masses. The subhalo mass function (SHMF) is found to be of the form $dN/dM\propto M^{-n}$ in the resolved mass range, where $n\in\left[1.9,2.0\right]$, depending on the specific simulation \cite{Springel+08, vlii_paper}. Together with the current resolution limit of MW-size simulations, this implies not only that the number of subhalos increases dramatically at lower masses, but also that the number of subhalos below the current resolution limit of Milky Way (MW) size simulations ($\sim 10^5$ M\textsubscript{\(\odot\)}, while the MW halo is $\sim 10^{12}$ M\textsubscript{\(\odot\)}) is still very uncertain.

For DM candidates with weak-scale masses and interactions, subhalos with masses from $10^{-11}-10^{-3}$ M\textsubscript{\(\odot\)} (depending on the model) up to roughly $10^{10}$ M\textsubscript{\(\odot\)} are expected to exist in a galaxy like our own \cite{Bertschinger06,Profumo+06,Bringmann09}. Dwarf spheroidal satellite galaxies (dSphs), such as Draco, are an example of the most massive members of this population \cite{Walker13}.
Yet, these dSphs are exceptional objects, in that they are massive enough to retain baryons (i.e., gas) and form stars.  Conversely, the vast majority of the Galactic DM subhalos are not expected to host baryons and therefore remain completely dark \cite{Gao2004}. Given their much larger number density, many of these small subhalos will be much closer to the Earth than the bigger ones, making them potentially interesting for dark matter searches.

Should the \textit{Weakly Interacting Massive Particle} (WIMP) DM model be correct (see, e.g., \cite{Roszkowski+17,Bertone10} for a review), these objects may be detectable in the gamma-ray data.  WIMPs can achieve the correct relic DM abundance (the so-called "WIMP miracle") through self-annihilation in the early Universe.  Self-annihilation of WIMPs gives rise to a Standard Model (SM) particle-antiparticle pair which, among other possible subsequent by-products, typically yields gamma-ray photons.  The ongoing self-annihilation of WIMPs in subhalos could be bright enough to be detectable.

Since its launch in 2008, the Large Area Telescope on board the NASA \textit{Fermi Gamma-ray Space Telescope} (\textit{Fermi}-LAT) has been surveying the sky searching for gamma-ray sources \cite{fermi_instrument_paper}. The \textit{Fermi}-LAT is a pair conversion telescope designed to observe the energy band from 20 MeV to greater than 300 GeV.  Several point-source {\it Fermi}-LAT catalogs have been released and contain hundreds to thousands of gamma-ray objects, many of them previously unknown \cite{3FGL_paper,2FHL_paper,3FHL_paper}.  The various catalogs cover different energy ranges and exposure times, and each was constructed with the best available astrophysical diffuse emission model and instrumental response functions (IRFs).

Many groups have used \textit{Fermi}-LAT data to constrain the WIMP DM parameter space: for example through observations of the diffuse extragalactic emission \cite{Ajello+15}, galaxy clusters \cite{fermi_cluster_paper}, gamma-ray lines \cite{Weniger12,fermi_gamma_lines_paper} and the previously mentioned dSphs \citep{dsphs_paper}, or to claim possible detection of DM in the Galactic Center (GC) \cite{fermi_gc_paper16,fermi_gc_paper17}.  These gamma-ray DM searches are complemented at larger WIMP masses by ground-based imaging atmospheric Cherenkov telescopes (IACTs) such as MAGIC, VERITAS and H.E.S.S. \cite{dm_magic_paper,veritas_paper,dm_hess_paper}.

An important fraction of objects in the \textit{Fermi}-LAT catalogs are unidentified sources (unIDs), i.e., objects with no clear single association or counterpart, to either a known object identified at other wavelengths, or to a known source type emitting only in gamma rays (such as certain pulsars).\footnote{Is important to note that in many cases there are actually multiple possible associations.  In all of these catalogs if a source was not uniquely associated it was categorized as unID.}  There is the exciting possibility that some of these unIDs may actually be DM subhalos. In this work, we will search for DM subhalos in three of the most recent \textit{Fermi}-LAT catalogs: namely the 3FGL \cite{3FGL_paper}, 2FHL \cite{2FHL_paper} and 3FHL \cite{3FHL_paper}. The number and fraction of the unIDs in each catalog is different; there are 1010 unIDs (33\% if the full catalog) in the 3FGL, 48 (13\%) in the 2FHL and 177 (11\%) in the 3FHL. Both the 3FGL and 2FHL catalogs have been used in previous works \cite{Bertoni+15, Bertoni+16, Calore+17, Schoonenberg+16,HooperWitte17, dsphs_paper}, while other works utilized previous catalogs \cite{BerlinHooper14, Zechlin+12, ZechlinHorns12, Belikov2012, BuckleyHooper10, fermi_dm_satellites_paper}. Since we do not know the distance to these unIDs, for a signal, the DM subhalo mass would be degenerate with distance, i.e., the same flux could be produced either by a massive, distant DM subhalo or by a less massive but closer one.  This implies that very nearby, low-mass subhalos, may potentially be excellent DM targets.

However, as previously mentioned, there is currently no simulation able to resolve the entire Galactic subhalo population. Thus, when we search for such objects in {\it Fermi}-LAT data, one of the biggest challenges is to find a good and reliable characterization of the low-mass subhalo population that allows us to make realistic predictions of the expected annihilation fluxes.  In our work, we will use a {\it repopulation} of the Via Lactea II (VL-II) N-body simulation with low-mass subhalos below the resolution limit \cite{aguirre2018}. We will do so by taking into account what is found above this limit for the abundance and distribution of subhalos, and by adopting state-of-the-art models to describe their structural properties \cite{Moline+17}.

We first perform an exhaustive {\it filtering} of unID sources from the LAT catalogs, based on the expected DM subhalo properties, in order to find subhalo candidates.   With this shortlist of potential DM subhalos, we set constraints on the DM annihilation cross section by comparing the number of observed subhalo candidates with predictions of N-body simulations.

With respect to previous efforts \cite{HooperWitte17, Bertoni+15, Bertoni+16, Schoonenberg+16, BerlinHooper14, BuckleyHooper10, fermi_dm_satellites_paper, Zechlin+12, Belikov2012, ZechlinHorns12, Calore+17}, our work includes a new catalog (3FHL), an N-body simulation repopulation, a precise characterization of the instrument sensitivity to DM subhalos, and a more extensive filtering of candidates in the catalogs based on diverse criteria to reject all those unIDs not being compatible with DM.

The structure of this paper is as follows. In Section \ref{sec:wimp_dm}, we describe the expected gamma-ray flux from annihilations in DM subhalos, covering the details of the repopulation of the VL-II N-body simulation with the low-mass subhalos, and how the expected J-factor is computed for each subhalo.  Section \ref{sec:catalogs} describes the \textit{Fermi}-LAT catalogs we considered, and discusses the criteria used to reject unIDs as being potential DM subhalos. The computation of the minimum detectable DM subhalo flux is described in Section \ref{sec:minimum_flux}. In Section \ref{sec:constraints}, we place constraints on the DM parameter space by comparing the number of unIDs that survive our selection criteria with the number of expected subhalos as obtained from our repopulated N-body simulation. We further discuss the impact of each of the rejection criteria used on the DM limits, and present the sensitivity reach of the method. We conclude in Section \ref{sec:conclusions}.

\section{Predictions of the gamma-ray annihilation flux from subhalos}
\label{sec:wimp_dm}
Within the WIMP model, the expected gamma-ray flux can be expressed \cite{Evans2004,Bergstroem1998} as:

\begin{equation}
F\left(E>E_{th}\right)=J\cdot f_{pp}\left(E>E_{th}\right),
\label{eq:flux}
\end{equation}

\noindent where $E_{th}$ is the threshold energy (set by the instrument), $J$ is the J-factor, which encloses all the astrophysical considerations, and $f_{pp}$ is the particle physics factor, which contains information on the underlying DM particle theoretical model (i.e., on the specific interaction properties of the DM particle considered).

The full expression for the J-factor is:

\begin{equation}
\label{eq:j_factor}
J=\frac{1}{D^2}\int_{\Delta\Omega}d\Omega\int_{l.o.s} \rho_{DM}^2\left[r\left(\lambda\right)\right]d\lambda,
\end{equation}

\noindent where $D$ is the distance to the target, the first integral is performed over the solid angle of observation ($\Omega$), the second one along the line of sight (l.o.s, $\lambda$), and $\rho_{DM}$ is the dark matter density profile of the object under consideration (in this paper that would be a single subhalo). Interestingly, subhalos are known to be more concentrated than field halos\footnote{Those that do not reside inside any other larger halo. Typically an isolation criterion is also applied (i.e. not having a massive neighbor located within a given distance), in which case they are also known as isolated halos.} of the same mass, e.g.~\cite{Moline+17}. Further discussion on this and other considerations pertaining the DM density profile of subhalos will be given in \cref{sec:jfactors}.

The particle physics factor, assuming Majorana DM, is given by:

\begin{equation}
\label{eq:f_pp}
f_{pp}\left(E_{th}\right)=\frac{1}{4\pi}\frac{\langle\sigma v\rangle}{2m_{\chi}^2}\sum_{f}B_{f}\int_{E_{th}}^E\frac{dN_{f}}{dE}dE,
\end{equation}

\noindent where $E_{th}$ is the threshold energy, the upper limit is $E=m_{\chi}$, the subscript $f$ refers to the annihilation channel, $B_f$ is the branching ratio to that channel (we will take $B_f$ as one for each of the considered channels, so \cref{eq:f_pp} will actually not be a sum), $dN_{f}/dE$ is the differential spectrum of gamma rays from the annihilation of a pair of DM particles via the channel {\it f}, $\langle\sigma v\rangle$ is the velocity-averaged annihilation cross section, and $m_{\chi}$ is the DM particle mass.

\subsection{Annihilation spectra}
\label{sec:ann_spectra}
In our work, we use the PPPC 4 DM ID tables to compute the DM annihilation spectra for different channels and DM masses \cite{Cirelli+12}. These tables were constructed using the \verb|PYTHIA 8| \cite{pythia8_paper} event generator to model the hadronization processes (including electroweak corrections) and obtain the $dN_{f}/dE$ spectra. The tables provide the annihilation spectra for DM masses ranging from 5 GeV up to 100 TeV.

We parametrize the DM annihilation spectra with a super-exponential cutoff power law, as done, for instance, by Ref. \cite{Calore+17}:

\begin{equation}\label{eq:parametric_spectra}
\frac{dN_{DM}}{dE}\left(E\right)=K\left(\frac{E}{E_{0}}\right)^{-\Gamma}e^{-\left(\frac{E}{E_{cut}}\right)^\beta},
\end{equation}

\noindent where $K$ is a prefactor, $E_{0}=10^3$ MeV is the pivot energy, $\Gamma$ is the spectral photon index, $E_{cut}$ is the cutoff energy and $\beta$ is the curvature index. To obtain these parameters for each channel and DM mass, we fit the up to 179 available spectral points from the PPPC 4 DM tables.  In each fit we leave the parameters $K, \Gamma, \beta$ and $E_{cut}$ free.

\begin{figure}[!ht]
\centering
\includegraphics[height=5.5cm]{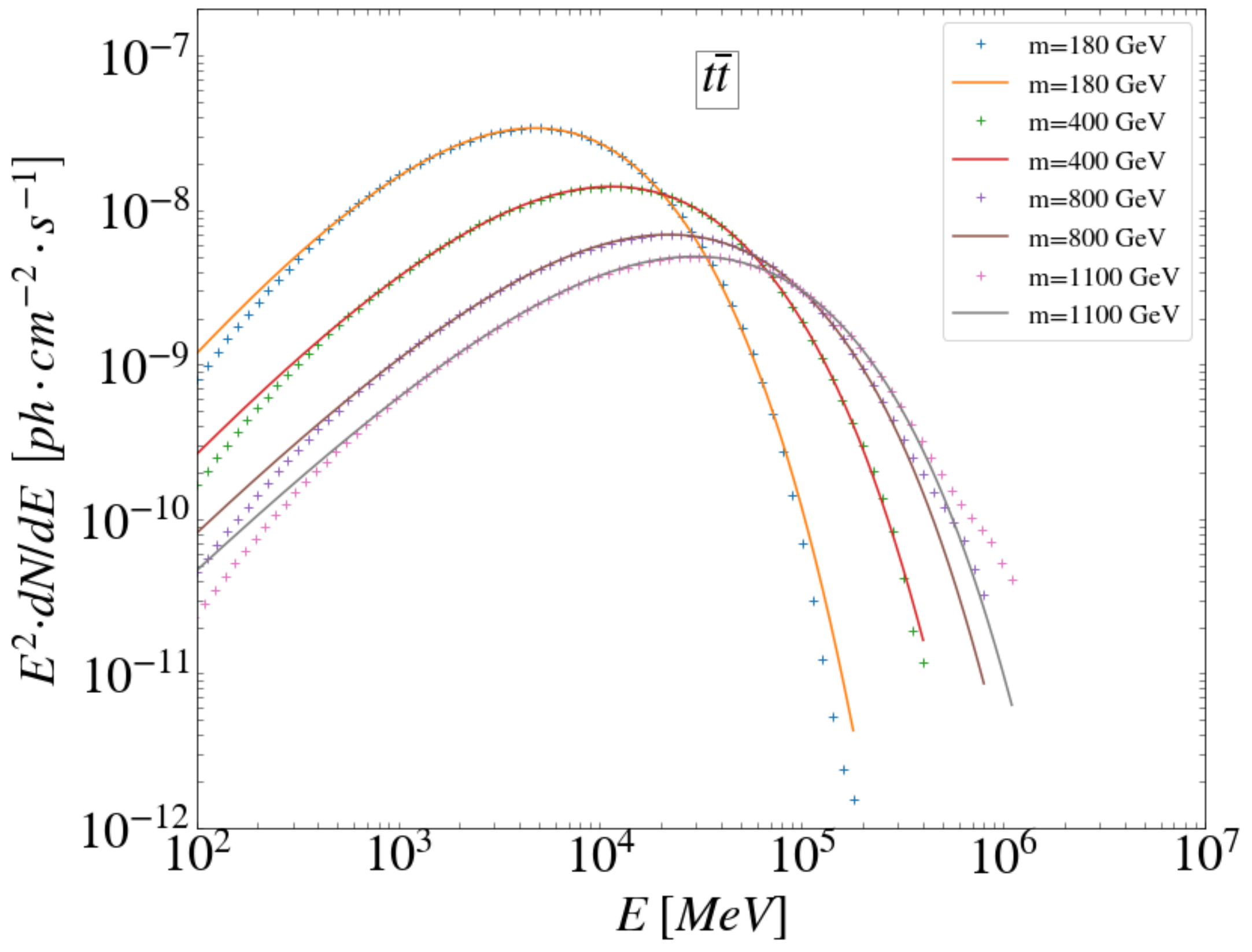}
\includegraphics[height=5.5cm]{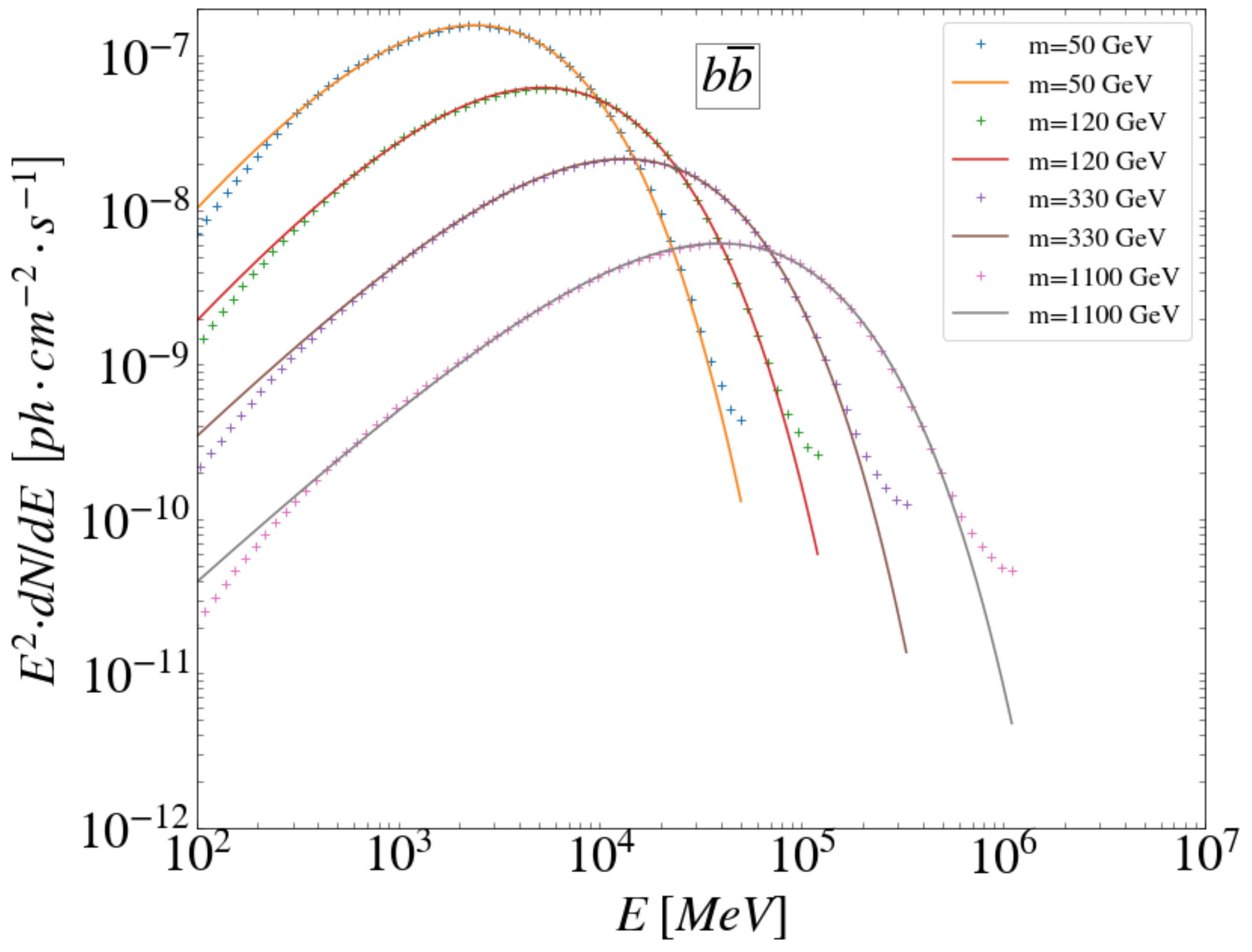}
\vfill
\includegraphics[height=5.5cm]{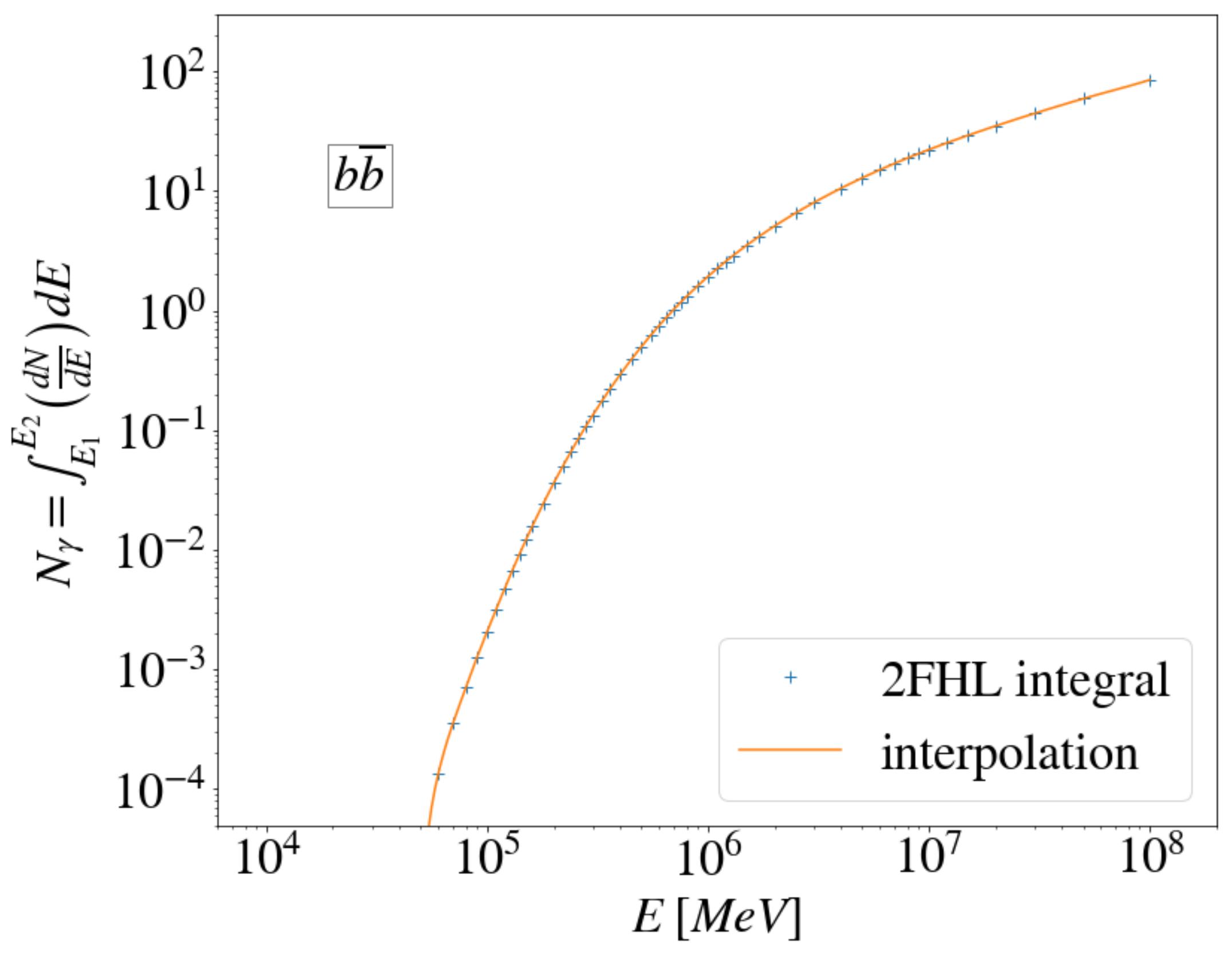}
\includegraphics[height=5.5cm]{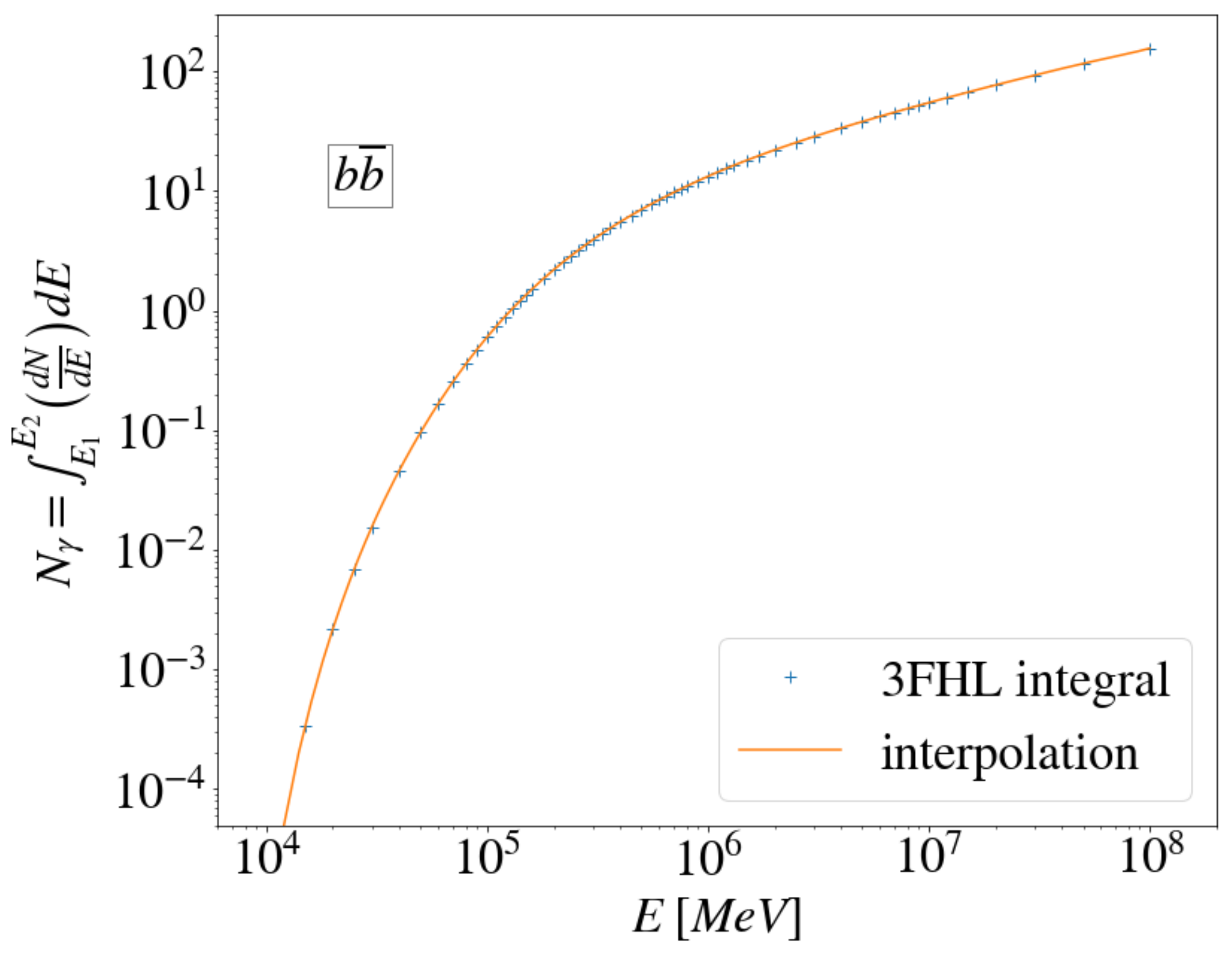}
\caption{\textbf{Upper panel}: DM annihilation spectra with overlaid eq. \ref{eq:parametric_spectra} parametric fit and interpolation for $t\overline{t}$ (left panel) and $b\overline{b}$ (right panel) for various DM masses. The points from \cite{Cirelli+12} are marked by crosses, while the fit is plotted as solid lines. Note that in the case of $t\overline{t}$ the plotted masses are larger because $m_{t}=172$ GeV. \textbf{Lower panel}: Integrated DM annihilation spectra $\left(N_{\gamma}\right)$ for 2FHL setup (left) and 3FHL (right), for $b\overline{b}$. Both setups differ in the threshold energy. Note that $N_{\gamma}=0$ if $E<50$ GeV ($E<10$ GeV) in the case of the 2FHL (3FHL) setup, where $E=E_{th}$. Blue crosses are the values of the integral, and orange line is the interpolation between the values of the integral, which is used to compute the DM constraints as described in \cref{sec:constraints}.}
\label{fig:spectra_cirelli}
\end{figure}

We perform this parametrization for each of the tabulated DM annihilation channels and masses. An example of DM tabulated spectrum and its corresponding parametric fit is shown in the upper panel of Figure \ref{fig:spectra_cirelli}. In \cref{sec:minimum_flux} we will use these parametrizations to facilitate computing the sensitivity of the LAT to DM sub-halos. Although the parametrization is not perfect, it is accurate to a few percent.

To compute the annihilation flux as given by \cref{eq:flux,eq:f_pp}, we integrate the DM annihilation spectrum above $E_{th}$:

\begin{equation}
\label{eq:integrated_spectra}
N_{\gamma}=\int_{E_{th}}^E\left(\frac{dN}{dE}\right)dE,
\end{equation}

\noindent where $N_\gamma$ is the number of gamma-rays per annihilation produced in the relevant energy range and the upper limit of the integral is $E=m_{DM}$. Scanning over all the tabulated masses from 5 GeV up to 100 TeV, we obtain the value of this integral for each mass.  These value can then be interpolated to find $N_\gamma$ for any given mass.  An example of this interpolation, which will be useful when computing DM limits in \cref{sec:constraints}, is plotted in the lower panel of Figure \ref{fig:spectra_cirelli}.

\subsection{Subhalo J-factors}
\label{sec:jfactors}
N-body cosmological simulations have become a powerful tool to study the formation and evolution of cold DM halos and their substructure. In particular, state-of-the-art N-body simulations of MW-size halos have provided the most valuable and accurate information on the properties of the present-day subhalo population \cite{Springel+08,vlii_paper,GHALO,Elvis}. Among these properties, the radial distribution within the MW halo, abundance, and internal subhalo structure are the most relevant ones when computing subhalo J-factors. In our work, we use publicly available results from the VL-II DM-only simulation at redshift zero to perform this task.\footnote{\url{http://www.ucolick.org/\~diemand/vl/}} The VL-II simulation follows the formation and evolution of a MW-like halo in a WMAP3 $\Lambda$CDM universe down to present time with a superb particle resolution. Specifically, the VL-II simulation algorithm tracks more than one billion DM particles, each with a mass of $4.1\times10^3~M_{\odot}$, to resolve the region within the virial radius of the simulated object, i.e., 402 kpc. More than fifty thousand subhalos are identified in the simulated VL-II volume, with masses from 10\% of the parent halo mass down to $\sim10^5~\msun$.  This means that VL-II is able to resolve the complex dynamics, merging and accretion histories of structures roughly spanning over 6-7 decades in mass. 

Although this is an outstanding result, the masses of the lightest VL-II subhalos are far above the minimum halo mass that is predicted to exist in standard $\Lambda$CDM cosmology, e.g. of the order of $10^{-6}\msun$ if the DM is made of WIMPs \cite{Profumo+06}. These {\it microhalos} are a natural expectation in the hierarchical structure formation scenario being at work in $\Lambda$CDM, and most of them, if not all, should survive to the present time given their early formation times and high DM concentrations, e.g.~\cite{Diemand2005,Anderhalden2013,Ishiyama2014}. As previously discussed, in making our predictions of the J-factors of the Galactic subhalo population it is important to include subhalos below the VL-II mass resolution limit, as some of them could yield large annihilation fluxes at Earth if they are nearby.  In fact, some of the lightest VL-II subhalos exhibited some of the highest J-factors of the entire subhalo population.

With this in mind, we {\it repopulate} the original VL-II simulation with subhalos well below its mass resolution limit.  To do this we first characterize the SHMF in the mass range well resolved in VL-II data, which we find to be above $M_{cut}=5\times10^6~\msun$. This value corresponds to the mass at which a departure from the power-law behavior of the SHMF is observed, i.e., below $M_{cut}$ the simulation is not complete as it starts to miss subhalos due to lack of resolution.  We obtain a power-law index of $n = 1.901 \pm 0.004$ for the SHMF above $M_{cut}$, which is slightly different than the result in \cite{jurg1}, where a value of $n = 1.97 \pm 0.03$ was reported, but it is still fully with other simulations \cite{Springel+08,garrison-kimmel} as well as with theoretical (hardly unavoidable) expectations from the Press–Schechter theory of structure formation in $\Lambda$CDM \cite{Press1974,shmf2,shmf3}. As for the  subhalo radial distribution, we find that the following parametrization:

\begin{equation}
\label{eq:radialdistrib}
n_{sub}(r)=  \left( \frac{r}{R_0}\right)^a \exp \left( -b\, \frac{r - R_0}{R_0} \right),
\end{equation}

\noindent with best-fit parameters $a =0.94 \pm 0.15$, $b = 10.0 \pm 0.4$, $R_0 = 785 \pm 60$ kpc, provides a reasonably good fit for the number of subhalos as a function of distance to the host halo center, $r$, within its virial radius. We note that we used {\it all} subhalos in the original simulation in our fit. We also note that eq.~(\ref{eq:radialdistrib}) does not correspond to any of the parametrizations traditionally used in the literature \cite{Pieri2011}, which we found to provide poorer fits to the simulation data. In particular, we note that the subhalo radial distribution given by Eq.~(\ref{eq:radialdistrib}) does not almost provide subhalos within the inner $\sim15-20$ kpc of the Galaxy, as expected from subhalo disruption due to tidal interactions with the host.

We then include additional subhalos with masses below $M_{cut}$ in the population by assuming that their abundance obeys the power-law found for the SHMF above $M_{cut}$, as expected, and that they distribute following Eq.~(\ref{eq:radialdistrib}). After we draw subhalos from both our derived SHMF and radial distribution, we place the Sun at a different random position along its Galactic orbit (which we assume to be circular and with a radius of 8.5 kpc) and compute the subhalo J-factors.  Finally, in order to derive statistically meaningful results, we repeat the whole exercise to produce 1000 realizations of the repopulated VL-II simulation. These realizations perfectly resemble the original VL-II data above $M_{cut}$ but now include subhalos down to $10^3~\msun$ as well. The latter number is found to be a good compromise between computing effort and relevance of low-mass subhalos for the purposes of this work, as we find that less massive subhalos are already not expected to be among the brightest ones.\footnote{We also apply the Roche criterium \cite{Binney2008} in our repopulated simulations in order to remove any subhalo that might have been included but should not have been. We find that, by doing so, we only remove, on average, an additional 1\% of subhalos within 10 kpc, meaning that our proposed fit in Eq.~(\ref{eq:radialdistrib}) is, indeed, a good representation of the actual VL-II subhalo radial distribution.}

The last ingredient needed for the computation of the subhalo J-factors is the description of their structural properties, for which we follow Ref.~\cite{Moline+17}. In that work, the inner subhalo structure is conveniently codified in terms of an alternative formulation of the so-called {\it concentration}\footnote{Formally defined as the ratio between the halo virial radius and its scale radius, i.e. the radius at which the logarithmic slope of the DM density profile is equal to -2.}, which does not depend on the assumed subhalo DM density profile as done previously.  Using data from both the VL-II and ELVIS simulations, the authors of Ref.~\cite{Moline+17} find that, when compared to field halos of the same mass, subhalos are typically a factor 2-3 more concentrated. They also find an important dependence of subhalo concentrations on their galactocentric distance: the closer the subhalo to the host halo center the more concentrated it is. These effects are mainly driven by the impact of tidal stripping on the subhalo population. We use the parametrizations in Ref.~\cite{Moline+17} to assign concentration values to each subhalo in the repopulated VL-II 1000 realizations. As we will see later below, the higher concentration values found in Ref.~\cite{Moline+17} will have a critical and direct impact on the J-factor values, as the latter roughly scale as the third power of the concentration.

\begin{figure}[!ht]
\centering
\includegraphics[height=8cm]{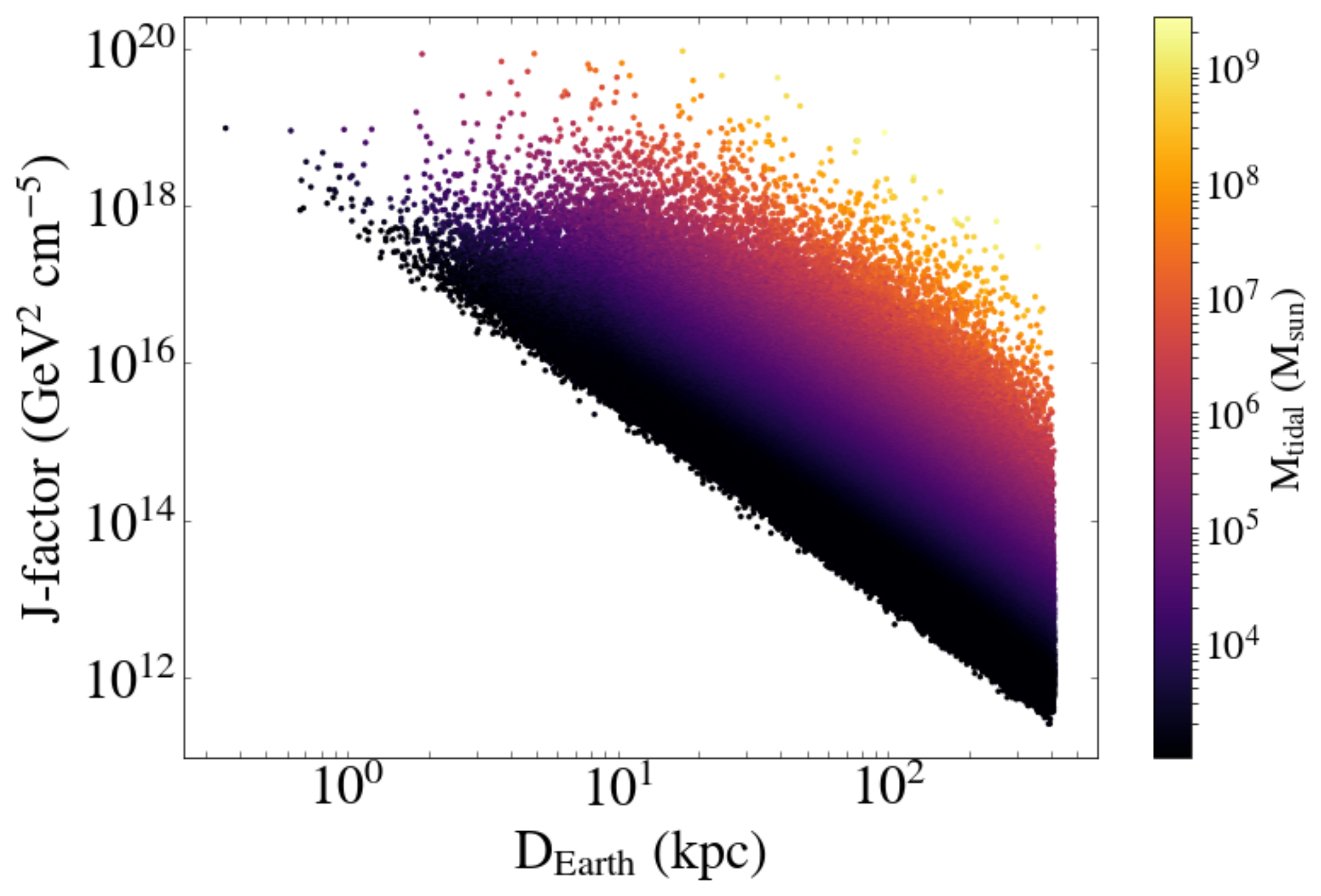}
\caption{Subhalo J-factors as a function of distance to the Earth for all subhalos in a random realization of the repopulated VL-II. The repopulation includes low-mass subhalos down to $10^3\msun$; see text for details. The color represents the subhalo mass in $M_\odot$.}
\label{fig:Jfactors}
\end{figure}

Our studies of VL-II subhalo abundance, radial distribution and structural properties, as well as our repopulation work with low-mass subhalos down to $10^3\msun$, finally allows us to derive the J-factors associated to the Galactic subhalo population, which, expanding upon Eq.~(\ref{eq:j_factor}), we compute using the following expression \cite{Moline+17}:

\begin{equation} \label{eq:jfacale}
\hspace{-0.3cm} J_{T}=\frac{1}{D^{2}}\frac{M_{sub}\,c^{\,3}_{sub}(M_{sub})}{[f(c_{sub}(M_{sub}))]^{2}}\frac{200\,\rho_{crit}}{9}\left(1-\frac{1}{(1+r_t(M_{sub},D)/r_s(M_{sub}))^{3}}\right),\end{equation} 

\noindent where $\rho_{crit}$ is the critical density of the Universe, $M_{sub}$ and $c_{sub}$ are, respectively, the mass and concentration of the subhalo, $r_t$ and $r_s$ refer to its tidal and scale radius, and $ f(c) = \log(1+c) - c/(1+c)$. Note that the above equation refers to the integrated J-factor of subhalos within their scale radii.\footnote{We note that this is a conservative estimate as it implicitly assumes that all subhalos are truncated at the scale radius due to tidal stripping, while this will only be the case for those in the host's innermost regions.} Our J-factor results are summarized in Figure \ref{fig:Jfactors}, which shows the J-factor of all subhalos in a random realization as a function of their distance to the Earth. The subhalo mass is also given by the color scale.  As can be seen, a number of the lighter subhalos in the repopulation yield some of the largest J-factor values that we inferred for the whole subhalo population.

With respect to previous work, e.g.~\cite{Calore+17}, the main difference is the significantly higher J-factors associated to low-mass subhalos in our analysis. We note that this is the first work on subhalo detectability where subhalos are assigned with ``proper'' subhalo concentrations that take into account both the fact that subhalos are more concentrated than field halos of the same mass, and the dependence of subhalo concentration with distance to host halo center.  Also, we note that although Ref.~\cite{Calore+17} used both a DM-only and a hydrodynamical simulation, their simulations did not include low-mass subhalos below $\sim10^7$ M\textsubscript{\(\odot\)}. However, the subhalo J-factor values presented in this section come with some important caveats. It is likely that the main limitation currently affecting every work on this topic is the knowledge of the actual abundance of low-mass subhalos in the innermost region of a galaxy like our own; different works provide very different answers about their survival probability, e.g.~\cite{vandenbosch,garrison-kimmel}. Closely related to this issue, the fraction of these small structures surviving from their formation until present time is not well known, due to the violent, non-linear processes that take place during their accretion and merging into larger halos, e.g.~ \cite{Berezinsky2006,Goerdt2007,Zhao2007}.  The unknown impact of baryons on the subhalo population adds additional uncertainty.  In the future, we will repeat our analyses using MW-size hydrodynamical simulations. Baryons are not expected to play a critical effect on the structural properties of subhalos below $\sim10^8\msun$; yet they could alter the subhalo abundance significantly, e.g.~\cite{Zhu2016,garrison-kimmel,Calore+17}.

Other second order uncertainties in our study come from the assumed cosmology (VL-II simulation algorithm was run with WMAP3 cosmological parameters instead of Planck), and the non inclusion of subhalos below $10^3\msun$ in the repopulation work (some of them could still yield relevant fluxes; see trend in Figure~\ref{fig:Jfactors}).

Although we cannot address these caveats with our current repopulation analysis, our work reflects the best knowledge of the low-mass subhalo population and addresses the uncertainties due to cosmic variance. Numerical work is already ongoing to shed further light on some of the mentioned limitations and open issues pertaining the subhalo population, which will be presented in a future publication.

\section{Search of DM subhalo candidates in \textit{Fermi}-LAT source catalogs}
\label{sec:catalogs}
A large number of the gamma-ray sources detected by \textit{Fermi}-LAT lack firm association with known astrophysical objects such as pulsars or blazars \cite{2FHL_paper,3FHL_paper,3FGL_paper}. Some of these unIDs can potentially be subhalos of DM whose annihilation produces gamma-rays.

With this in mind, we searched for potential DM subhalo candidates among the unIDs in three LAT catalogs: the 2FHL and 3FHL (second and third catalog of hard \textit{Fermi}-LAT sources), containing sources detected above 10 and 50 GeV and observation time of 6.7 and 7 years, respectively, and the 3FGL (third \textit{Fermi}-LAT source catalog), with a detection threshold of 100 MeV and 4 years of observation time. While the 3FGL is composed mainly of blazars and Galactic pulsars, the high-energy catalogs (2FHL, 3FHL) are composed of an even larger fraction of AGNs \cite{2FHL_paper,3FHL_paper,3FGL_paper}. These catalogs contain large numbers of unIDs, 13\% of the full catalog in the 2FHL, 11\% in the 3FHL, and 33\% in the 3FGL. We summarize in \cref{tab:fermi_catalogs} the most important features of the three catalogs for this particular study.

\begin{table}
  \begin{center}
    \begin{tabular}{|M{1.5cm}||M{1.5cm}|M{2.6cm}|M{2cm}|M{1.6cm}|M{1.5cm}|}
    \hline 
    Catalog & Year & Energy (GeV) & Observation time (yr) & Sources & UnIDs\\
    \hline
    3FGL & 2015 & 0.1-300 & 4   & 3033 & 1010\\
    2FHL & 2015 & 50-2000 & 6.7 & 360  & 48\\
    3FHL & 2017 & 10-2000 & 7   & 1556 & 177\\
    \hline
    \end{tabular}
    \caption{Key properties of the {\it Fermi}-LAT catalogs used in this study.}
    \label{tab:fermi_catalogs}
  \end{center}
\end{table}

Our first step will be to build up a list of potential subhalo candidates among the pool of unIDs in these catalogs. 
Importantly, the smaller the number of potential DM subhalo candidates is, the stronger the DM constraints will be.  This will be discussed in further detail in \cref{sec:impact_limits}. 
 We define several rejection criteria based on the expected properties of DM subhalos versus sources of astrophysical origin, namely:

\begin{itemize}
  \item Astrophysical association: unIDs already associated to conventional astrophysical sources in follow-up observational campaigns after the catalogs publication will be removed.
  \item Galactic latitude: a cut around the Galactic plane to decrease contamination due to Galactic sources will be applied.
  \item Variability: DM subhalos are expected to be steady, so any variable source will be removed.
  \item Machine learning: we will use information from source classification algorithms to reject unIDs.
  \item Multiwavelength emission: DM subhalos are expected to emit only in gamma rays; thus unIDs exhibiting emissions at other wavelengths will be removed.
  \item Complex regions: those unIDs lying on regions of mismodeled diffuse emission, considered as potential artifacts, will be dropped as well.
\end{itemize}

These criteria and their effect on the number of subhalo candidates are summarized in the flow chart shown in Figure \ref{fig:flowchart}. Unless specified, the number of rejected unIDs in step $n$ is over the clean list, i.e., taking into account all the cuts in the $n-1$ previous steps. Note also that there is no filter in the list above that explicitly uses either spectral or spatial information of the sources, that is, we did not perform a dedicated {\it Fermi}-LAT spectral and spatial analysis of the unIDs.  That work is ongoing and will be presented in a future publication.

In the following, we describe each rejection criterion in further detail:

\begin{figure}[!ht]
\centering
\includegraphics[height=15cm]{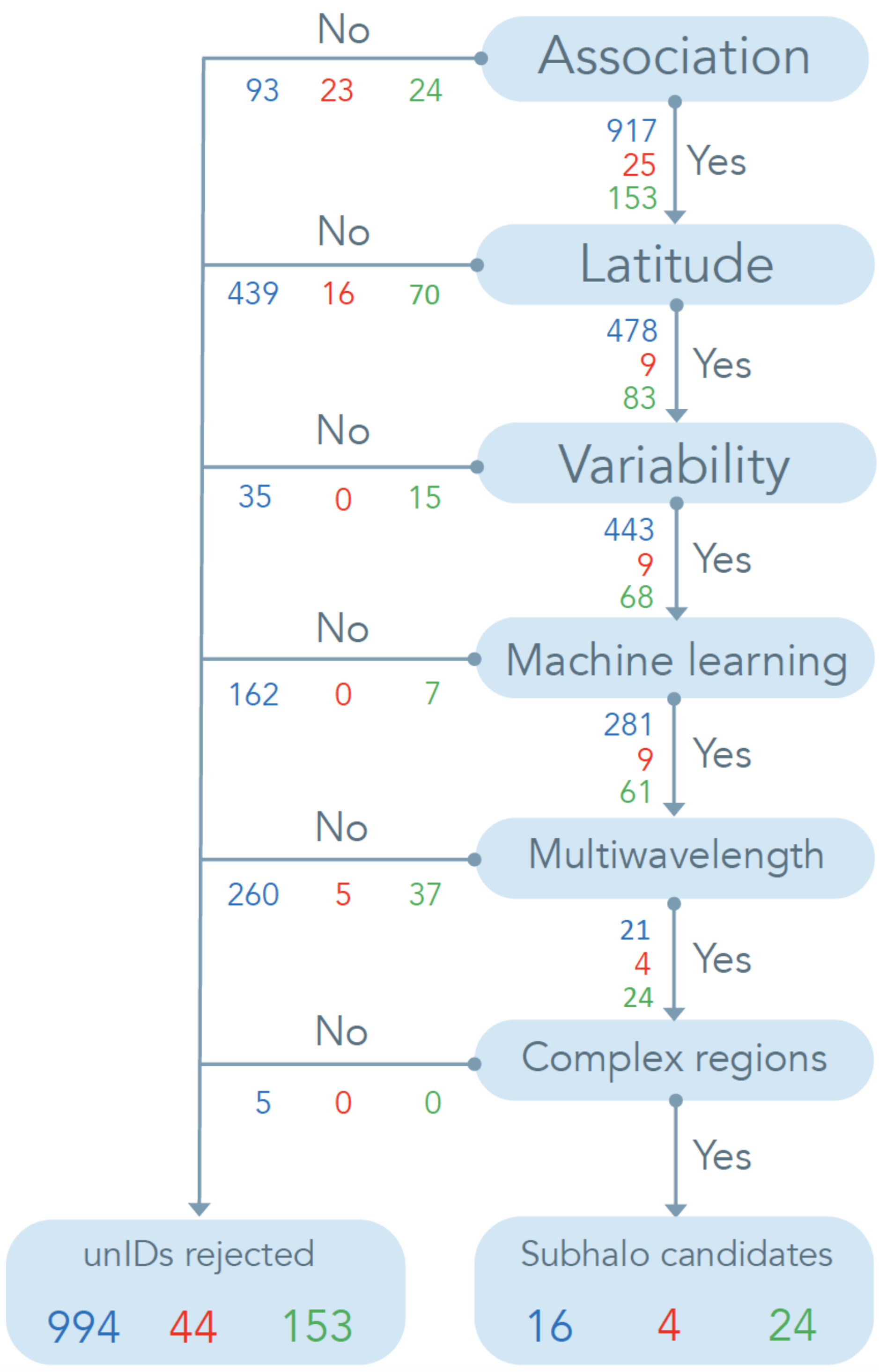}
\caption{Flowchart of the rejection criteria. Blue, red and green correspond to numbers in the 3FGL, 2FHL and 3FHL catalogs, respectively. The "yes" flow lists the unIDs that pass a certain filter, while the "no" flow indicates  sources that do not, for each criteria. See text for details.}
\label{fig:flowchart}
\end{figure}

 \begin{enumerate}
 \item \textbf{Association}: Many of the unID sources from early catalogs have subsequently received high probability associations, in part because of targeted observation campaigns \cite{saz-parkinson_pulsars_review,Laffon2015,Smith2017}. This is the case for the 2FHL and 3FHL catalogs, where 22 unIDs of the former appeared later as associated in the latter and therefore are rejected from our sample. One additional source is flagged as unknown rather than as associated; it exhibits emission in X-rays and therefore is discarded as well (see step 4).
\\When comparing 3FGL and 3FHL unIDs, although the latter was published almost 2 years later than the former, 7 of the 3FHL unIDs have firm association in the 3FGL and are also removed from our list.
\\Seven additional 3FGL sources are associated with blazars in \cite{Crespo2016_1}, although they are all at low latitude (see second criterium below); 31 more in \cite{Pena-Herazo2017}, associated with blazars, quasars and galaxies, 10 more blazars in \cite{Paiano17}, and 18 more miscellaneous AGNs collected in \cite{Massaro+16} are rejected. Also, 13 more 3FGL-only sources are removed with a blind search of Einstein@Home \cite{Clark+17} as they are associated to low-latitude pulsars. Within the same project, one more source, a radio-quiet pulsar, is rejected \cite{Clark+18}. Additionally, 5 high latitude sources are removed from the 3FGL candidate sample; one is associated in the 3FHL, and the other four are pulsar associations, and present multiwavelength emission (see criterium number 5): 3FGL J1544.6$-$1125 \cite{BogdanovHalpern15}, 3FGL J2039.6$-$5618 \cite{Salvetti+15,Romani15}, 3FGL J1946.4$-$5403 and 3FGL J1744.1$-$7619 \cite{Camilo+15}.
\\We are also able to discard 13 3FGL sources for being recently discovered millisecond pulsars (MSPs) \footnote{\url{http://astro.phys.wvu.edu/GalacticMSPs/}}. One of these sources, 3FGL J1016.6$-$4244, is also present in the 3FHL catalog.
\\Recent improvements, including updated positional error ellipses based on 8 years of LAT data and/or a broader energy range compared to the initial catalogs, along with an updated association procedure developed in the context of the future 4FGL catalog, have revealed counterparts to a number of sources initially unidentified in the 3FGL, 2FHL and 3FHL catalogs \cite{4fgl_paper_prel}. This improvement is especially relevant in the case of the 3FGL catalog, as the new data has almost the double of exposure time and makes use of Pass 8 events. There are two sources which have also been discarded for this reason. Also, there are three 3FGL sources, three 3FHL and four remaining 2FHL sources with no counterpart in the 8-year LAT data, which are likely to be flaring objects or artifacts. Yet, they are conservatively not discarded from our list. Further observations will be needed in order to determine their true nature.

As a final remark, we note that there exists the possibility of a misassociation. This rate of false positives is expected to be under 5\% according to the Bayesian analysis in the considered catalogs, although this number can be significantly smaller, e.g. a 1\% in the 3FHL \cite{3FHL_paper}.

 \item \textbf{Latitude}: Galactic astrophysical sources present in {\it Fermi}-LAT catalogs, such as pulsars, pulsar wind nebulae or supernova remnants, are observed to be strongly concentrated along the Galactic plane.  Since many objects of this class are expected to also be hidden among the pool of unIDs awaiting higher confidence classification, we expect the distribution of unIDs to peak around zero Galactic latitudes as well. Indeed, this can clearly seen in Figure \ref{fig:latitude_hist}, where the latitude histograms for the all the unIDs in the catalogs are shown.
\begin{figure}[!ht]
\centering
\includegraphics[height=9cm]{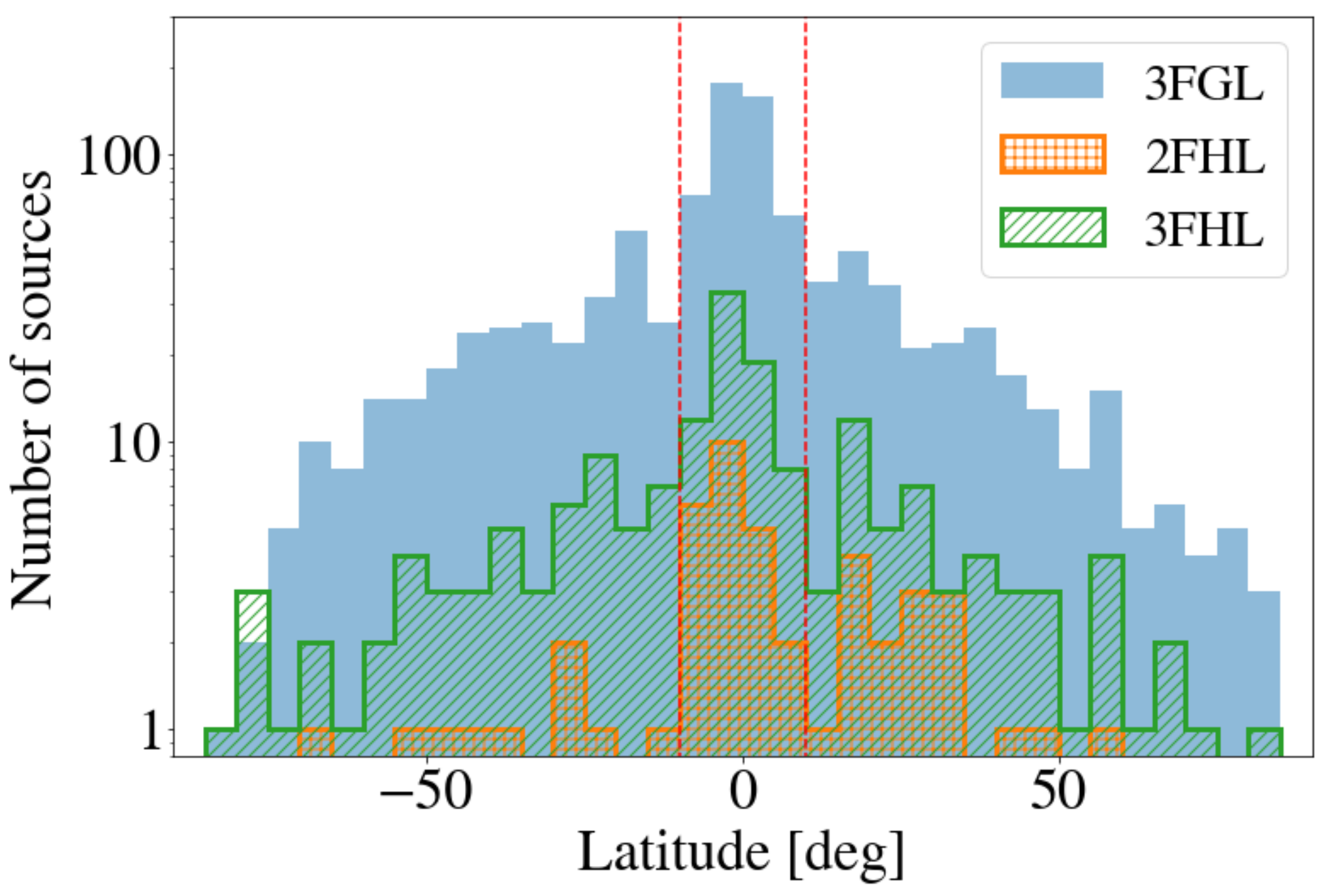}
\caption{Galactic latitude distribution of unIDs in the 2FHL, 3FHL, and 3FGL catalogs. Those sources between the red dashed lines are excluded from our analysis $\left(|b|\leq10^\circ\right)$.}
\label{fig:latitude_hist}
\end{figure}
\\For our purposes, these low-latitude sources (especially pulsars, which can fake DM annihilation spectra at low energies \cite{fermi_dm_satellites_paper}) are not interesting and only add contamination to our sample of potential DM subhalo candidates.  Furthermore, from N-body simulations we expect the subhalos to be isotropically distributed over the sky; thus we apply a cut at Galactic latitudes $\left|b\right|\leq10^\circ$. \footnote{The same cut is also be applied on the N-body simulation data.  Assuming an isotropic distribution of subhalos, the cut removes, on average, 11\% of the simulated subhalos.}
\\After having applied this latitude cut to the catalogs, 16 2FHL, 70 3FHL and 439 3FGL sources are removed from our list. More of the 3FGL sources are rejected by the Galactic latitude cut because that catalog has a lower energy threshold and thus contains a large fraction of Galactic (though unidentified) sources.  Conversely, the harder sources found in both the 2FHL and 3FHL catalogs, are mostly AGNs.
 
 \item \textbf{Variability}: DM annihilation in subhalos is expected to be steady, i.e., to not display flux variability over time. We will use two different methodologies to identify and eliminate variable sources: the variability reported in the catalogs and our own studies performed with the the Fermi All-sky Variability Analysis (FAVA) \cite{FAVA_paper} tool. Regarding the first one, in order to be conservative, we only remove sources marked as variable at 99\% C.L. in the catalogs. Specifically, the variability statistics in the catalogs are $\chi^2$-distributed reported, with different degrees of freedom depending on the catalog.  For each catalog we cut on the variability statistics at a value corresponding to 99\% C.L.  This removes 2 unIDs from the 3FHL sample (note that one of these is marked as not variable in the 3FHL but as variable in the 3FGL), 16 in the 3FGL (note that there are 9 more variable sources, but already rejected by latitude) and none in the 2FHL.
\\These flux variations are computed taking into account the Galactic diffuse emission model \cite{fermi_galactic_diffuse_paper}. Yet, ideally, we would be interested in systematic deviations around the mean flux of the source, without relying on any background model to model the region around the sources. As previously mentioned, our second methodology is based upon FAVA. By default, FAVA scans the sky in a weekly time lapse searching for deviations from the median detected flux in each direction, presenting lightcurves refered to the collection of pixels within the corresponding PSF centered on the source position. It performs this analysis in a low-energy band (100-800 MeV) and a high-energy band (800 MeV-10 GeV), where there is still enough photon statistics to perform a proper variability analysis.
\\In order for us to reject an unID due to variability, and with the intention to be conservative, we require the presence of a flare at 5$\sigma$ significance in the data, searching for flares at the reported unIDs sky positions. 

Note that in some cases the mean of the difference between the flux and average flux may be null, yet an hypothetical flare lasting a tiny fraction of the whole dataset would reject automatically a DM subhalo origin. 
When performing variability studies with FAVA a flaring source can induce fake variability in its surrounding sources/regions. This is so because of the spill-over of photons coming from the point spread function (PSF) of the LAT (highly energy-dependent and of the order of $2^\circ$ at few hundred MeV)\footnote{\url{http://www.slac.stanford.edu/exp/glast/groups/canda/lat_Performance.htm}}. Therefore, should we find a variable unID with FAVA, an inspection of the significance map of the region around the unID is necessary in order to conclude that the variability is real and not induced by spill-over from a nearby, flaring source. An example of this effect can be seen in figures \ref{fig:fava_lightcurves} and \ref{fig:fava_tsmap}. Having checked that the variability is not induced and, thus, that the source presents a flare over 5$\sigma$ significance (see Figure \ref{fig:fava_lightcurve_clean}), the source is rejected.
\\The weekly time binning used by FAVA, in comparison to the catalog monthly binning, proves to be very useful when dealing with short flaring episodes, and allow us to search in narrower time bins than those in the catalogs, and to detect very short flares which would get diluted on longer time scales. Our study can also find additional variable sources with respect to the catalogs as we extend beyond the observation times of the catalogs, using all available data up to March 1st, 2018 (MET 541555205).
\\In summary, the FAVA analysis allows us to discard 13 3FHL sources and 11 3FGL sources, with no source overlap between catalogs.
\\Interestingly, we also found two source coincidences or \textit{duplicities}, i.e., sources very near to each other and present in different catalogs which, apparently, are not related but nevertheless possess highly correlated lightcurves and similar spectra, and therefore seem to be actually the same source. These are the pairs (2FHL J0738.6+1741, 3FGL J0738.1+1741) and (2FHL J1630.0+7644, 3FGL J1628.2+7703).

\begin{figure}[!ht]
\centering
\includegraphics[height=5cm]{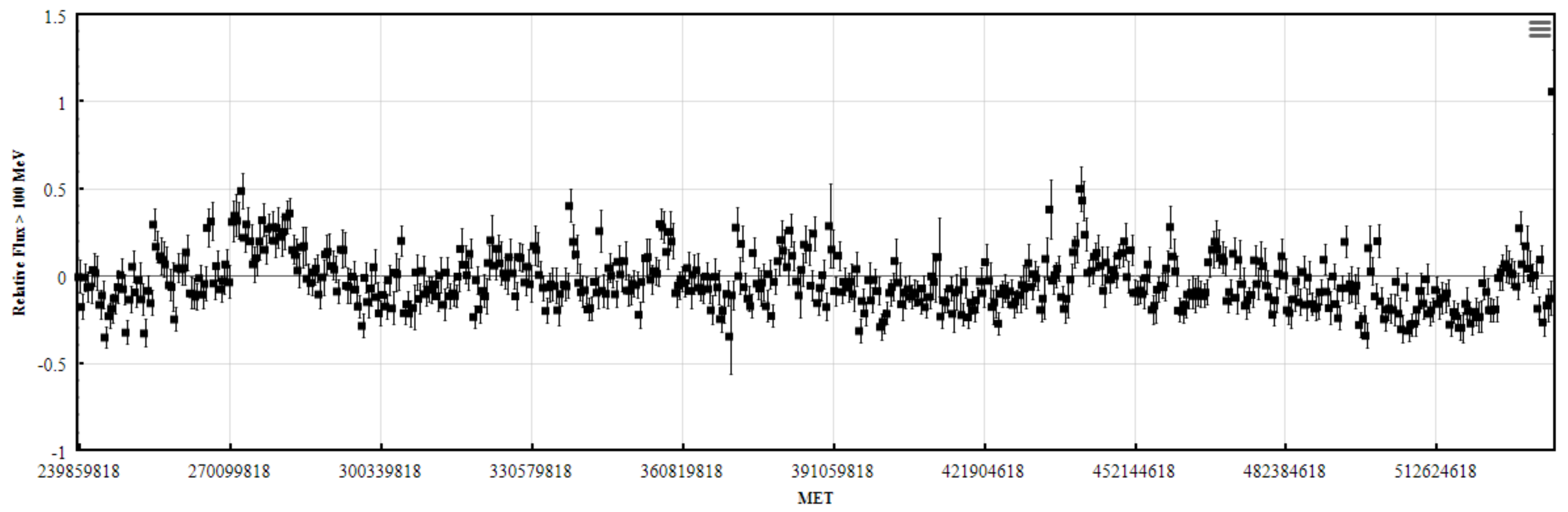}
\vfill
\includegraphics[height=5.1cm]{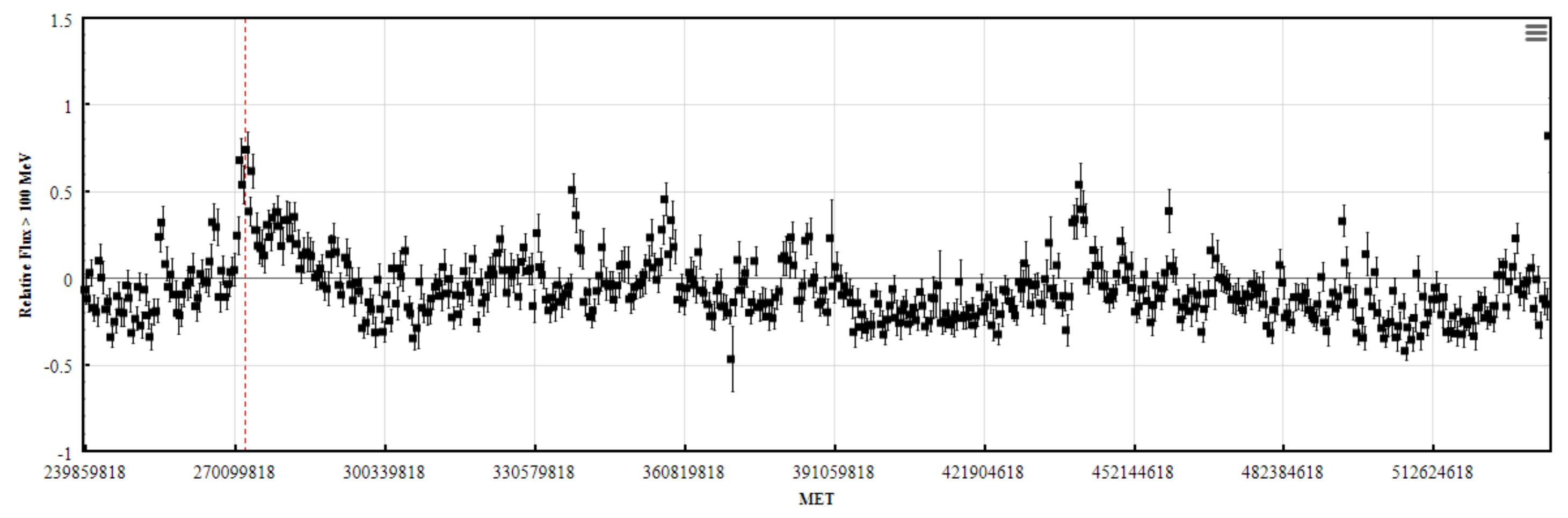}
\caption{Example of ``fake'' variability induced by the spill over of photons due to the PSF. \textbf{Top panel}: FAVA lightcurve of 3FGL J2043.8-4801. \textbf{Low panel}: FAVA lightcurve of 3FGL J2056.2-4714 (associated with a flat-spectrum radio quasar) flare in week 55 (mission elapsed time (MET) 272519018), indicated by the dashed red line. Note the tight correlation between the variations in both curves. These two sources are 2.3$^\circ$ from each other in the sky.}
\label{fig:fava_lightcurves}
\end{figure}

\begin{figure}[!ht]
\centering
\includegraphics[height=8cm]{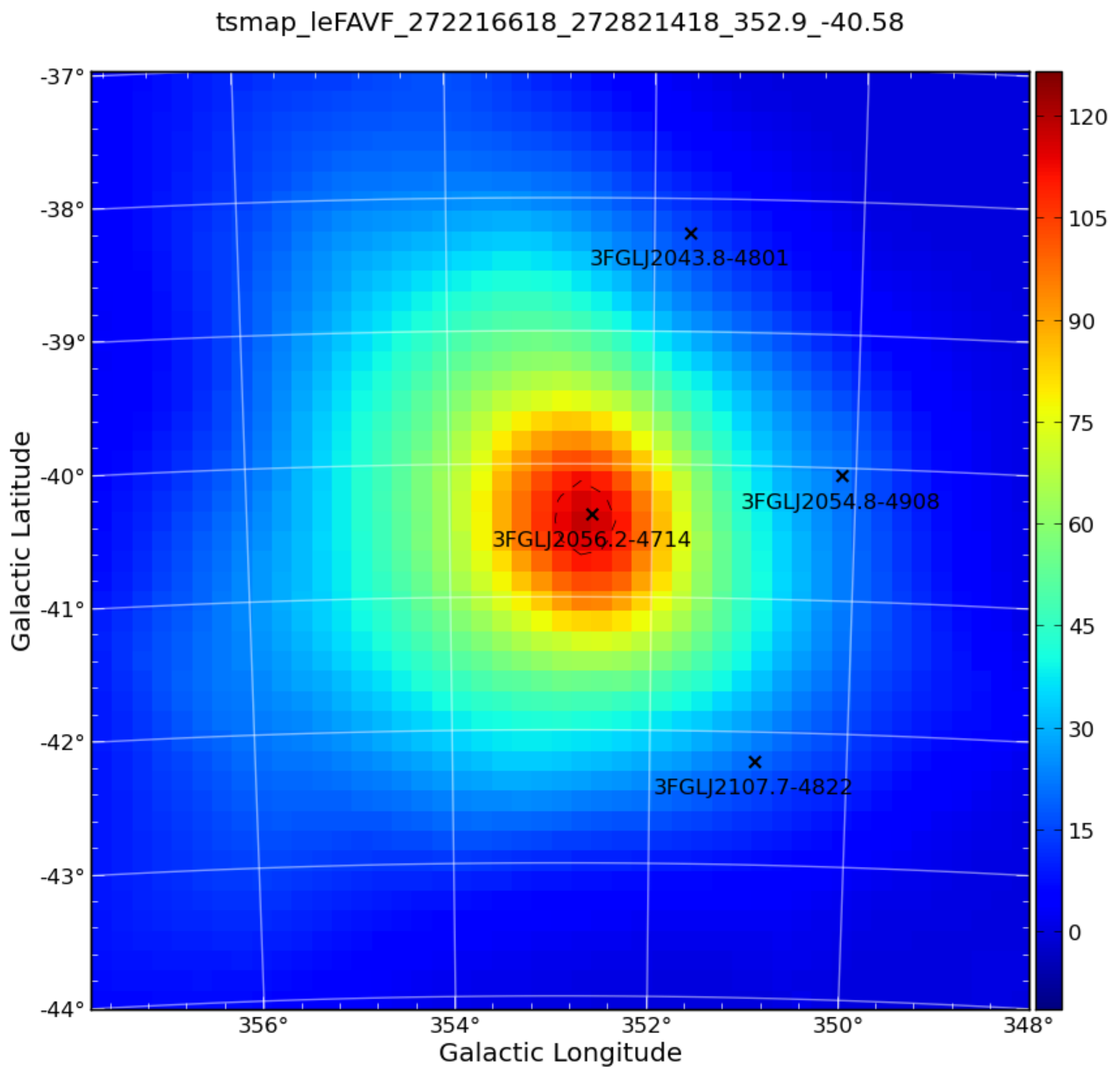}
\caption{TS map of 3FGL J2056.2$-$4714. This object shows a flare in week 55 (MET 272519018).  There is still a relatively high significance ($TS=17$) at the position of 3FGL 2043.8$-$4801, which lies only 2.3$^\circ$ away from the flaring source. Thus, we can conclude that the flare of 3FGL J2056.2$-$4714 is causing the variability observed for 3FGL 2043.8$-$4801 (the timing coincides as well, see Figure \ref{fig:fava_lightcurves})}
\label{fig:fava_tsmap}
\end{figure}

\begin{figure}[!ht]
\centering
\includegraphics[height=5cm]{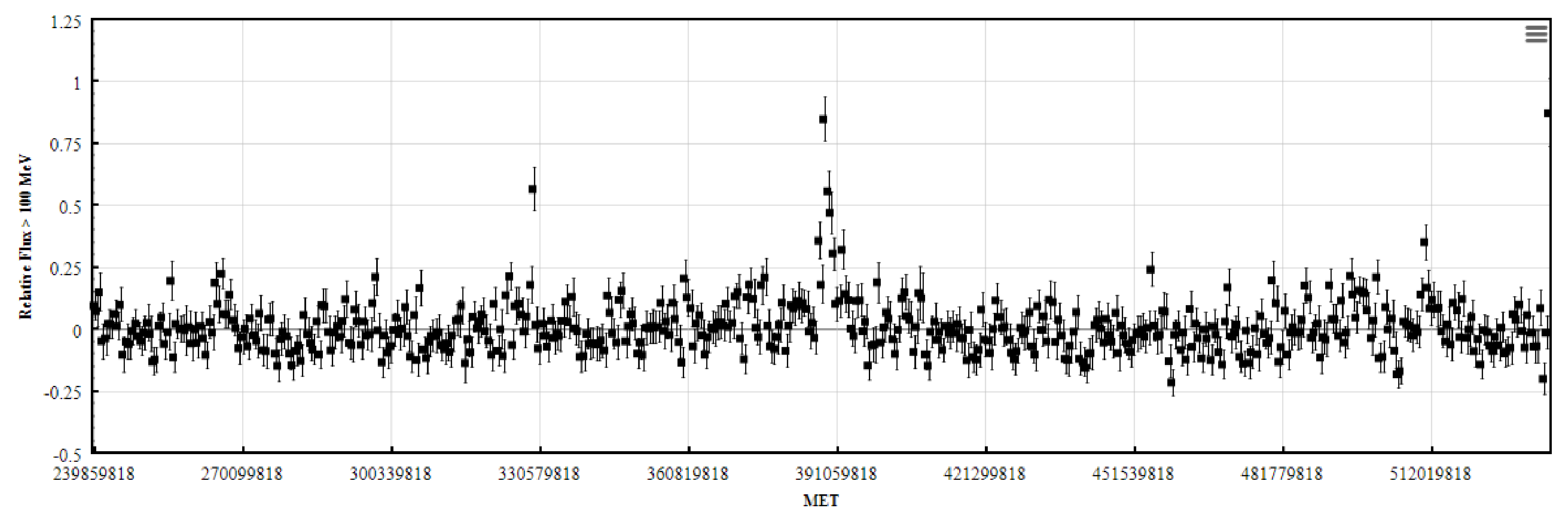}
\caption{FAVA lightcurve of 3FHL J0500.6+1903. No other source flare is temporally and spatially coincident with the flare centered on week 247 (MET 388640618), and therefore this object is rejected as DM subhalo due to its intrinsic variability.}
\label{fig:fava_lightcurve_clean}
\end{figure}

 \item \textbf{Machine-learning algorithms}: these have been used to classify unIDs into different types of sources. Typically, the machine learning algorithm is trained with the associated sources and then is run over the unidentified sample, assigning each of the unIDs a probability to be a particular type of source. We use the results from \cite{3fglzoo_paper, LefaucheurPita17}, both of them trained and run over the 3FGL. Since this is the catalog with the largest number of sources we expect the most accurate results, because the training sample is the largest. In both works, the algorithms use time-series and spectral information to search for AGN candidates, so the classification is made and optimized for this type of sources. Many of the sources classified as AGNs are common to both of the mentioned works. Note a source being in just one of them is discarded, i.e., we do not require an "and" criterium, but an "or". We reject a total of 162 3FGL sources with this method, which are not discarded by previous filters (the total number is 559 for \cite{3fglzoo_paper} and 595 for \cite{LefaucheurPita17}). Also, by cross-checking the results of these works for the 3FGL with the 2FHL and 3FHL catalogs, 7 3FHL sources are discarded.
\\The expected false positive ratio is estimated to be of order 4\% in both works. As we use 162 sources for the 3FGL catalog, a 4\% are approximately 6 false positives. We will be taking into account this uncertainty when setting the DM constraints in \cref{sec:constraints}. Further information about the propagation of this uncertainty can be found in \cref{app:err_comp}.
\\Another machine learning work \cite{Mirabal+16} used a similar algorithm but for pulsars instead of AGNs. In this case, it is not possible to discard the obtained pulsar candidates, as they have a spectrum compatible with DM annihilation. Yet, for this same reason, they actually are interesting sources for our purposes. From their 34 reported pulsar candidates, we are left with only 3 sources in the 3FGL which have not been rejected by any other of our selection criteria (3FGL J0336.1+7500, 3FGL J0953.7$-$1510 and 3FGL J1225.9+2953). These sources may be especially interesting for future observational follow-up campaigns.

 \item \textbf{Multi-wavelength emission}: If the gamma-ray emission of the unIDs is indeed produced by DM annihilation, we do not expect to see these objects emitting in other wavelengths. Therefore, the observation of any of these sources in IR, Optical, UV or X-ray would represent another reason for rejection. Note that subhalos with masses larger than $\gtrsim10^{7}$ M\textsubscript{\(\odot\)} are expected to have baryonic content and consequently they could emit and be observed at other wavelengths (see \cref{sec:jfactors} and \cref{sec:impact_limits}). As said, in this work we will focus on the search for less massive DM subhalos with no visible astrophysical counterparts.
\\We follow three different approaches when looking for multiwavelength (MWL) emission. First of all, we manually search for additional emission in the ASDC repository\footnote{\url{https://tools.asdc.asi.it/}}. To reject a source, we require  i) that any additional emission to be within a 5 arcmin radius of the catalog source position for the 2FHL and 3FHL sources\footnote{The typical LAT PSF at these high energies is below 0.1$^\circ$, i.e., 6 arcmin. Therefore, we can conservatively exclude a source showing any kind of emission within this area.} and ii) no other known source is present in this region. In the case of 3FGL sources, we search within their corresponding 95\% confidence level positional error ellipses. We are able to discard 4 2FHL, 12 3FHL and 7 3FGL sources. For these, there are observations in the search area by WISE \cite{wise_paper}, 2MASS \cite{2mass_paper}, USNO \cite{usno_paper}, SDSS \cite{SDSS_paper} and NVSS \cite{NVSS_paper}.
\\When considering the 3FGL sources, we also made use of the observational campaign performed by authors in Ref. \cite{StrohFalcone13} using the \textit{Swift} X-ray telescope\footnote{The results of this survey can be retrieved at \url{www.swift.psu.edu/unassociated/}}. In this case, we reject 3FGL sources that have been detected at least once by \textit{Swift}. This removes 16 additional sources, 2 of which are also in the 3FHL catalog.
\\We also search for associated \textit{Fermi}-LAT sources in the \textit{Swift} Master Catalog on HEASARC\footnote{\url{https://heasarc.gsfc.nasa.gov}}. The search is extended to other X-ray telescopes as well, such as Chandra, Hitomi, NICER, NuSTAR, ROSAT, RXTE, Suzaku and XMM-Newton. By using this tool, we remove 207 3FGL, 1 2FHL and 30 3FHL sources. 25 out of these 30 are common to the 3FGL.

 \item \textbf{Complex regions}: As reported in the 3FGL and previous low-energy threshold catalogs (such as 2FGL), sources with moderate Test Statistics (TS) are labeled with a "c" in case they lie in complex regions with a poor modeling of the diffuse emission, thus having significantly higher probability of being artifacts (see \S ~3.8 in \cite{3FGL_paper}). These sources are mostly concentrated along the Galactic plane, and very close to the Galactic center, although some of them are also at higher latitude. We remove them from our list of potential subhalo candidates as they are probably star-forming regions or just mere artifacts, as described by \cite{3FGL_paper}. From our remaining candidate list, this filter removes 5 additional sources in the 3FGL catalog, none of which are present in the other catalogs.  The fraction of the sky affected is by this cut is negligible, so we do not attempt to mask these regions when analyzing the N-body simulation data, or to correct the subhalo sensitivity predictions for this cut.
 
\end{enumerate}

 After the above filtering procedure, we are left with 4 2FHL (8\% of the original unIDs sample), 24 3FHL (14\%) and 16 3FGL (2\%) candidates. We note that no source flagged as extended in the catalogs survived our cuts. In particular, the source 3FGL J2212.5+0703, which is marked as possible DM subhalo by \cite{Bertoni+16}, is rejected by its multiwavelength emission, and 3FGL J1924.8-1034, which is also marked as a possible extended DM subhalo by \cite{Xia+17}, is classified as an AGN by the machine-learning algorithms.

Table \ref{tab:filtering_efficiency} summarizes the expected efficiency, defined as the fraction of each filtering effect which is univocally applied (i.e., with no possibility of a false positive), and how we dealt in each case with the fact of not having 100\% efficiency. 

\begin{table}
  \begin{center}
    \begin{tabular}{|M{3.5cm}||M{2cm}|M{8.6cm}|}
    \hline 
    Filter & Efficiency & Treatment\\
    \hline
    Association & >95\% & Neglected\\
    Latitude & $\sim$89\% & Applied to N-body simulations\\
    Variability & 99.9999\% & Neglected\\
    Machine learning & >96\% & Used to correct number of observed candidates\\
    Complex regions & >99\% & Neglected\\
    \hline
    \end{tabular}
    \caption{Estimated efficiency for each of the applied filters in  section \cref{sec:catalogs}. The third column refers to the treatment adopted in each case for the remaining unIDs ``beyond'' the efficiency of our cuts. Note that we omit the multi-wavelength emission filter in this table given the difficulty to provide a realistic estimate as of today: among other ingredients, one would need to compute the expected source number density of each of the considered catalogs at other wavelengths, which is beyond the scope of this work.}
    \label{tab:filtering_efficiency}
  \end{center}
\end{table}

\section{\textit{Fermi}-LAT sensitivity to DM subhalos}
\label{sec:minimum_flux}
Most of previous works \cite{Bertoni+15,Bertoni+16,Schoonenberg+16,BerlinHooper14,BuckleyHooper10,Zechlin+12} assumed that the sensitivity of the LAT to DM subhalos is the same across the sky and equal to the detection threshold of the catalog under consideration. Yet, the reality is that the minimum flux required to have a $\sim 5\sigma$ (where $\sigma$ is the significance expressed in standard deviations) detection with the LAT, $F_{min}$, depends on the source spectrum and position in the sky.  In fact, the threshold for detection is strictly performed in terms of another parameter, i.e., the Test Statistic:

\begin{equation}
\label{eq:TS_equation}
TS=-2\cdot \textrm{log}\left[\frac{\mathcal{L}(H_1)}{\mathcal{L}(H_0)}\right],
\end{equation}

\noindent where $\mathcal{L}(H_0)$ and $\mathcal{L}(H_1)$ are the likelihood under the null (no source) and alternative (existing source) hypotheses, respectively.  We note that $TS \sim \sigma^2$, and each of the catalogs we used applied a TS $\ge$ 25 threshold.  Thus, when it comes to DM subhalo detection, $F_{min}$ will depend on the adopted annihilation channel and WIMP mass, as well as on the position of the object in the sky.  Indeed, this $F_{min}$ could significantly differ from the characteristic detection threshold of the catalog, which is typically computed assuming a power-law spectrum of spectral index $\Gamma =2$ ($dN/dE \propto E^{-\Gamma}$), a fairly typical spectrum for conventional astrophysical emitters.  This is not the case for DM annihilation spectra, as seen in \cref{sec:ann_spectra}, where curvature and a cut-off are present.

Also, there is a strong dependence on the latitude of the considered source due to the variations of the diffuse emission:  at high Galactic latitudes, the Galactic diffuse emission is much lower when compared to regions along the Galactic plane.  Thus, a low-latitude source needs a comparatively larger flux in order to reach the TS $\ge$ 25 detection threshold.

To compute the dependence of the minimum detection flux, we use \textit{Fermipy}\footnote{\url{http://fermipy.readthedocs.io/en/latest/}} v0.17.0 \cite{Fermipy_paper}, a \verb|PYTHON|-based code to analyze \textit{Fermi}-LAT data.
We begin by generating all-sky sensitivity maps for each annihilation channel and DM mass. This must be done for each catalog, as they cover different energy ranges. The sensitivity at each (l,b) Galactic coordinate is obtained by placing a putative (point-like) DM subhalo there and computing the integrated flux needed to reach TS $=$ 25.

\begin{figure}[!ht]
\centering
\includegraphics[height=8.5cm]{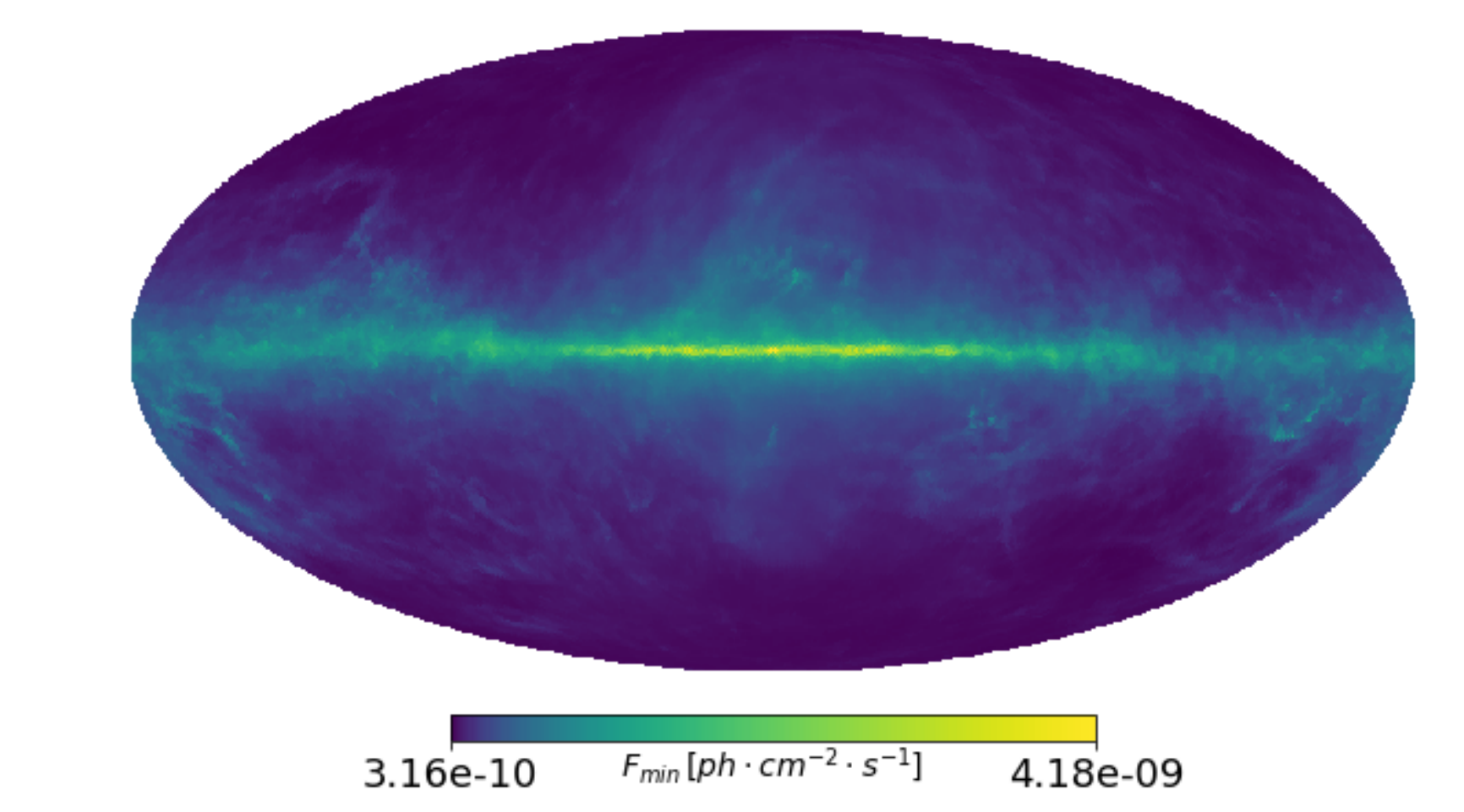}
\vfill
\includegraphics[height=8.5cm]{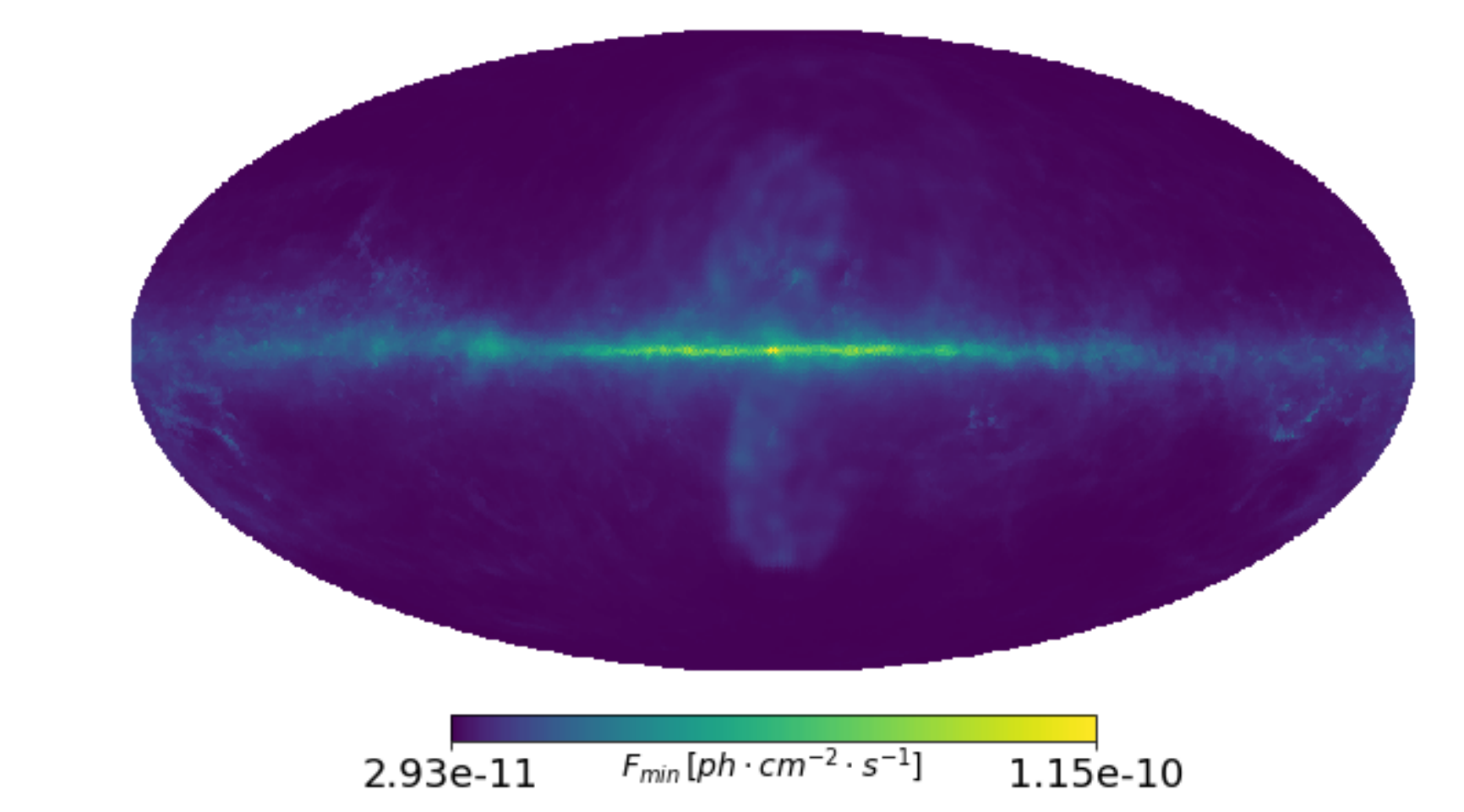}
\caption{LAT sensitivity to DM subhalos for $\tau^{+}\tau^{-}$ annihilation and the 3FGL setup, for $m_{\chi}=10$ GeV (top) and $m_{\chi}=1$ TeV (bottom).}
\label{fig:min_flux_maps}
\end{figure}

The input to the sensitivity maps generation procedure consists of the spectral parameters of the DM subhalo annihilation emission (implicitly set by annihilation channel and particle mass, which are obtained with the parametrization of \cref{sec:ann_spectra}), a threshold for the test statistics (set to 25), a minimum number of photon counts (set to 3), diffuse and isotropic templates (catalog dependent), energy range (depending on the catalog), spatial extension (set to point-like) and type of Healpix \cite{healpix_paper} pixelisation (NESTED ordering, NSIDE=64). The Galactic diffuse emission templates are {\tt gll\_iem\_v06.fits} (2FHL, 3FHL) and {\tt gll\_iem\_v05\_rev1.fit} (3FGL), while the isotropic templates are {\tt iso\_P8R2\_SOURCE\_V6\_v06.txt} (2FHL, 3FHL) and {\tt iso\_source\_v05.txt} (3FGL). The output is an all-sky map for each annihilation channel, particle mass and source catalog. As an example, the minimum detection flux ($F_{min}$) for the $\tau^+\tau^-$ channel and two different DM masses is plotted in Figure \ref{fig:min_flux_maps} for the 3FGL catalog.

At this point we remove the 10$^\circ$ above and below the Galactic plane, following our latitude cut described in \cref{sec:catalogs}, and take the mean and standard deviation of the $F_{min}$ values over the remaining sky.  

We note that the differences in $F_{min}$ between the North and South hemispheres is found to be always under 1\%. This whole procedure allows us to characterize in detail the $F_{min}$ function for each catalog, channel and mass, which will be later used to place the constraints in \cref{sec:constraints}.

\begin{figure}[!ht]
\centering
\includegraphics[height=5.2cm]{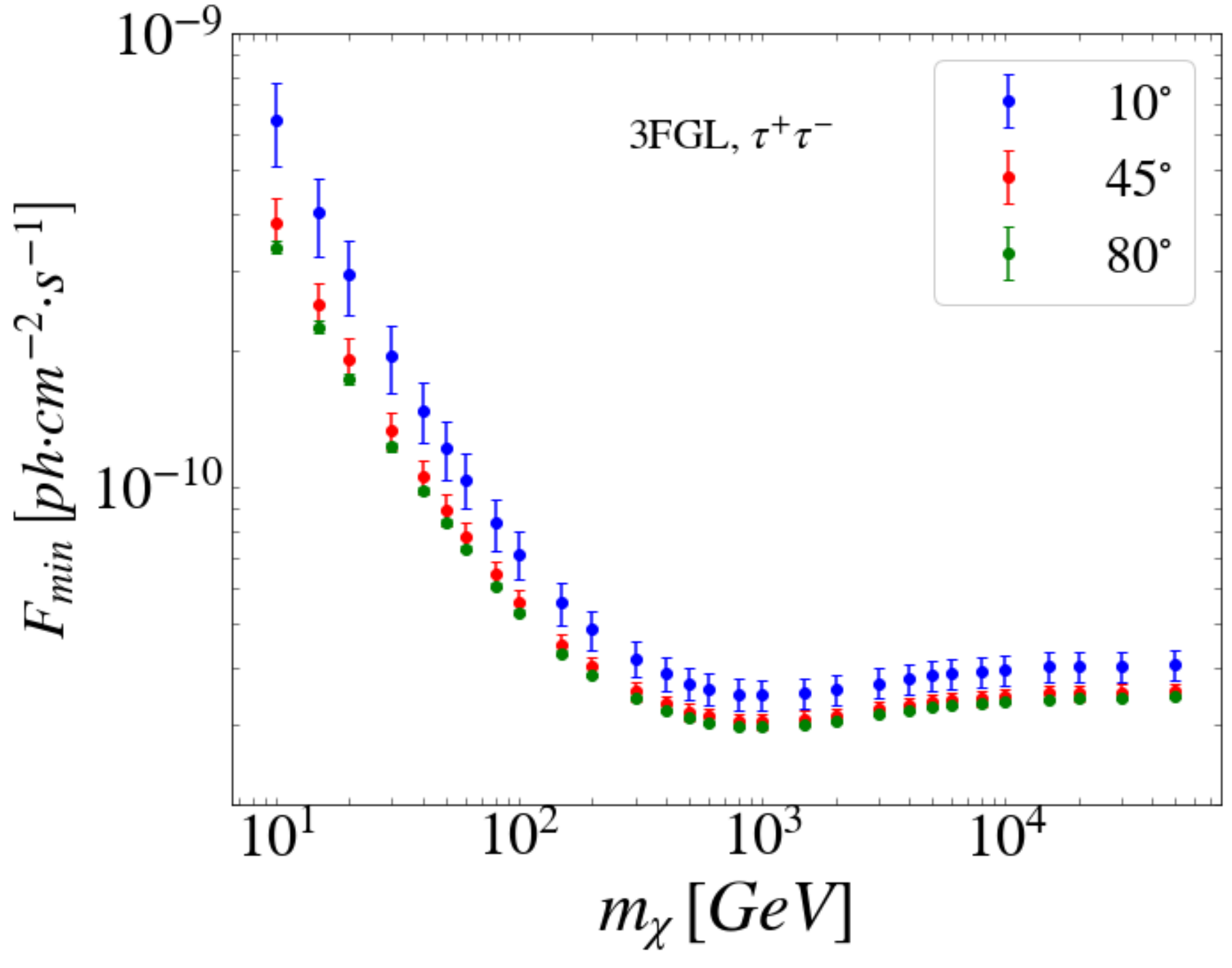}
\includegraphics[height=5.2cm]{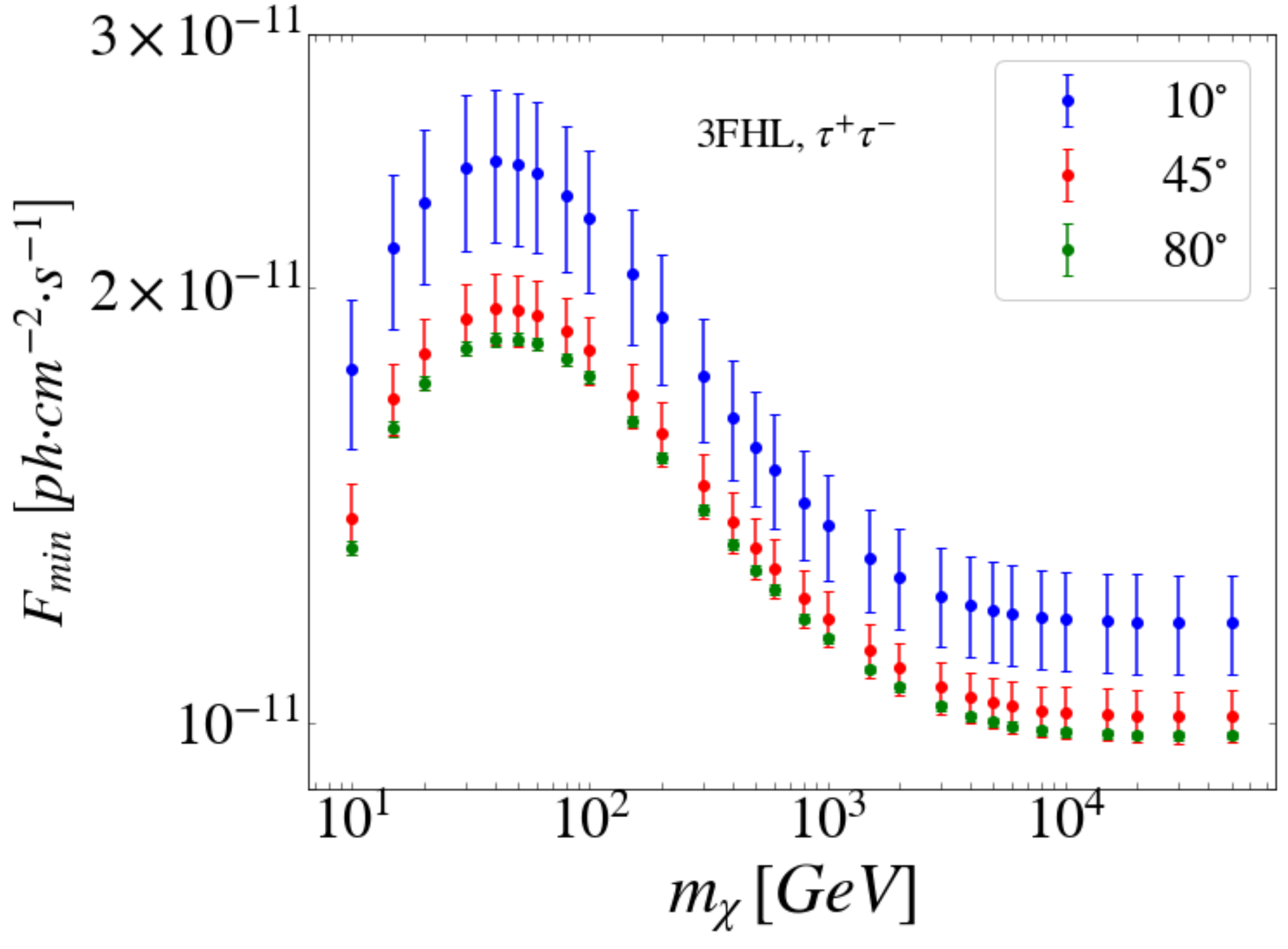}
\vfill
\includegraphics[height=5cm]{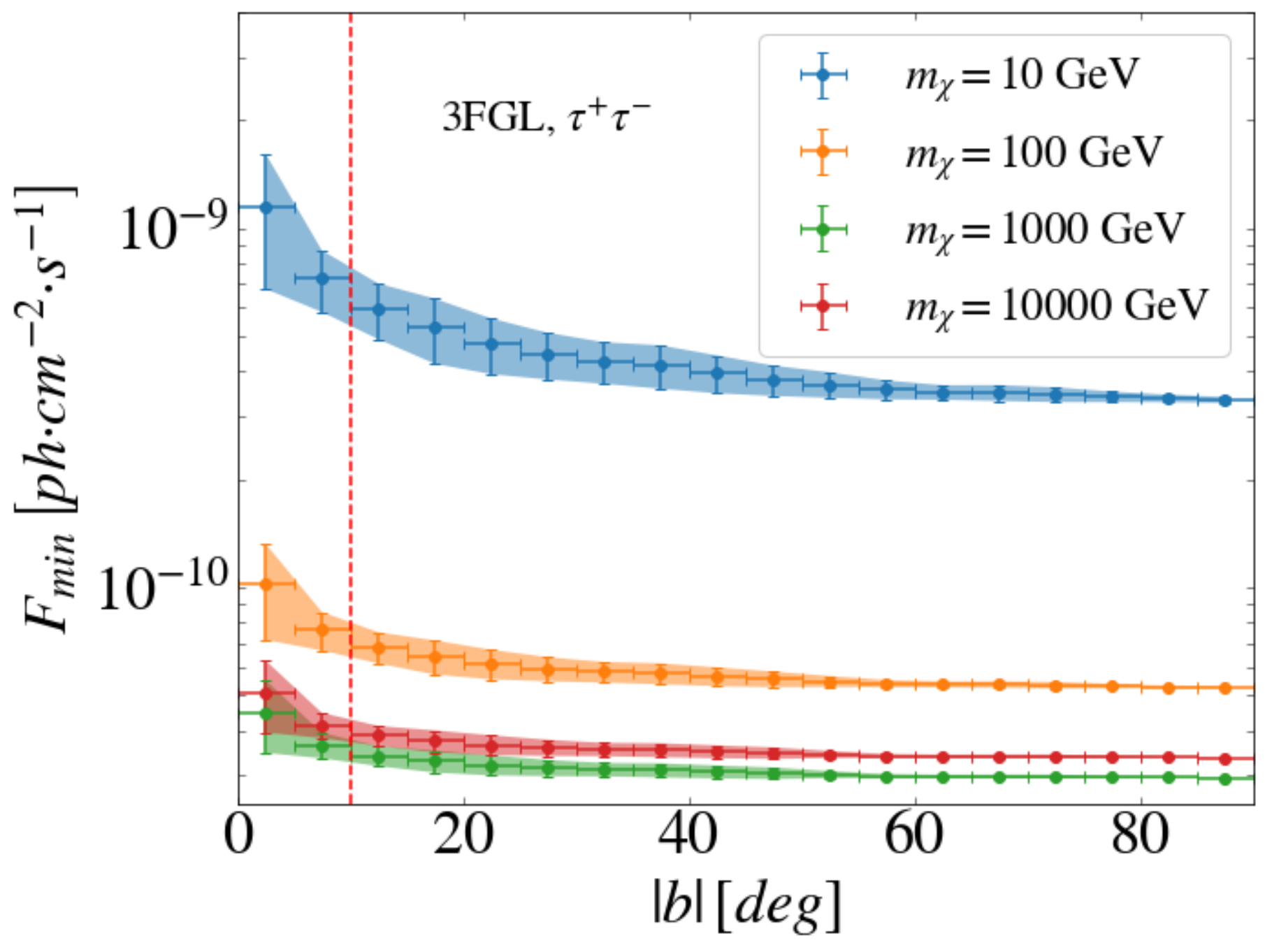}
\includegraphics[height=5cm]{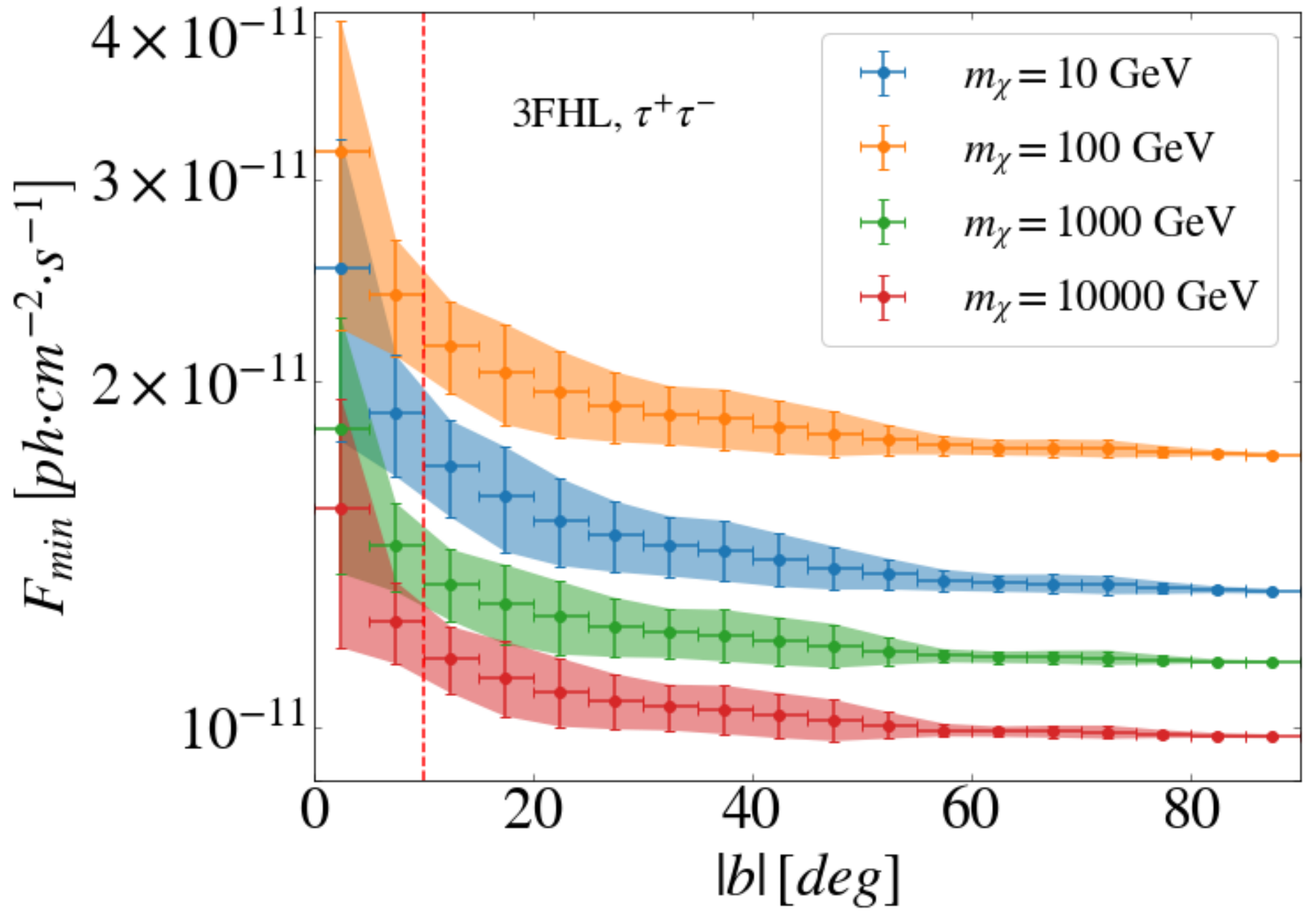}
\caption{LAT sensitivity to DM subhalos annihilating to $\tau^+\tau^-$ as a function of DM mass for various absolute Galactic latitudes (top panels) and as a function of absolute Galactic latitude for various DM masses (bottom panels). Left and right panels refer, respectively, to the 3FGL and 3FHL setups. A vertical red dashed line is plotted in the bottom panels to mark our latitude cut, $|b|<10^\circ$.}
\label{fig:min_flux_graphics}
\end{figure}

Some examples of the mean and standard deviation of $F_{min}$ and its dependencies are plotted in Figure \ref{fig:min_flux_graphics}. As it can be seen, for the $\tau^+\tau^-$ channel, the behavior of $F_{min}$ is significantly different depending on the adopted catalog. For example, for 3FGL and at a fixed latitude, $F_{min}$ decreases rapidly as the mass increases, reaching a minimum and then remaining almost constant at the largest DM masses. For 3FHL, $F_{min}$ slightly worsen (i.e. larger flux values) as the mass increases to slowly decrease at larger masses. The maximum variation in this case is approximately a factor 2, while in the case of the 3FGL setup it is roughly an order of magnitude. In all cases, the associated uncertainties become smaller at higher latitudes. This is expected because of the morphology of the Galactic diffuse emission (see also Figure \ref{fig:min_flux_maps}).

A comparison between the size of the uncertainties associated to both the J-factor and $F_{min}$ is described in \cref{app:err_comp}.

\section{DM Constraints}
\label{sec:constraints}
After the unIDs filtering performed in \cref{sec:catalogs}, we are left with a certain number of DM subhalo candidates in each {\it Fermi}-LAT point-source catalog.  In the absence of a conclusive answer about the nature of these remaining unIDs,\footnote{A careful spectral and spatial scrutiny is ongoing for these sources, which will be presented in further work.} we will place constraints on the WIMP annihilation cross section.  To do so, we make use of \cref{eq:flux,eq:j_factor,eq:f_pp}, to set a relation between $\langle\sigma v\rangle$ and $m_{\chi}$,

\begin{equation}
\label{eq:master_formula}
\langle\sigma v\rangle=\frac{8\pi\cdot m_{\chi}^2\cdot F_{min}}{J\cdot N_{\gamma}},
\end{equation}

\noindent where $F_{min}$ is the minimum detection flux (see \cref{sec:minimum_flux}), $J$ the J-factor (see \cref{sec:jfactors}) and $N_{\gamma}$ the integrated DM spectra (see Eq. \ref{eq:integrated_spectra}). Note that $\langle\sigma v\rangle$ is proportional to the DM particle mass squared, thus the constraints will be weaker for larger masses if $F_{min}$ was constant, yet $F_{min}$ exhibits a dependence on the energy, so at the end the exact shape of the constraints will depend on a combination of these parameters.

\subsection{Procedure to set DM limits}
\label{sec:impact_limits}
Our procedure to set limits is based on a comparison of the number of surviving candidates with the predictions from our N-body simulation work (see \cref{sec:jfactors}). More precisely, we allow for the possibility that the DM subhalo candidates we identified in {\it Fermi}-LAT catalogs are actually subhalos, and that they correspond to our brightest subhalos in the simulations.\footnote{We only refer here to subhalos with no baryonic components and thus completely “dark”. In our work, we will adopt an upper mass limit of $M<10^7M_{\odot}$ for these objects. Above this limit, we assume that subhalos host visible dwarf satellites and thus are observed. Indeed, some dwarfs are known with inferred masses as light as few times $10^7M_{\odot}$, e.g., Ref. \cite{Simon2011}. Also, we note that our mass cut is conservative, since the adoption of a larger value would translate into even stronger DM limits.}  To do this, we sort the simulated subhalos by J-factor, and confront this list with the remaining unIDs number. For example, in the case of 3FGL, where we are left with 16 subhalo candidates, we take the 16th largest J-factor subhalo in each realization of the simulation as the subhalo J-factor to be used in \cref{eq:master_formula}. 

\begin{figure}[!ht]
\centering
\includegraphics[height=9cm]{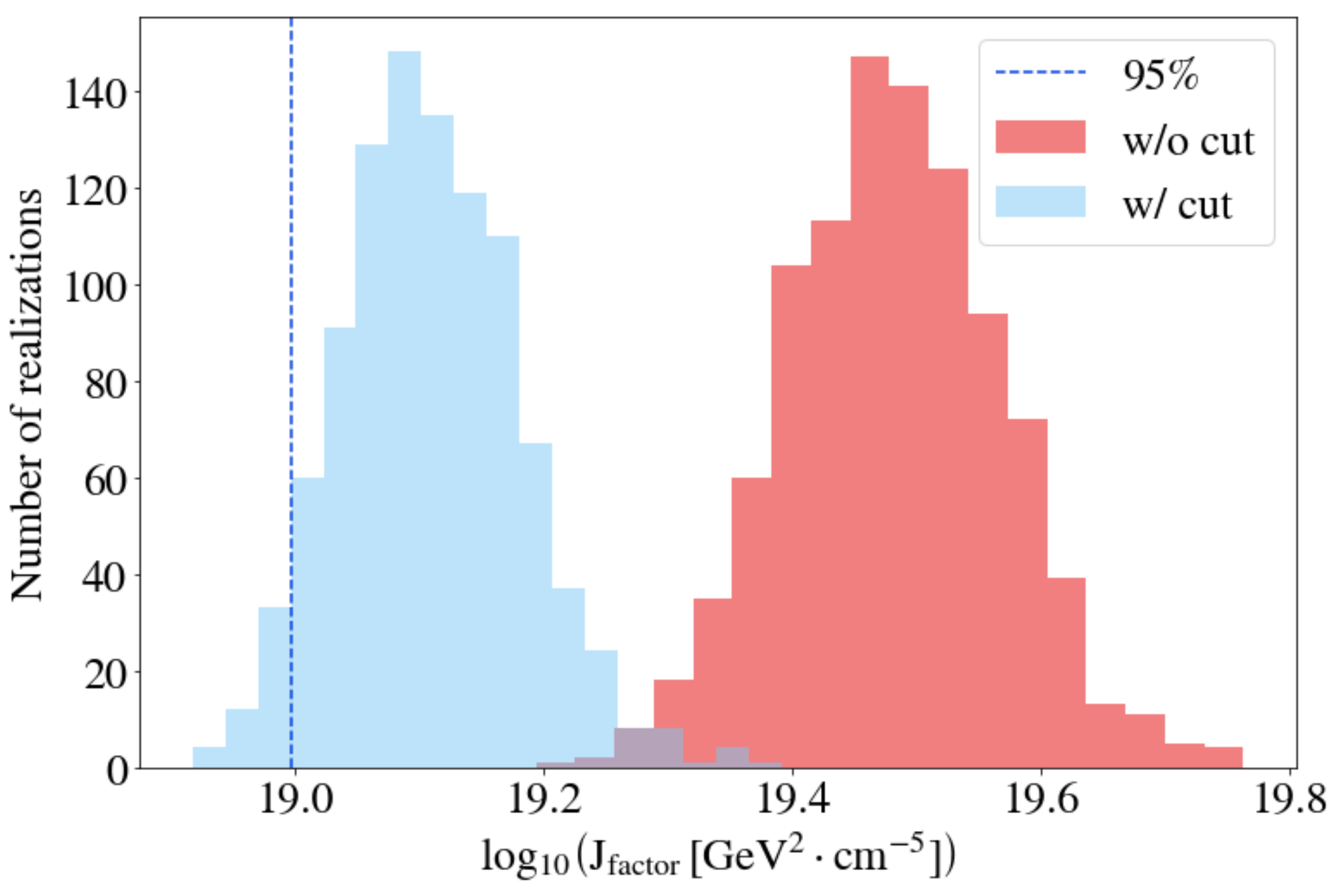}
\caption{Distribution of the 16th (i.e. the number of remaining DM candidates in the 3FGL catalog) brightest subhalo J-factor in each of the 1000 VL-II repopulation realizations. The red (blue) distribution refers to the case without (with) a cut in both subhalo mass ($M\geq10^7M_{\odot}$) and latitude ($|b|\leq10º$); see text for details. The dark blue vertical dashed line marks the J-factor above which 95\% of the blue J-factor distribution is contained. This is the J-factor we would adopt for the computation of the 95\% C.L. limits in this particular case.}
\label{fig:jfact_95dist}
\end{figure}

 More precisely, as we are interested in obtaining the limits at 95\% confidence level (C.L.), what we actually do for the generic case of having $n$ remaining candidates, is to draw the distribution of J-factors corresponding to the $n^{th}$ brightest subhalo across all the N-body realizations, and pick the J-factor value above which 95\% of this distribution is contained. To be consistent with our filtering procedure of \cref{sec:catalogs}, this J-factor distribution is obtained with a cut at $|b|\leq10º$ (to match the latitude filter) and $M\geq10^7M_{\odot}$ (to ensure that we only consider dark subhalos). As an example, we show in Figure \ref{fig:jfact_95dist} the J-factor distribution for the 3FGL setup, i.e. 16 remaining unIDs, with and without the mentioned cuts.

\begin{figure}[!ht]
\centering
\includegraphics[height=10cm]{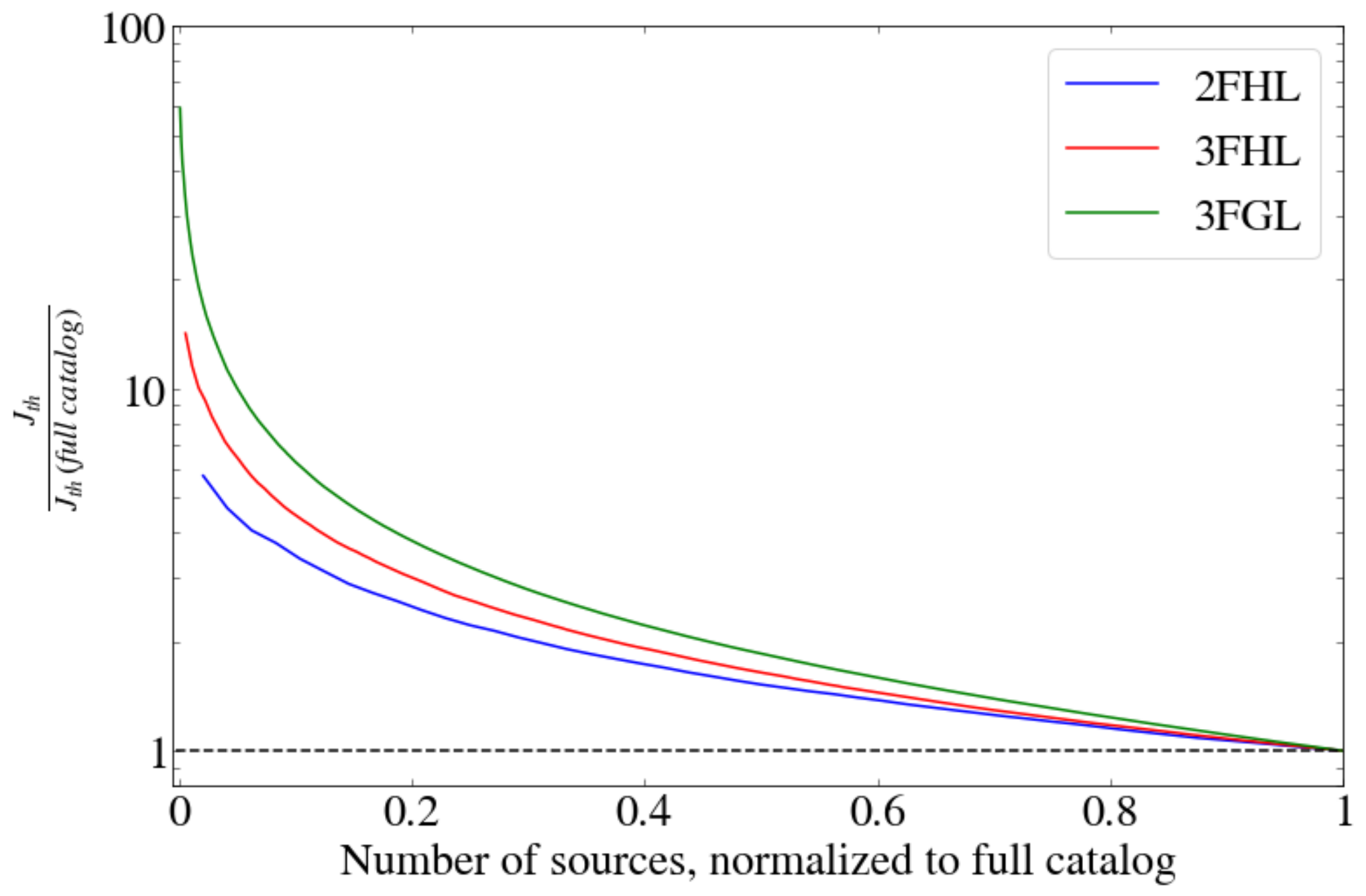}
\caption{Improvement in the DM constraints as a function of number of remaining unIDs in the catalogs. The improvement is codified in terms of a ratio of J-factors: the one to be used for the number of remaining sources after filtering, and the one corresponding to the full unID catalog (i.e. no unID rejection). Both J-factors refer to the values above which 95\% of the corresponding J-factor distributions, for each number of candidates, are contained; see text and Figure \ref{fig:jfact_95dist} for details. The horizontal dashed line shows no improvement at all due to (lack of) unIDs filtering, and it is computed with respect to the case of having a single remaining unID (sensitivity reach of the method; see Figure \ref{fig:comparison_one}). Note the exponential rise of constraining power once less than $\sim$20\% of sources are left.}
\label{fig:improvement}
\end{figure}

Clearly, the number of surviving DM subhalo candidates has a direct impact on the constraints: the fewer the candidates, the larger the adopted J-factor in \cref{eq:master_formula} will be, as the value to be used will correspond to a brighter simulated subhalo. As $\langle\sigma v\rangle\propto J^{-1}$, by lowering the number of surviving subhalo candidates we will be improving the derived limits on $\langle\sigma v\rangle$. This can be seen in Figure \ref{fig:improvement}, which shows the ratio between the J-factor that corresponds to having a certain number of remaining unIDs as subhalo candidates, and the J-factor to be used when the full list of unIDs is used instead. Both of them refer to the values above which 95\% of the corresponding J-factors distributions are contained, for each number of candidates. This ratio of J-factors is shown versus the normalized number of unIDs we are left with, for the three considered catalogs.

The figure summarizes the need for decreasing the number of unIDs as potential subhalo candidates, for which the careful filtering of sources performed in \cref{sec:catalogs} becomes critical. Indeed, from Figure \ref{fig:improvement} it can be seen that by rejecting up to $\sim$80\% of unIDs in a given catalog the improvement with respect to the full catalog (i.e., no filtering) is not relevant (around a factor 3-4 in the constraints), yet for bigger rejection rates the gain becomes exponential: every additional source we are able to remove will have a significant impact on the limits.

The maximum improvement that is possible to achieve depends on the catalog: for small catalogs, such as the 2FHL (48 unIDs), discarding 30 to 40 sources is almost equal to discarding the full catalog, while for the 3FGL (1010 unIDs) this same number represents just a tiny fraction. In other words, removing 80\% of unIDs in a catalog (to lie on the 20\% turnover of Figure \ref{fig:improvement}) is more difficult the more sources we have on the catalog. This explains e.g. that for the case of considering the minimum number of sources (1) in Figure \ref{fig:improvement} we obtain a different percentage for each catalog ($1/48 = 0.02$ for 2FHL; $1/177 = 0.006$ for 3FHL; $1/1010 = 0.001$ for 3FGL). And this is why the maximum improvement (i.e., removing all sources but one) is different between catalogs and equal to a factor 16 for 2FHL, 38 for 3FHL and 140 for 3FGL unIDs, with respect to having the full catalog of unIDs.

We note that, after the filtering performed in \cref{sec:catalogs} we are left with 8\%, 12\% and 2\% of the full list of unIDs in the 2FHL, 3FHL and 3FGL catalogs, respectively. These numbers already lie in the exponential-growing part of the curve shown in Figure \ref{fig:improvement}, showing the power of the unID filtering we performed. Since the 3FHL is the most recent catalog, a smaller fraction of its unIDs have been associated yet. Thus the corresponding DM constraints are not so competitive with respect to the other catalogs.

In short, we must reduce the number of potential subhalos to improve the DM constraints. The maximum potential of this method corresponds to the scenario in which we derive constraints for zero DM subhalo candidates, i.e., no unIDs compatible with DM subhalos. In the following, we will derive DM constraints to the annihilation cross section by i) adopting the number of potential DM subhalos in each catalog after our filtering work, and ii) by assuming that only 1 unID survives our cuts and is therefore compatible with being a subhalo. The latter case will be close to the sensitivity reach of the method as mentioned above.

\subsection{Current DM limits}
\begin{figure}[!ht]
\centering
\includegraphics[height=8.5cm]{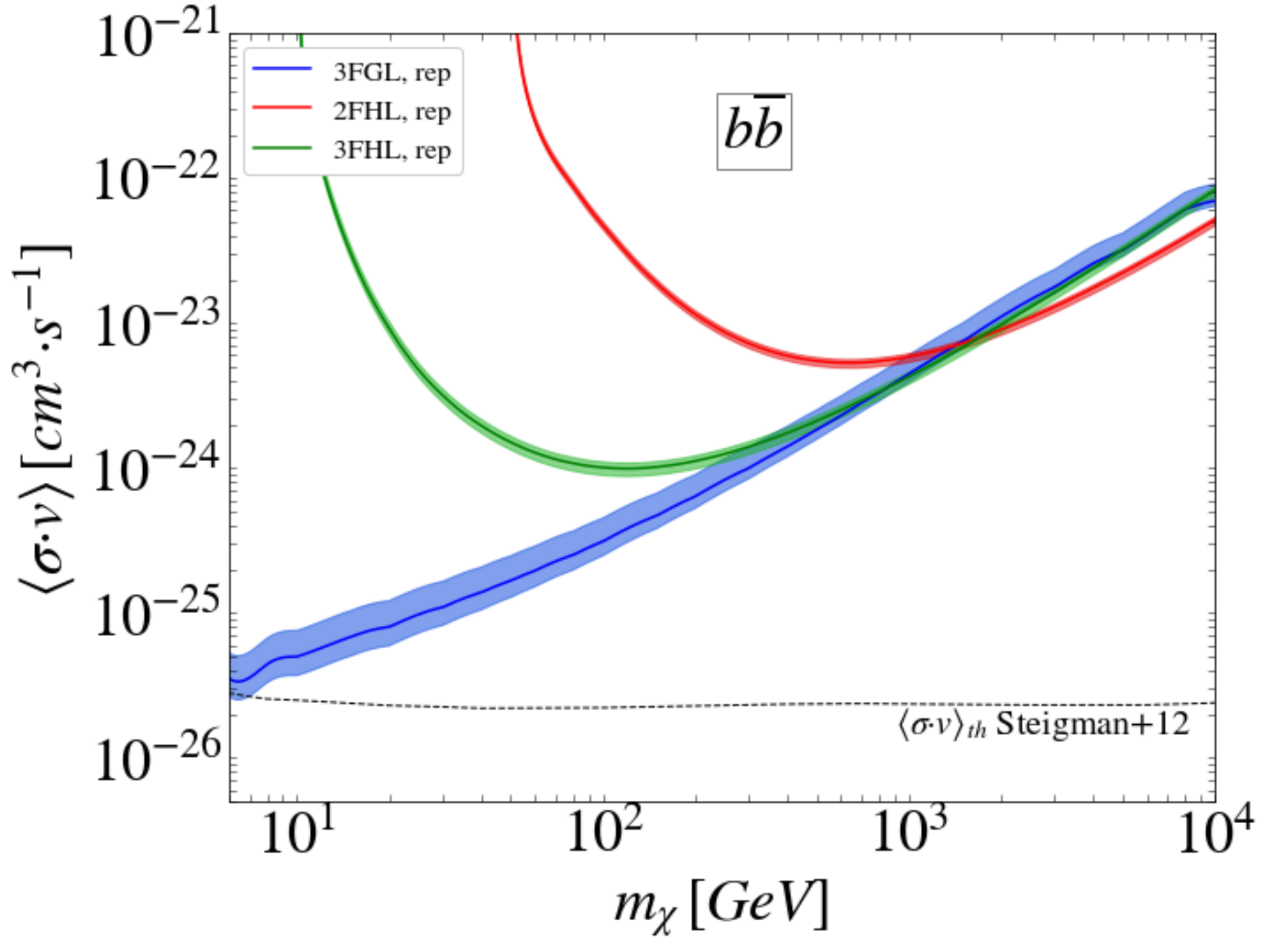}
\vfill
\includegraphics[height=8.5cm]{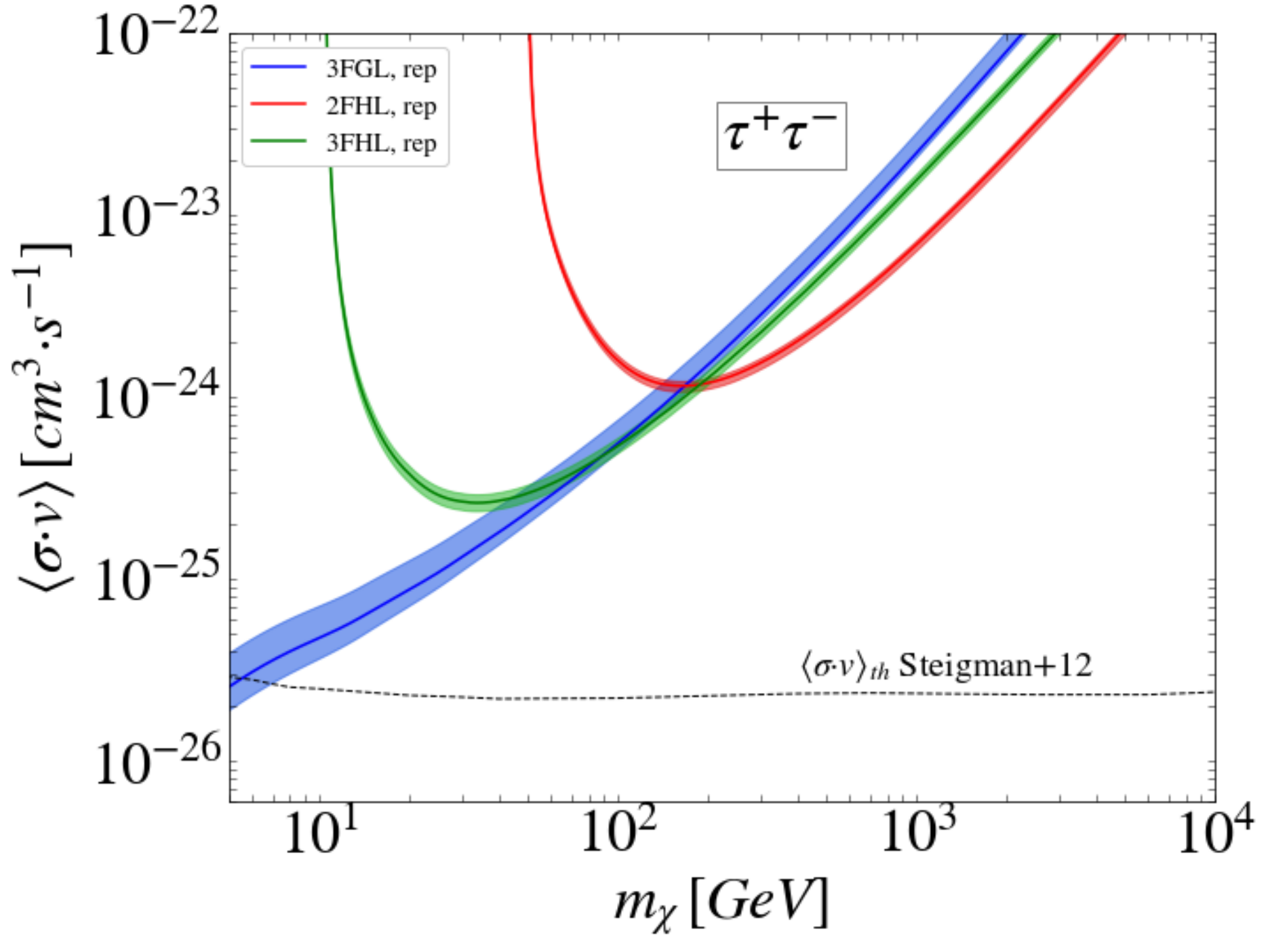}
\caption{Limits on the DM annihilation cross section for $b\overline{b}$ (top) and $\tau^+\tau^-$ (bottom) for the three LAT catalogs used in this work, and once the unID filtering detailed in \cref{sec:catalogs} has been applied to each of them. More precisely, 16, 4 and 24 unIDs remain in the 3FGL, 2FHL and 3FHL catalogs, respectively. The shaded bands refer to the 1-$\sigma$ uncertainty band coming from $F_{min}$; see text for details. The dashed line represents the thermal value of the annihilation cross section \cite{Steigman+12}. The "rep" label stands for repopulated.}
\label{fig:comparison_real}
\end{figure}

We first derive the constraints for the most realistic scenario, i.e., the one in which we consider all the unIDs that survive our proposed cuts in \cref{sec:catalogs} as potential DM subhalos.  This means that we are left with 16 candidates in the 3FGL, 4 in the 2FHL and 24 in the 3FHL. The results are shown in Figure \ref{fig:comparison_real} for both the $b\overline{b}$ and $\tau^+\tau^-$ annihilation channels and for the three catalogs.

We show in Figure \ref{fig:comparison_real} the 95\% C.L limits computed as explained in \cref{sec:impact_limits}. We also show the 1-$\sigma$ uncertainty band coming from the $F_{min}$ uncertainties, due to the average over the whole considered sky. Additionally, we conservatively include the 4\% false-rate positive as the error associated to the machine learning classification algorithms. We discuss in further detail on the computations of these errors in Appendix \ref{app:err_comp}. In Appendix \ref{app:constraints_no_repop} the limits are compared with those obtained with no repopulation.

As expected, the 3FGL catalog provides the best constraints in the low-mass range, while the 2FHL setup dominates the high-mass end covered by the LAT. More precisely, for $b\overline{b}$ ($\tau^+\tau^-$) the 2FHL provides the best constraints above $\sim$1 TeV (200 GeV). On the other hand, the 3FHL does not improve significantly the constraints for medium masses with respect to the other two due to the more inefficient filtering. Note also that the 2FHL and 3FHL constraints go to infinity at their respective energy thresholds, 50 and 10 GeV. This is an expected result according to \cref{eq:master_formula}, because $N_{\gamma} = 0$ below the catalog thresholds.

\subsection{Sensitivity reach of the method}
\label{sec:max_potential}
The maximum potential of the method is reached for the case in which no unID is compatible with the DM subhalo scenario.  However, even in this case we must adopt a J-factor in order to set constraints.  We do this by using the J-factor of the brightest object in the simulation. This may look similar to the case in which still one unID is compatible with DM. However, it is conceptually different: in the latter case the resulting sensitivity curve refers to the cross section needed to have one subhalo detected, while in the zero unID case this same sensitivity curve will indeed represent an upper limit to the allowed values. Note, also, that for this reason the sensitivity reach obtained this way will be conservative.

Although it may seem overly optimistic and unlikely that we are able to eliminate all the unIDs in a given catalog as potential DM subhalos, in fact many efforts are currently ongoing to associate the largest possible number of unIDs with known objects. Thus, it may be possible to actually reach this goal in the future.

\begin{figure}[!ht]
\centering
\includegraphics[height=8.5cm]{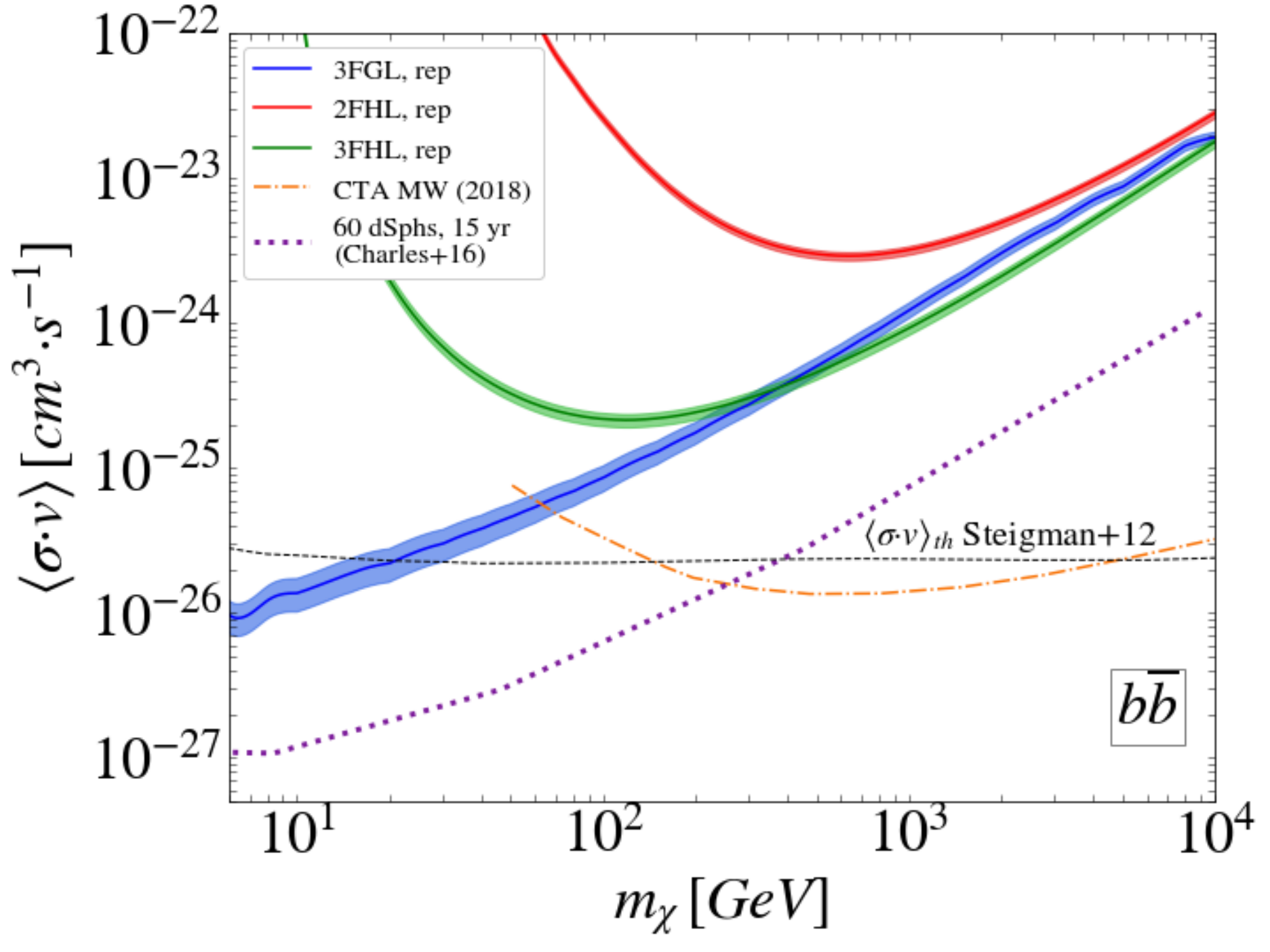}
\vfill
\includegraphics[height=8.5cm]{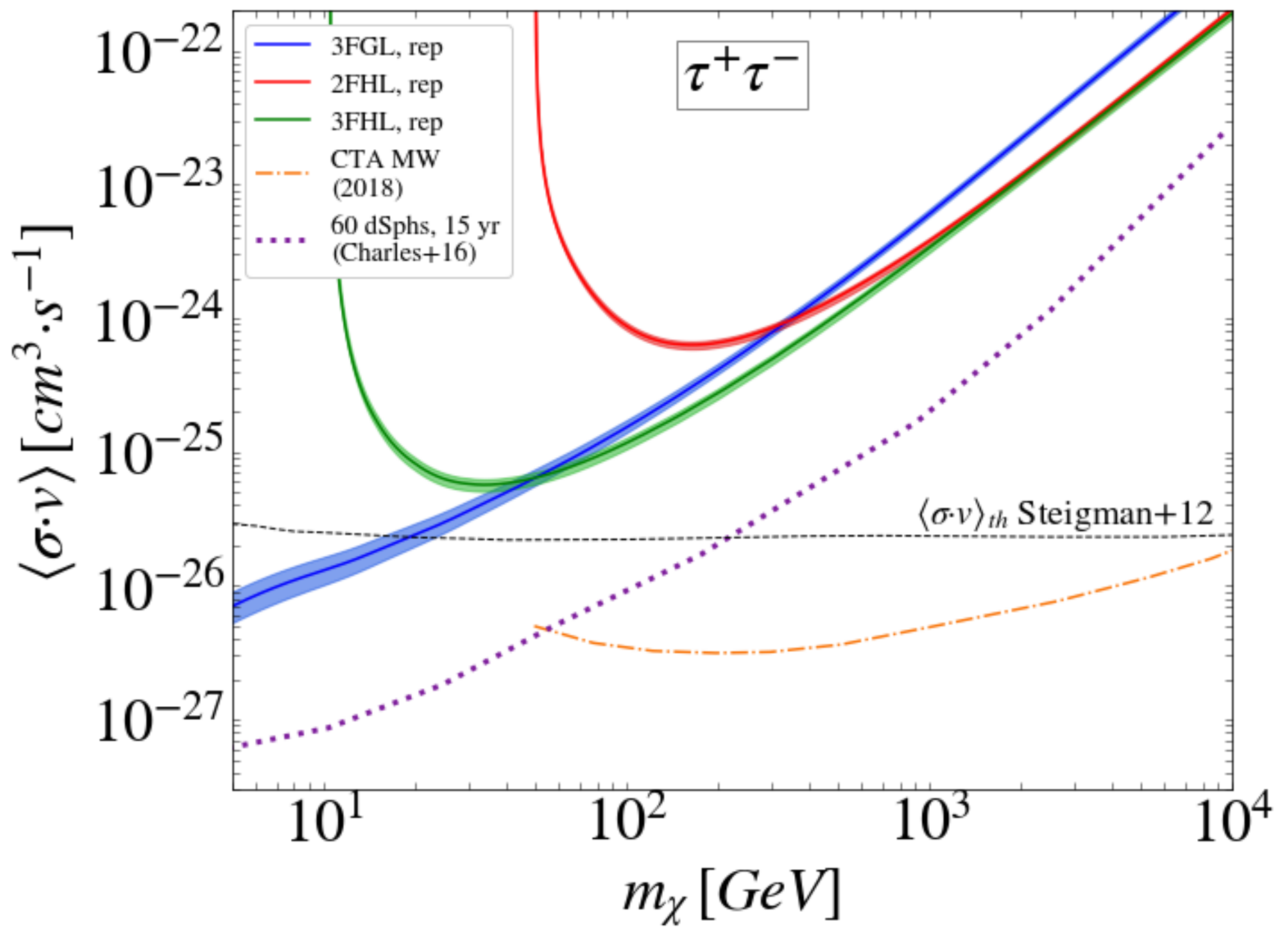}
\caption{Same as Figure \ref{fig:comparison_real}, but for the sensitivity reach scenario presented in \cref{sec:max_potential}, where only 1 unID is left in each corresponding catalog. We also compare our projections with the predictions for dwarfs with the Fermi LAT \cite{dsphs_projections} and from the Milky Way halo with CTA \cite{Lefranc+15}.}
\label{fig:comparison_one}
\end{figure}

Figure \ref{fig:comparison_one} shows the sensitivity reach for the three catalogs under consideration. In the figure, we also include the projected limits for \textit{Fermi}-LAT 60 dwarf galaxies in 15 years of operation \cite{dsphs_projections}, and the latest CTA prospects for the Milky Way halo \cite{CTA_science_paper} (note that these assume no uncertainties).

\section{Discussion and conclusions}
\label{sec:conclusions}

We have presented a competitive method to constrain the nature of DM, assuming the WIMP model. Our method is based on the search for DM subhalo candidates among the sample of unID sources in \textit{Fermi}-LAT point-source catalogs. We performed an exhaustive unID filtering in \cref{sec:catalogs} with the intention to remove all those unIDs with features not compatible with being a DM subhalo. Our selection cuts included: (1) astrophysical associations; (2) a Galactic latitude cut to avoid issues with the bulk of the Galactic astrophysical sources and diffuse contamination; (3) variability studies in the official catalogs and with the FAVA tool; (4) machine learning classification algorithms; (5) unambiguous multiwavelength emission of the unID; (6) previously identified complex regions. These selection criteria did not include a dedicated LAT spectral and spatial analysis. This work is ongoing and will be presented elsewhere.

Lacking knowledge on the actual nature of the remaining unIDs after our filtering process (16, 4 and 24 in the 3FGL, 2FHL and 3FHL catalogs, respectively), and in the absence of a clear hint of DM annihilation, we set 95\% upper limits on the $\langle\sigma v\rangle-m_{\chi}$ parameter space. Three basic ingredients were needed to do so. First, we computed the DM annihilation spectra for different annihilation channels and WIMP masses by making use of the PPPC4ID tables of Ref.~\cite{Cirelli+12}. Second, we characterized the LAT sensitivity to DM with unprecedented accuracy by simulating a population of (point-like) DM subhalos at each sky position, annihilation channel and DM mass, and by computing the minimum integrated flux needed to to have a $TS=25$ signal in the LAT, i.e., a source detection. This was done for each considered catalog by mimicking its exact configuration, like exposure time, and diffuse and isotropic templates. 
Third, we relied on results from the VL-II N-body cosmological simulation to compute subhalo J-factors. Since our work focused on low-mass, completely ``dark'' subhalos, we performed an additional simulation work to {\it repopulate} the original VL-II with low-mass subhalos well below its formal resolution limit so as not to miss any potential bright dark subhalo. This was done by assuming the same radial distribution and mass function for subhalos as observed in the mass range well resolved in the parent simulation. Subhalo structural properties were modeled using the state-of-the-art subhalo mass-concentration relation by Ref.~\cite{Moline+17}. This repopulation exercise was repeated 1000 times to derive statistically meaningful results. The final outcome was a prediction of the distribution of expected subhalo J-factors in $\Lambda$CDM.

\begin{figure}[!ht]
\centering
\includegraphics[height=8.5cm]{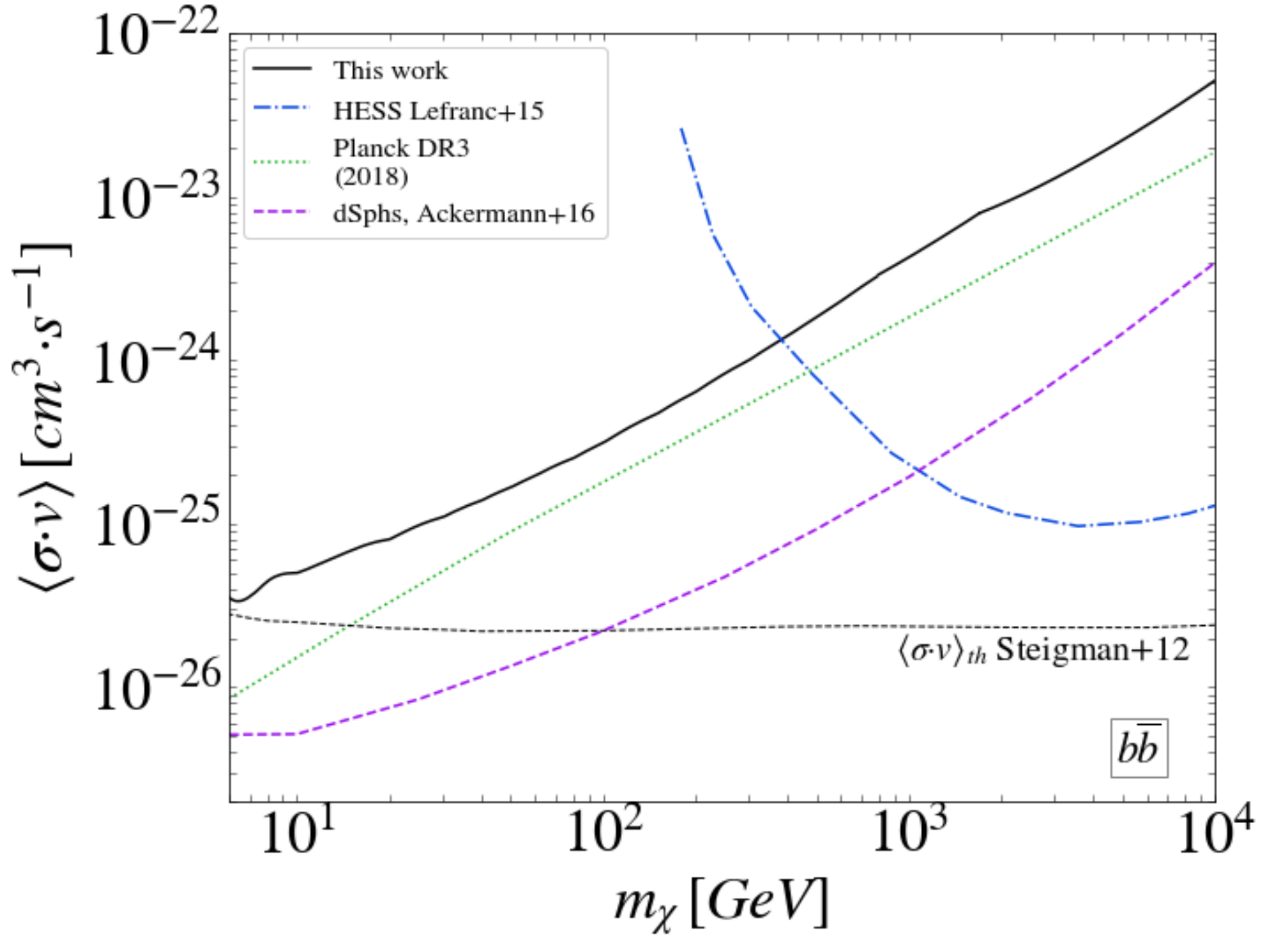}
\vfill
\includegraphics[height=8.5cm]{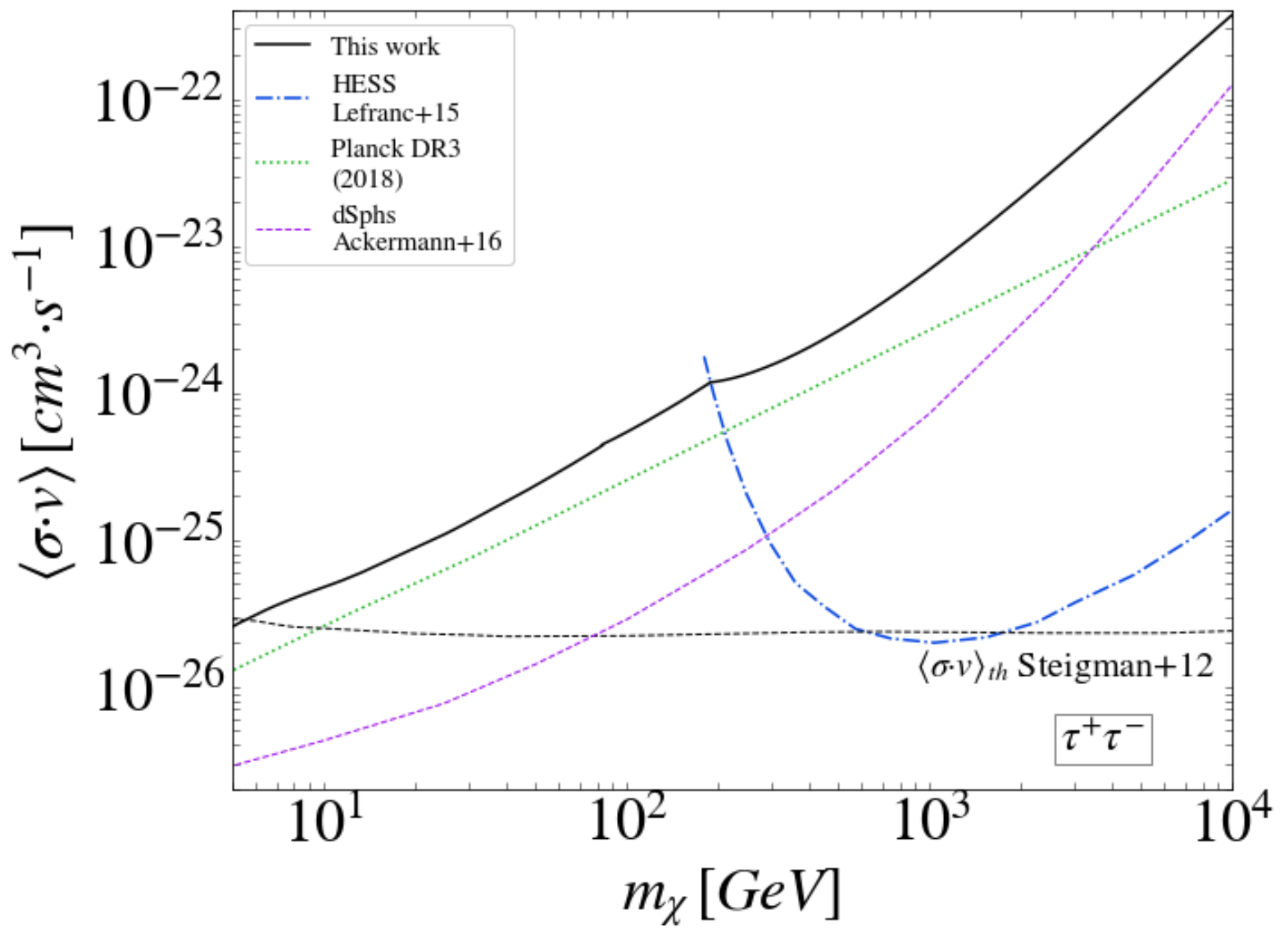}
\caption{Comparison of DM limits for different targets and probes. Solid line corresponds to the envelope of the three sets of limits shown in Figures \ref{fig:comparison_real} and \ref{fig:comparison_one} for the realistic scenario (\cref{sec:constraints}). dot-dashed, dashed and dotted lines are, respectively, the DM constraints derived by H.E.S.S. for the Milky Way halo \cite{Hess_Lefranc+15}, by the \textit{Fermi}-LAT for dSphs \cite{dsphs_paper}, and by Planck using the CMB DR3 latest release \cite{Planck_dr3}. Top panel is for $b\overline{b}$ and bottom panel for $\tau^+\tau^-$ annihilation channel.}
\label{fig:comparison_envolvente}
\end{figure}

With a precise characterization of these three basic ingredients at hand, we were able to set conservative yet competitive constraints in Figure \ref{fig:comparison_real}. We also put our results in context by comparing them with other state-of-the-art constraints coming from H.E.S.S., Planck and LAT in Figure \ref{fig:comparison_envolvente}. The obtained DM limits reach the level of the thermal relic cross section for both $b\overline{b}$ and $\tau^+\tau^-$ at the smallest WIMP masses considered, indeed ruling out thermal WIMPs up to 6 GeV in the case of $\tau^+\tau^-$. Our constraints are complementary and independent to the ones obtained by means of other targets such as dSphs \cite{dsphs_paper}.

The results of this paper can be compared with those found in Ref.~\cite{Calore+17}, where a study of LAT sensitivity to DM subhalos was also performed. The $F_{min}$ results for $b\overline{b}$ and $\tau^+\tau^-$ are fully compatible. We here extend the calculation to many other channels and use a much finer grid. In our work, we performed a filtering of unIDs that allowed us to derive a conservative number of catalog sources that cannot be discarded as DM subhalos at present time, and set {\it realistic} DM limits according to this number. In contrast, Ref.~\cite{Calore+17} did not perform an unID filtering and simply fixed the number of remaining candidates (namely 20, 5, 0) when presenting their limits\footnote{Indeed, they implicitly adopted one remaining source instead of zero for their sensitivity reach; V. de Romeri, private communication.}. For the same number of considered sources, we improve their results roughly one order of magnitude once our low-mass subhalo repopulation work and the latest subhalo mass-concentration relation are implemented. Yet, we must bear in mind that the repopulation and characterization of the subhalo population was done with VL-II, a DM-only, WMAP-cosmology N-body cosmological simulation. In the next years, new Milky-Way-size simulations should be available that will adopt the Planck cosmology and will include baryonic physics as well. All together, this may significantly alter the current subhalo predictions. Work is already ongoing to help clarify the properties of the subhalo population that are expected to be more relevant for DM annihilation searches (abundance, radial distribution, inner structure).

We also made an effort to quantify the involved uncertainties in our limits, which we summarized in \cref{app:err_comp}. In short, for a small number of sources we are dominated by the uncertainty in the J-factor, which is implicitly taken into account in the computation of the 95\% limits, while for large unID samples the uncertainty in $F_{min}$ is the largest.

In order to put the DM limits derived in this work into a more general context,  we compare them in Figure \ref{fig:comparison_envolvente} to those obtained by the \textit{Fermi}-LAT from dSphs \cite{dsphs_paper}, by the H.E.S.S. Collaboration from the Milky Way halo \cite{dm_hess_paper} and by the Planck Collaboration using the Cosmic Microwave Background (CMB) DR3 recent results \cite{Planck_dr3}. For this figure, we only show the envelope of the (mean) limits presented in Figure \ref{fig:comparison_real} for the three catalogs. This allows a single, best-curve representation of the three sets of limits. Figure \ref{fig:comparison_envolvente} shows, once again, that DM subhalos can yield very competitive limits compared to other targets and probes.

We also studied the maximum potential or sensitivity reach of our method to set limits. More precisely, we obtained that by having one single DM subhalo candidate left in each of the catalogs, the method would potentially yield the best limits to the annihilation cross section so far, discarding the thermal relic cross section value up to $\sim$20 GeV in the case of annihilation into $b\overline{b}$ and $\tau^+\tau^-$. This sensitivity reach scenario is indeed a plausible one, as the increasing quantity and quality of available data, as well as several ongoing observational campaigns at other wavelengths could definitely unveil the true nature of the remaining unID sources. Also, as already mentioned, a dedicated spectral and spatial analysis is currently ongoing, which will hopefully be able to either reject additional sources as DM subhalos (thus improving the current constraints significantly, as illustrated in Figure \ref{fig:improvement}), or point out very interesting sources from the DM perspective. At larger WIMP masses, the Cherenkov Telescope Array (CTA) in the near future \cite{CTA_science_paper} will provide results at higher energies than \textit{Fermi}-LAT. A combination of both instruments may be able for the first time to test the WIMP paradigm over more than 4 decades in energy. \\

The most relevant results from this work are being made public for community's use. In particular, the tabulated DM constraints for both the conservative and sensitivity reach scenarios; tables of $F_{min}$ for the whole considered WIMP masses and channels; and the full list of rejected unID sources containing the rejection criteria applied in each case.

\acknowledgments
The authors would like to thank Rolf B\"uhler, Elizabeth Ferrara, Dan Kocevski and Benoit Lott for their valuable help on the different aspects of this work.

JCB and MASC are supported by the {\it Atracci\'on de Talento} contract no. 2016-T1/TIC-1542 granted by the Comunidad de Madrid in Spain. AD thanks the support of the Ram{\'o}n y Cajal program from the Spanish MINECO. AAS is very grateful to the IFT Centro de Excelencia ``Severo Ochoa'' Spanish program under reference SEV-2016-0597. MDM acknowledges support by the NASA {\it Fermi} Guest Investigator Program 2014 through the {\it Fermi} multi-year Large Program N. 81303 (P.I. E.~Charles) and by the NASA {\it Fermi} Guest Investigator Program 2016 through the {\it Fermi} one-year Program N. 91245 (P.I. M.~Di Mauro). DN wants to acknowledge support by the Spanish Ministry of Economy, Industry, and Competitiveness / ERDF UE grant FPA2015-73913-JIN and partial support by NASA's Fermi Guest Investigator Program (Cycle 7) NNX14AQ70G. The work of JCB, MASC and AAS was additionally supported by the Spanish Agencia Estatal de Investigación through the grants PGC2018-095161-B-I00, IFT Centro de Excelencia Severo Ochoa SEV-2016-0597, and Red Consolider MultiDark FPA2017-90566-REDC.

The Fermi LAT Collaboration acknowledges generous ongoing support from a number of agencies and institutes that have supported both the development and the operation of the LAT as well as scientific data analysis. These include the National Aeronautics and Space Administration and the Department of Energy in the United States, the Commissariat `a l’Energie Atomique and the Centre National de la Recherche Scientifique / Institut National de Physique Nucl\'eaire et de Physique des Particules in France, the Agenzia Spaziale Italiana and the Istituto Nazionale di Fisica Nucleare in Italy, the Ministry of Education, Culture, Sports, Science and Technology (MEXT), High Energy Accelerator Research Organization (KEK) and Japan Aerospace Exploration Agency (JAXA) in Japan, and the K. A. Wallenberg Foundation, the Swedish Research Council and the Swedish National Space Board in Sweden. Additional support for science analysis during the operations phase is gratefully acknowledged from the Istituto Nazionale di Astrofisica in Italy and the Centre National d'Etudes Spatiales in France. This work performed in part under DOE Contract DE- AC02-76SF00515

This research made use of Python, along with community-developed or maintained software packages, including IPython \cite{Ipython_paper}, Matplotlib \cite{Matplotlib_paper}, NumPy \cite{Numpy_paper}, SciPy \cite{scipy_paper} and Healpix \cite{healpix_paper}. This work made use of NASA’s Astrophysics Data System for bibliographic information.


\appendix

\section{Impact of uncertainties}
\label{app:err_comp}

In this paper we have considered those uncertainties associated to $F_{min}$, J-factor, and machine learning techniques. We assumed the DM annihilation spectra (see \cref{sec:ann_spectra}) to have negligible errors.

Here, we perform a study to understand how these uncertainties behave with the considered catalog, DM mass and annihilation channel. Note that the J-factor uncertainty does not depend on the DM mass, as it only depends on the number of unID sources left (see \cref{sec:impact_limits}).

The $F_{min}$ relative error can be computed as:

\begin{equation}
\label{eq:error_fmin}
\varepsilon_{F}=100\cdot\frac{\sigma_F}{F_{min}^{avg}},
\end{equation}

\noindent where $F^{avg}_{min}$ is the average minimum flux and $\sigma_F$ is the standard deviation of the corresponding log-normal distribution, both computed over the whole sky but the $|b|\leq 10^\circ$ band. We note that in some cases this log-normal distribution may significantly differ from being Gaussian. Also, we note that there may be other second-order effects in the characterization of this quantity, such as inhomogeneities or slight deviations with respect to the mean values across the sky, yet they are expected to be subdominant.

The J-factor uncertainty comes from the cosmic variance across 1000 realizations from the N-body simulations, and may be computed as:

\begin{equation}
\label{eq:error_jfact}
\varepsilon_{J}=100\cdot\frac{\sigma_J}{J^{avg}}
\end{equation}

\noindent where $J^{avg}$ is the average of the J-factors over the different realizations of the repopulated simulation, and $\sigma_J$ is their standard deviation, using the $n^{th}$ J-factor for the considered $n$ subhalos. This implicitly assumes that the J-factor distribution is well represented by a Gaussian; however we warn that for a very low number of sources ($n<3$) this distribution may exhibit significant departures from it.

\begin{figure}[!ht]
\centering
\includegraphics[height=6.4cm]{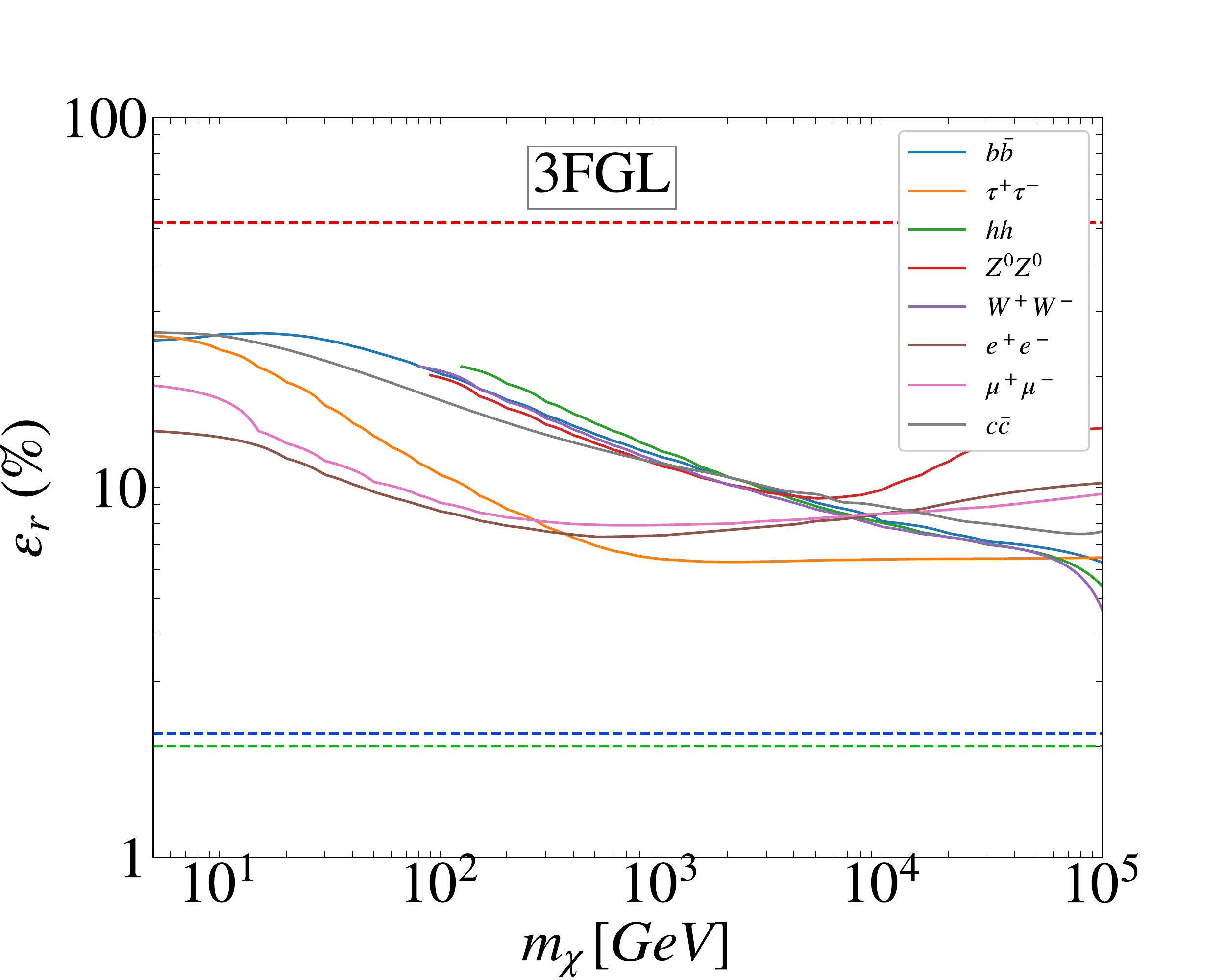} 
\vfill
\includegraphics[height=6.4cm]{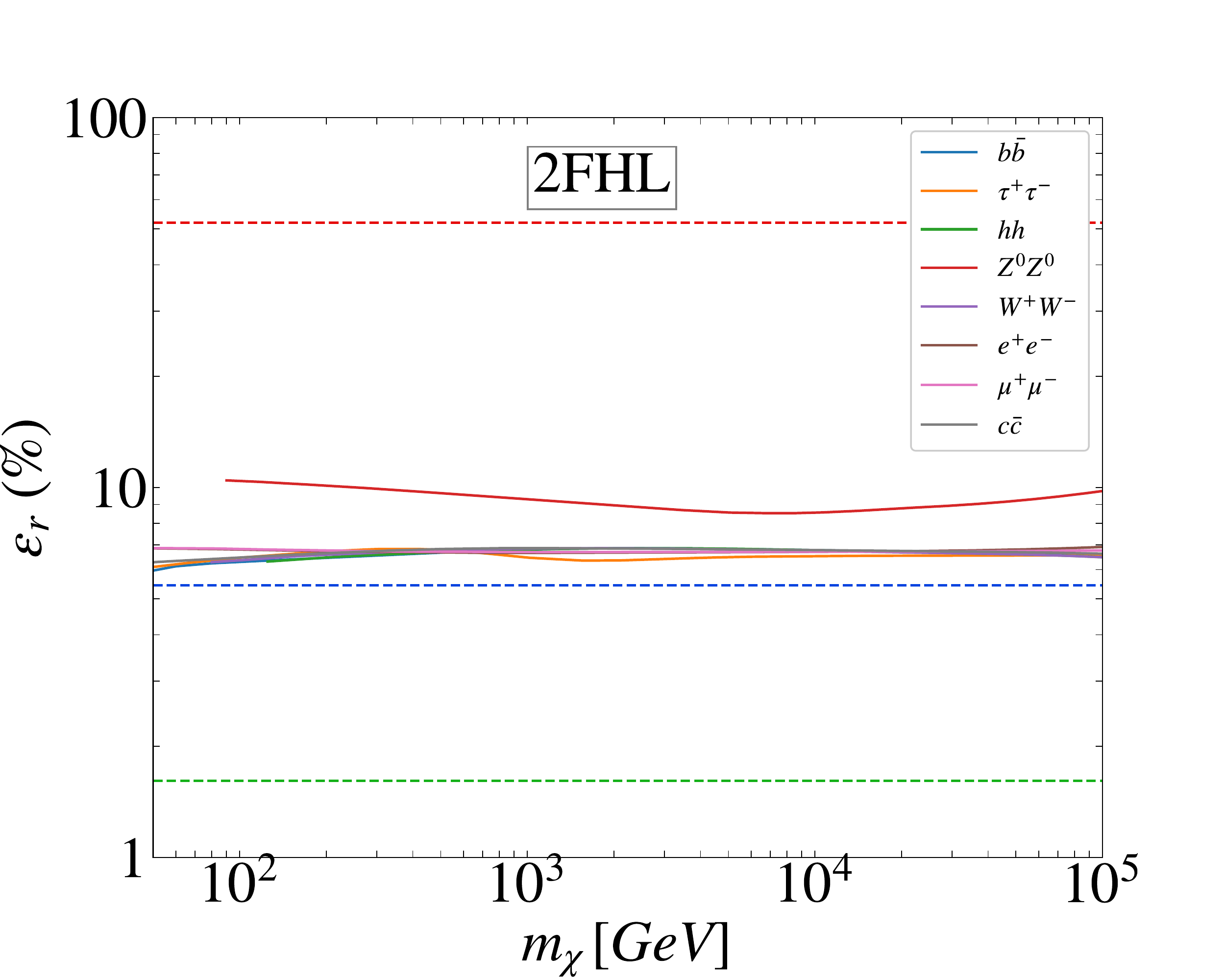}
\vfill
\includegraphics[height=6.4cm]{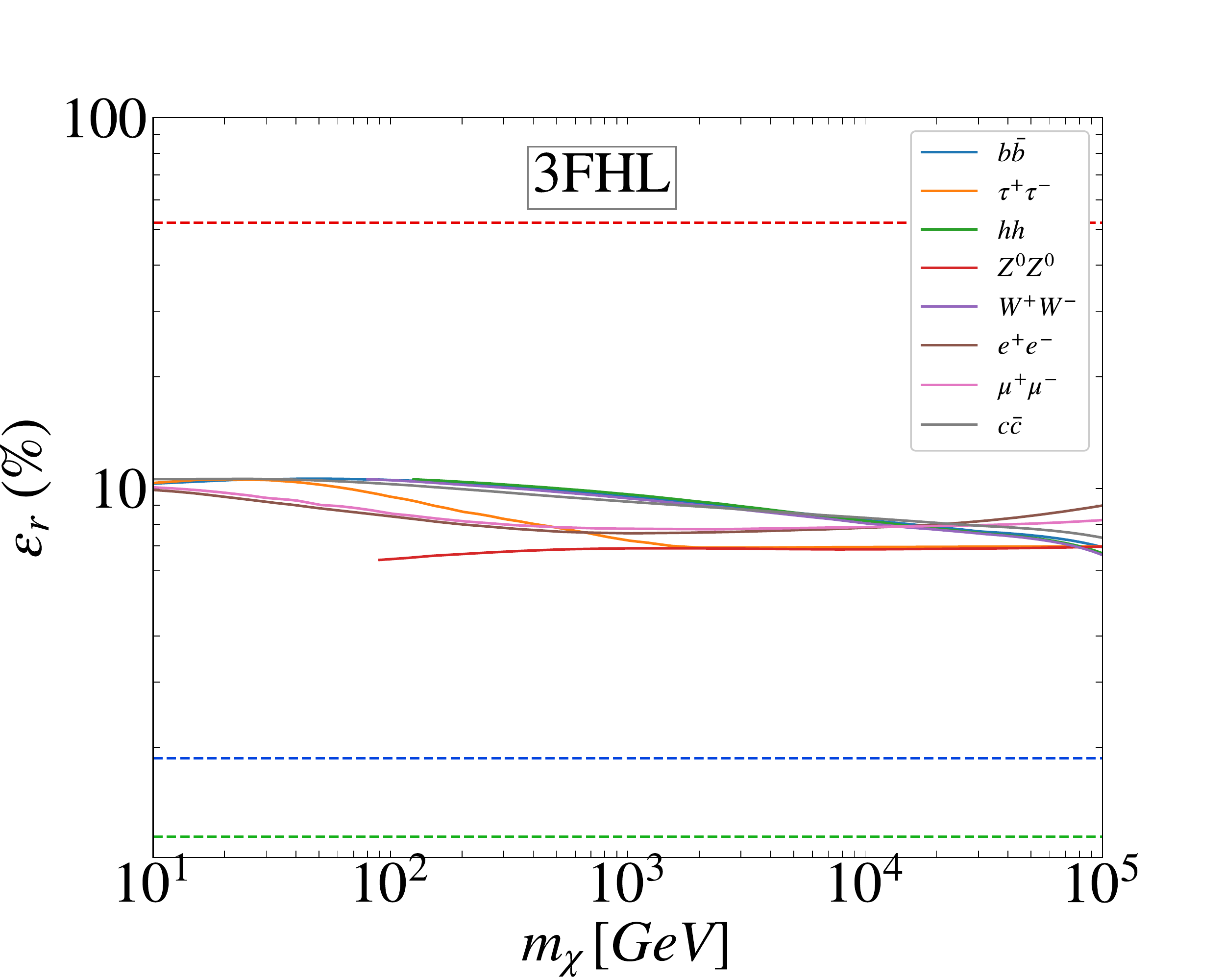}
\caption{Comparison of $\varepsilon_F$ versus $\varepsilon_J$, i.e. the relative errors associated to the $F_{min}$ (solid lines) and J-factor (dashed lines), respectively, as a function of the DM particle mass and different annihilation channels. From bottom to top, for the 3FGL, 2FHL and 3FHL catalogs. The dashed lines are, from bottom to top, the $\varepsilon_J$ of the full number of unIDs in the corresponding catalog (green), the number of unIDs in the realistic scenario detailed in \cref{sec:constraints} (blue) and the 1-source sensitivity reach of \cref{sec:max_potential} (red).}
\label{fig:errors_fmin_jfact_compared}
\end{figure}

We show a comparison between both $\varepsilon_F$ and $\varepsilon_J$ as given by the above expressions in Figure \ref{fig:errors_fmin_jfact_compared}. We stress that $\sigma_J$ refers to the mean J-factor ($J^{avg}$) and not to the one we used to set the 95\% C.L. upper limits (see \cref{sec:constraints}).
As it can be seen, $\varepsilon_J$ is sub-dominant with respect to $\varepsilon_F$ when the number of unIDs is large. Then, below $\sim$15 unIDs, $\varepsilon_J$ comes comparable or even larger than $\varepsilon_F$. 

As for the dependency of these uncertainties with annihilation channel: the general trend is a decrease of $\varepsilon_F$ as the DM mass increases. There are some exceptions: $Z^0Z^0$, $\mu^+\mu^-$ and $e^+e^-$. For these, $\varepsilon_F$ increases mildly for the heaviest DM masses. It is worth noting that the $Z^0Z^0$ channel behaves slightly different under the 2FHL setup, while the other channels behave always very similarly independently of the catalog.

Concerning the different catalog setups, $\varepsilon_F$ is typically larger for the 3FGL setup because of the larger variations of $F_{min}$ in this case compared to the other catalogs (the diffuse emission is more intense at lower energies, and the 3FGL energy threshold is 100 MeV so it is the most affected one).

Finally, the 4\% false association rate from the use of the machine learning criterium in our work has also been implemented in the DM constraints by adopting all of this 4\% rate as an uncertainty. The net effect is a broadening of the J-factor (upper) uncertainty band, since a slightly larger number of unIDs cannot be discarded. This effect is only non-negligible in the case of the 3FGL catalog, where we applied the machine learning to 162 sources and therefore this 4\% false rate uncertainty translates into 6 potentially wrongly rejected sources. Namely, the number of unIDs left after our filtering procedure increases from 16 to a maximum of 22.

\section{Impact of the low-mass subhalo repopulation on DM limits}
\label{app:constraints_no_repop}
\begin{figure}[!ht]
\centering
\includegraphics[height=5.65cm]{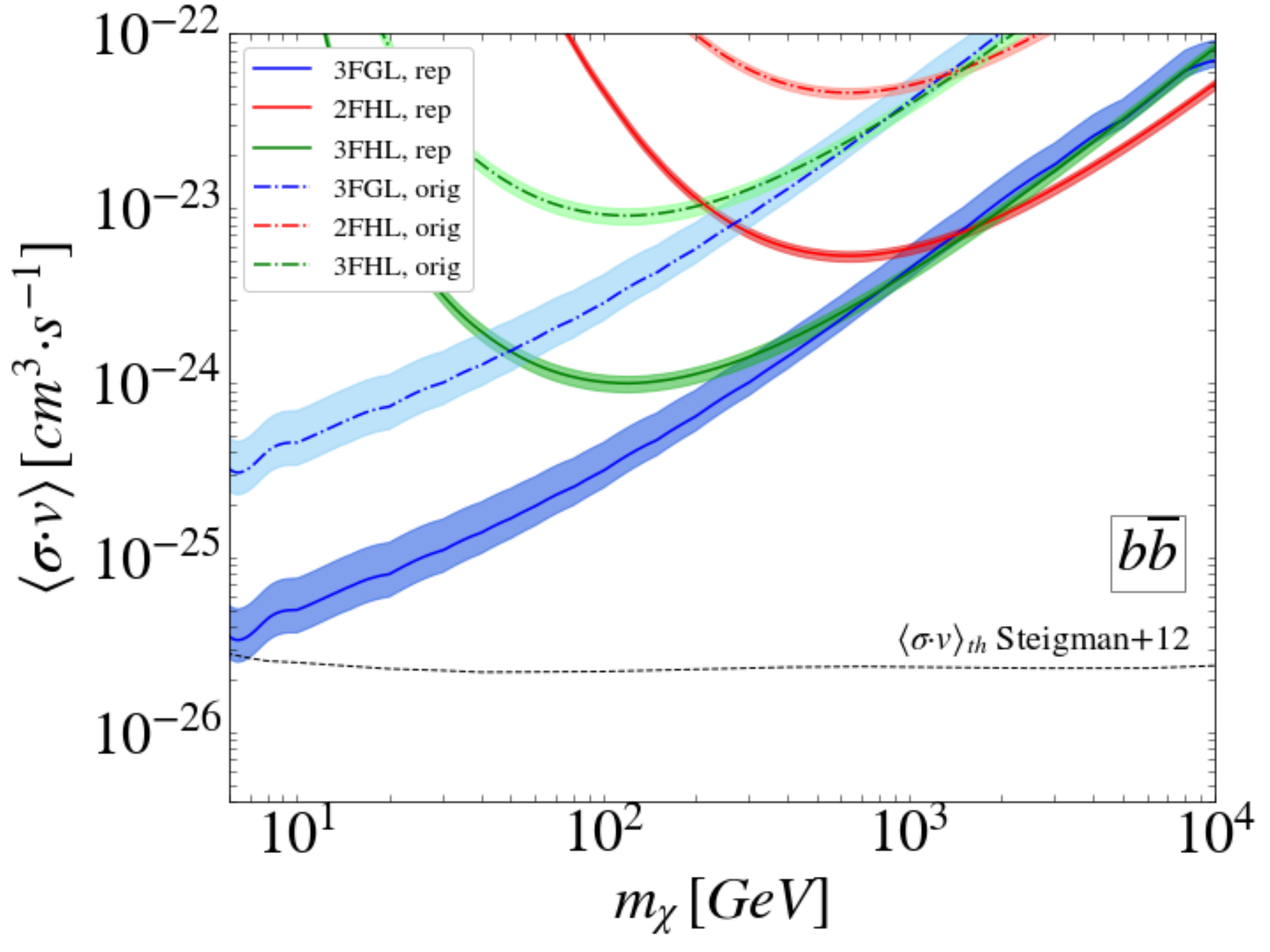}
\hfill
\includegraphics[height=5.5cm]{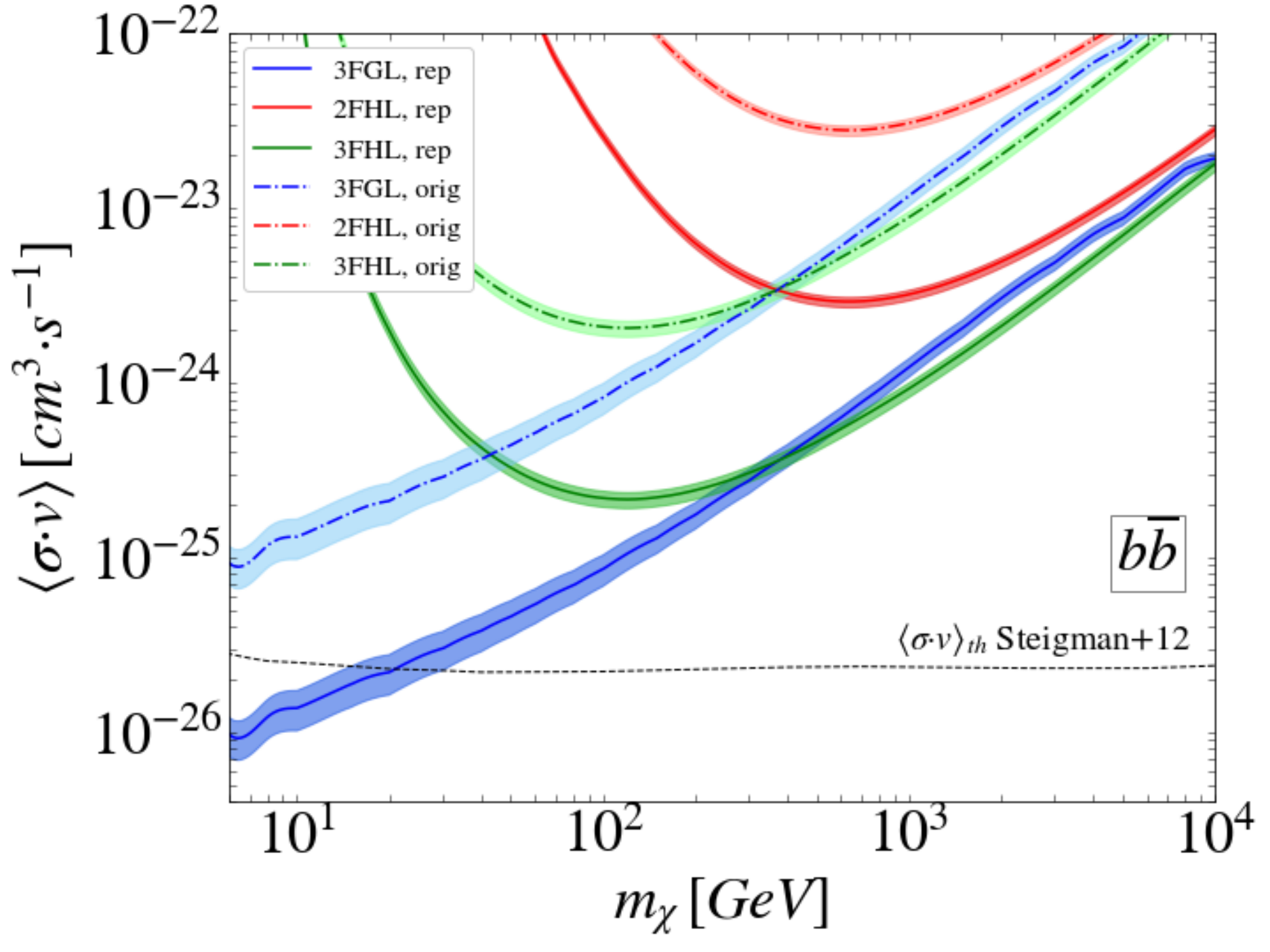}
\caption{Comparison of DM 95\% C.L. upper limits found with and without low-mass subhalo repopulation below the resolution limit of the original VL-II simulation. Both panels are for $b\overline{b}$, in the three catalog setups. Dot-dashed lines are the constraints from the original VL-II simulation (``orig''), while the solid lines use the repopulated simulation (``rep''). Left: current number of DM subhalo candidates in the catalogs (see Figure \ref{fig:comparison_real}); right: sensitivity reach of the method (see Figure \ref{fig:comparison_one}).}
\label{fig:comparison_repop}
\end{figure}

Here we compare the DM constraints presented on \cref{sec:constraints} to those without repopulating VL-II with low-mass subhalos below the resolution limit of the simulation. Figure \ref{fig:comparison_repop} shows such a comparison for both the case that adopts the current remaining number of DM subhalo candidates in the catalogs (\cref{sec:catalogs}), and the sensitivity reach case (\cref{sec:max_potential}) as well as the constraints both with and without the repopulation.

The differences in the J-factor distribution between the original and repopulated VL-II simulation are large enough to improve the limits by a factor $\sim 10$. At high masses, limits worsen due to the loss of sensitivity of the \textit{Fermi}-LAT, and all three catalog setups converge to approximately the same value. Similar improvements are found when considering either the realistic scenario or the sensitivity reach.

From these results, we find low-mass subhalos to be especially relevant for this work, as a significant number of them are expected to exhibit similar annihilation fluxes than resolved, more massive objects in the original simulation. We note that there may be still room for some further improvement by extending our repopulation work to include even smaller subhalos masses (i.e., below $\gtrsim 10^3$ M\textsubscript{\(\odot\)}). This additional numerical work will be done elsewhere.

\section{Full set of DM constraints}
\label{app:full_set}

\begin{figure}[!ht]
\centering
\includegraphics[height=8cm]{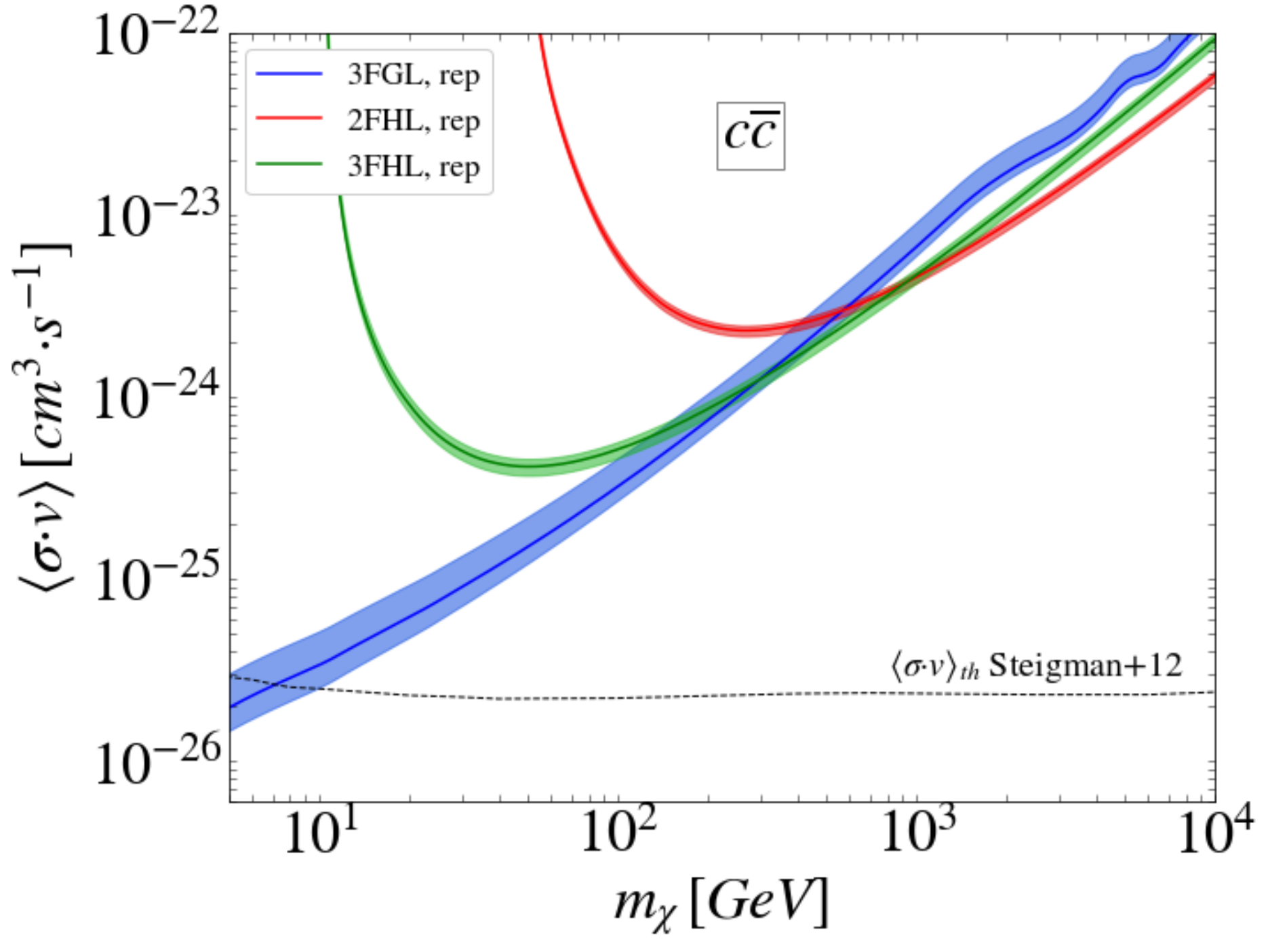}
\caption{Same as Figure \ref{fig:comparison_real} but for $c\overline{c}$ annihilation channel.}
\label{fig:constraints_real_cc}
\end{figure}

\begin{figure}[!ht]
\centering
\includegraphics[height=8cm]{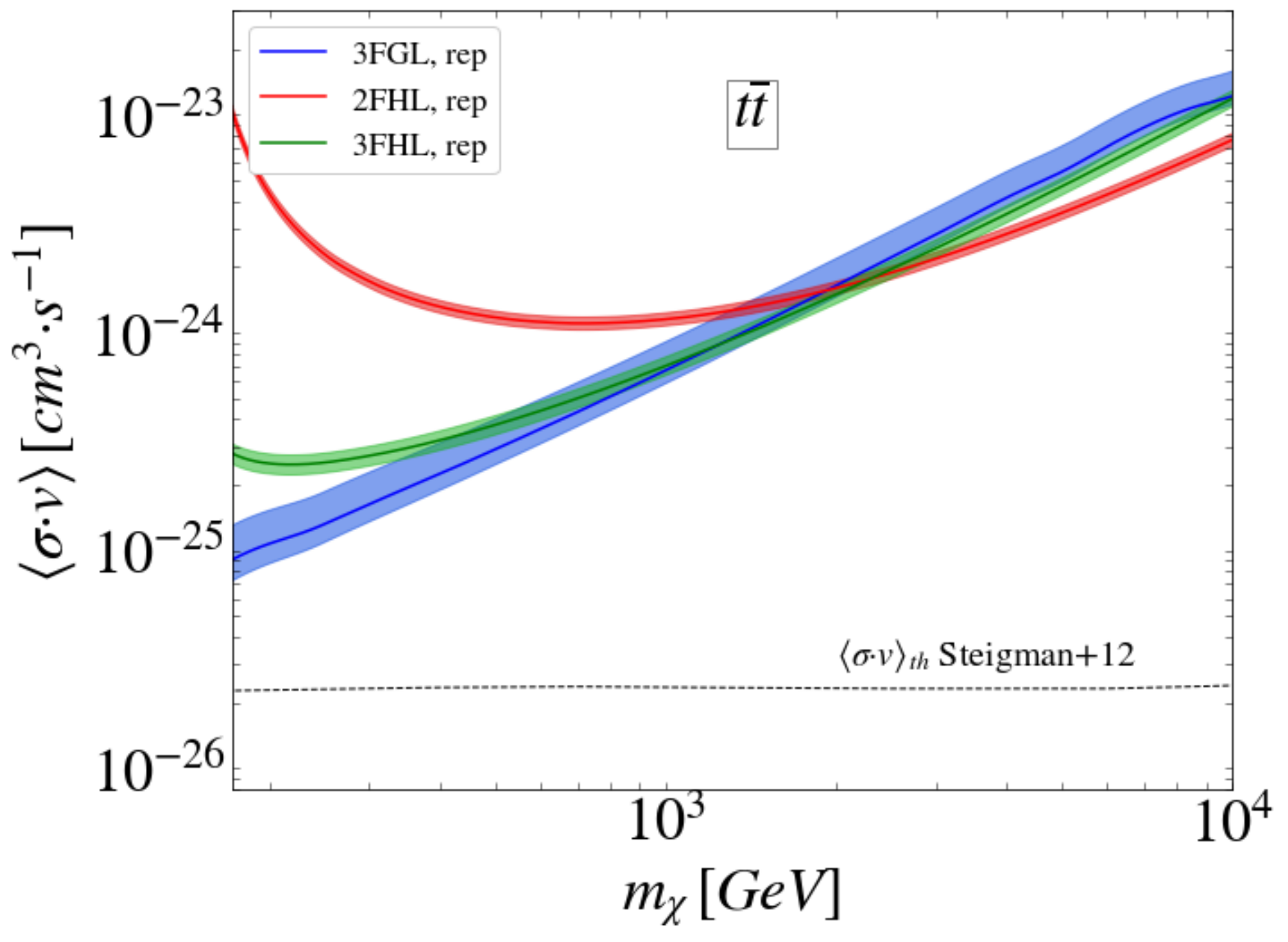}
\caption{Same as Figure \ref{fig:comparison_real} but for $t\overline{t}$ annihilation channel.}
\label{fig:constraints_real_tt}
\end{figure}

\begin{figure}[!ht]
\centering
\includegraphics[height=8cm]{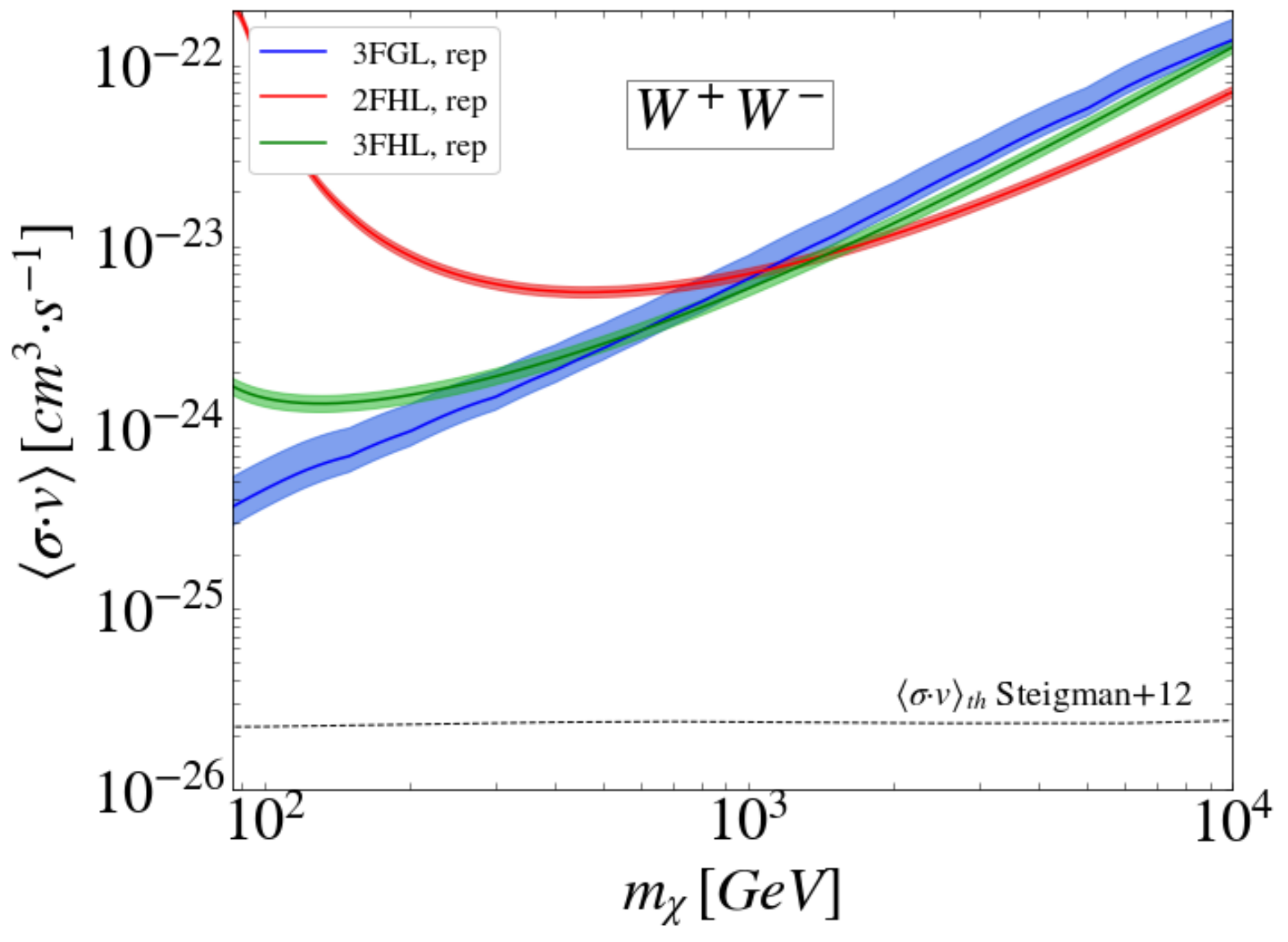}
\caption{Same as Figure \ref{fig:comparison_real} but for $W^+W^-$ annihilation channel.}
\label{fig:constraints_real_ww}
\end{figure}

\begin{figure}[!ht]
\centering
\includegraphics[height=8cm]{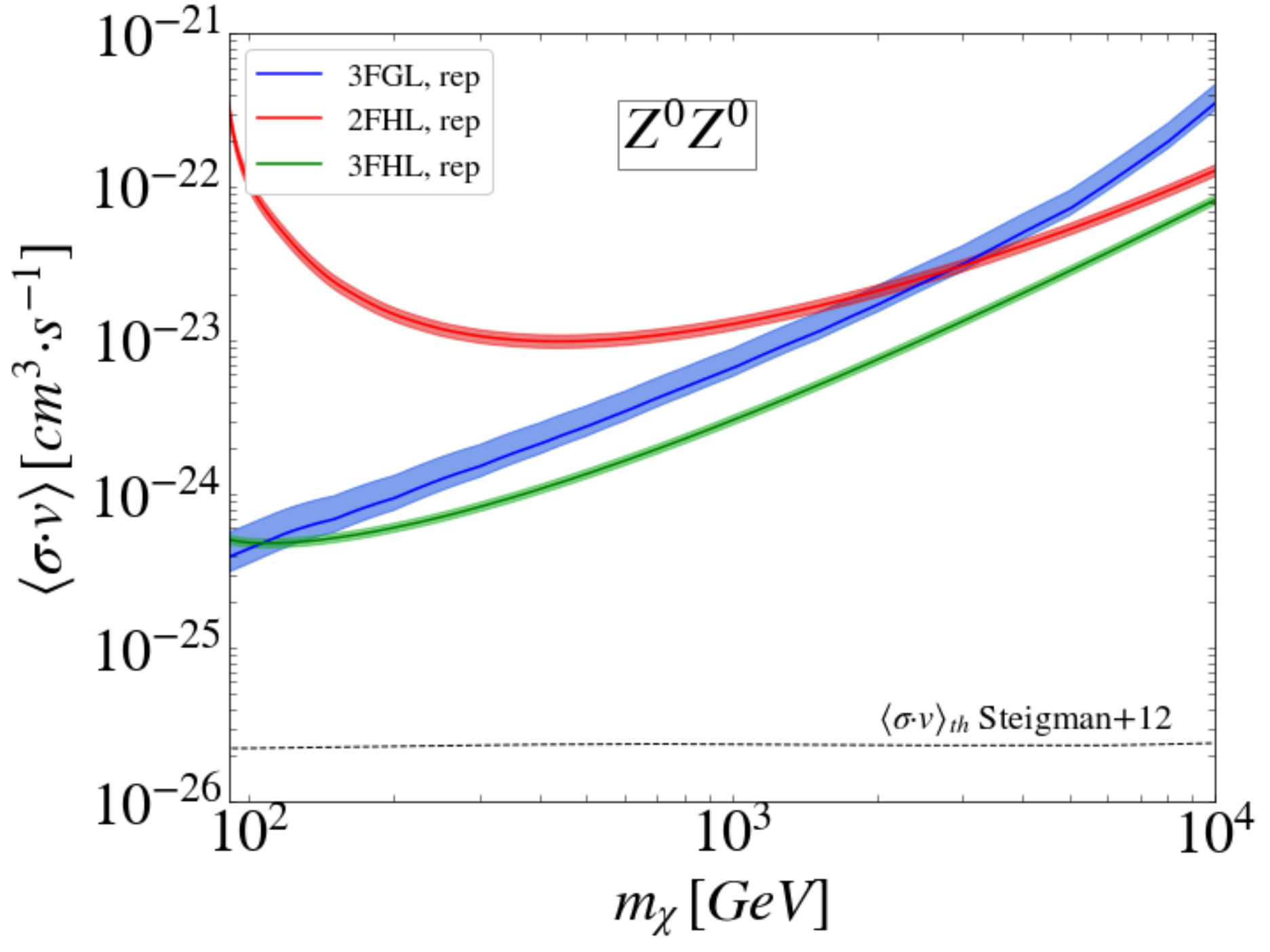}
\caption{Same as Figure \ref{fig:comparison_real} but for $Z^0Z^0$ annihilation channel.}
\label{fig:constraints_real_zz}
\end{figure}

\begin{figure}[!ht]
\centering
\includegraphics[height=8cm]{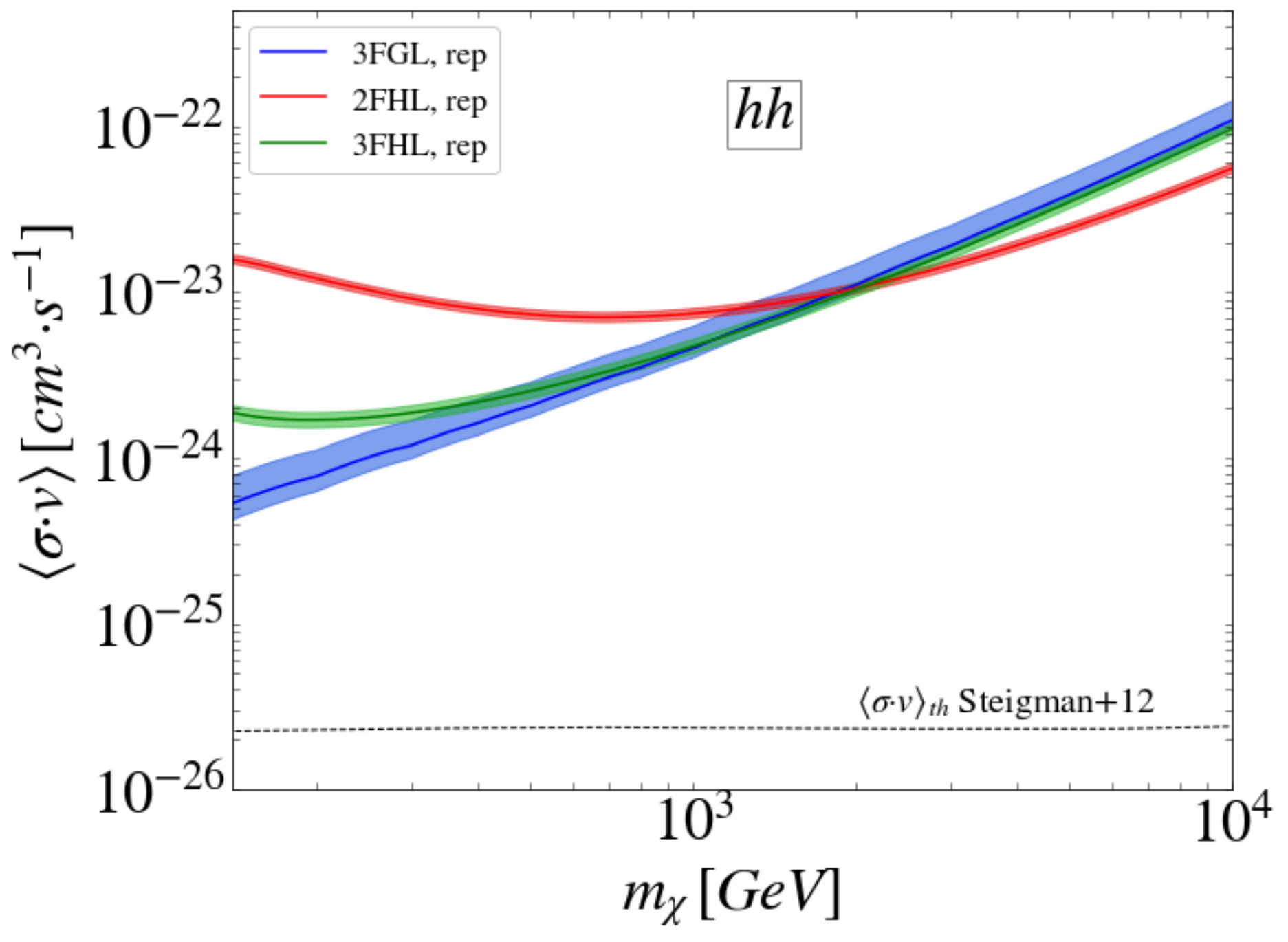}
\caption{Same as Figure \ref{fig:comparison_real} but for $hh$ annihilation channel.}
\label{fig:constraints_real_hh}
\end{figure}

\begin{figure}[!ht]
\centering
\includegraphics[height=8cm]{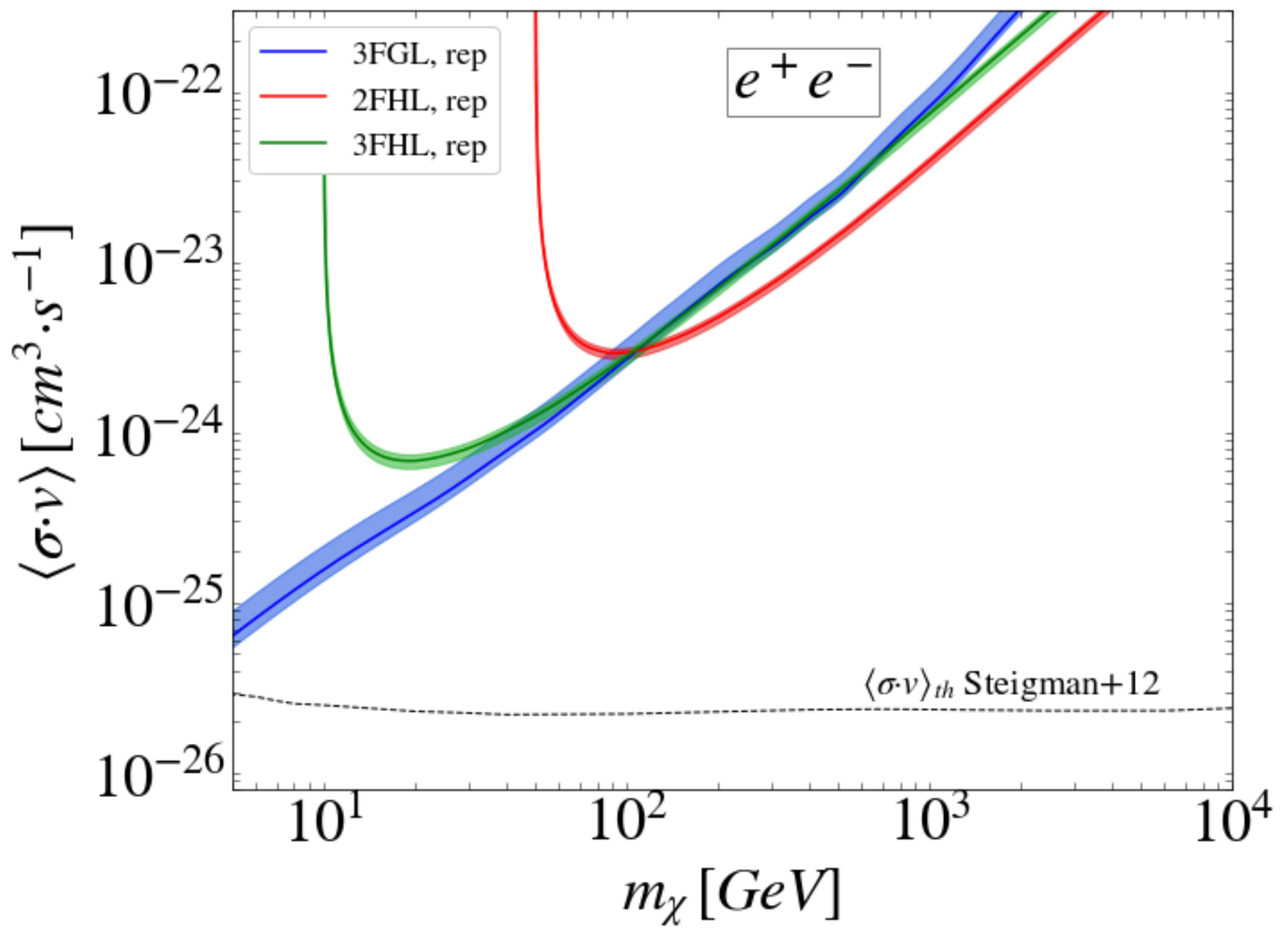}
\caption{Same as Figure \ref{fig:comparison_real} but for $e^+e^-$ annihilation channel.}
\label{fig:constraints_real_ee}
\end{figure}

\begin{figure}[!ht]
\centering
\includegraphics[height=8cm]{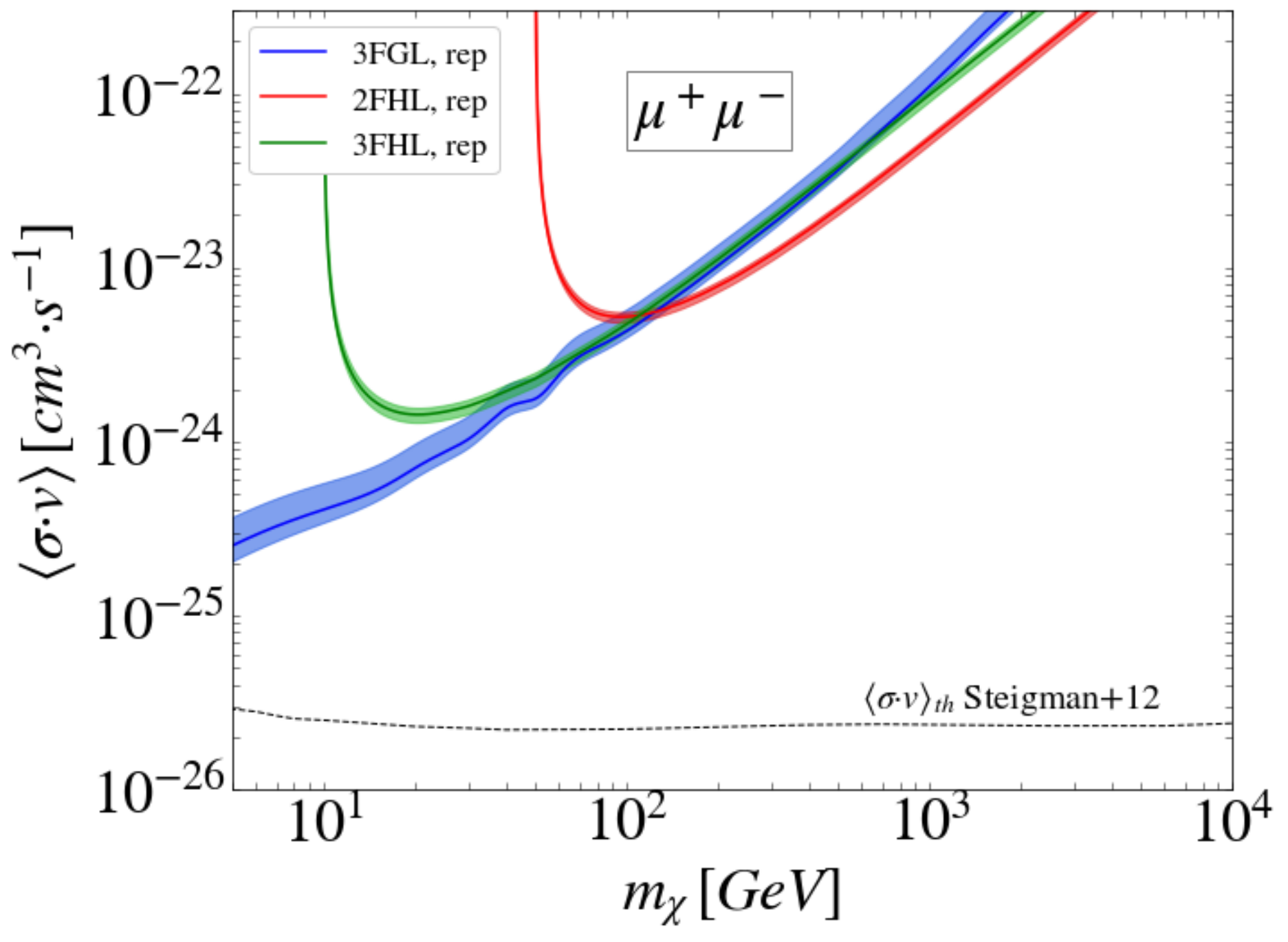}
\caption{Same as Figure \ref{fig:comparison_real} but for $\mu^+\mu^-$ annihilation channel.}
\label{fig:constraints_real_mumu}
\end{figure}

In this Appendix, we provide the DM constraints for various channels, namely $c\bar{c}$ (Figure \ref{fig:constraints_real_cc}), $t\bar{t}$ (Figure \ref{fig:constraints_real_tt}), $W^+W^-$ (Figure \ref{fig:constraints_real_ww}), $Z^0Z^0$ (Figure \ref{fig:constraints_real_zz}), $hh$ (Figure \ref{fig:constraints_real_hh}), $e^+e^-$ (Figure \ref{fig:constraints_real_ee}), $\mu^+\mu^-$ (Figure \ref{fig:constraints_real_mumu}), in addition to the $b\bar{b}$ and $\tau^+\tau^-$ channels, which were shown in Figure \ref{fig:comparison_real}. We note that all the plots have the VL-II low-mass subhalo repopulation implemented.

\bibliographystyle{JHEP.bst}
\bibliography{References.bib}

\end{document}